\documentclass[14pt]{article} % Computer Modern font calls
\usepackage{amsmath}
\usepackage{ifpdf}
\usepackage{hyperref}
\usepackage{amsmath}
\usepackage{graphicx}

\newcommand\mysection{\setcounter{equation}{0}\section}

\def\baeq{\begin{appeq}}     \def\eaeq{\end{appeq}}
\def\baeeq{\begin{appeeq}}   \def\eaeeq{\end{appeeq}}
\newenvironment{appeq}{\beq}{\eeq}
\newenvironment{appeeq}{\beeq}{\eeeq}

\newcounter{Ahran}

\renewcommand{\theequation}{\thesection.\arabic{equation}}
\newcounter{hran}

\def\kps{\relax\ifmmode{{k_\perp^2}}\else{$k_\perp^2${ }}\fi}

\def\stot{\sigma_{\mbox{\scriptsize tot}}}

\def\ben{\begin{enumerate}}  \def\een{\end{enumerate}}
\def\bit{\begin{itemize}}    \def\eit{\end{itemize}}
\def\beq{\begin{equation}}   \def\eeq{\end{equation}}
    
\def\beeq{\begin{eqnarray}}  \def\eeeq{\end{eqnarray}}
\def\bq{\begin{quote}}       \def\eq{\end{quote}}

\def\eqref#1{(\ref{#1})}

\newskip\humongous \humongous=0pt plus 1000pt minus 1000pt
\def\caja{\mathsurround=0pt}

\newif\ifdtup

\def\eqal2#1{\,\vcenter{\openup1\jot
\caja   \ialign{\strut \hfil$\displaystyle{##}$&\hfil$
\displaystyle{{}##}$\hfil &$
\displaystyle{{}##}$\hfil\crcr#1\crcr}}\,}

\def\eV{{\rm e\kern-0.12em V}}  \def\GeV{{\rm G}\eV} \def\TeV{{\rm T}\eV}
\def\MeV{{\rm M}\eV}

\def\be{\relax\ifmmode\beta\else{$\beta${ }}\fi}

\def\LQCD{\Lambda_{\mbox{\scriptsize QCD}}}

\def\cO#1{{\cal{O}}\!\left(#1\right)}

\def\al{\alpha}
\def\as{\al_s}
\def\abs#1{\left|#1\right|}
\def\half{{\textstyle \frac12}}
\def\threehalf{{\textstyle {\frac32}}}
\def\third{{\textstyle {\frac13}}}
\def\twothird{{\textstyle {\frac23}}} 
\def\fourthird{{\textstyle {\frac43}}}
\def\quart{{\textstyle {\frac14}}}

\def\br{brems\-strah\-lung}

\def\ee{\relax\ifmmode{e^+e^-}\else{${e^+e^-}${ }}\fi}
\def\qq{\relax\ifmmode{q\overline{q}}\else{$q\overline{q}${ }}\fi}
\def\cR{{\cal{R}}}
\def\cF{{\cal{F}}}

\def\lrang#1{\left\langle #1 \right\rangle}
\def\kp{\relax\ifmmode{k_\perp}\else{$k_\perp${ }}\fi}

\def\la{\mathrel{\mathchoice {\vcenter{\offinterlineskip\halign{\hfil
$\displaystyle##$\hfil\cr<\cr\sim\cr}}}
{\vcenter{\offinterlineskip\halign{\hfil$\textstyle##$\hfil\cr
<\cr\sim\cr}}}
{\vcenter{\offinterlineskip\halign{\hfil$\scriptstyle##$\hfil\cr
<\cr\sim\cr}}}
{\vcenter{\offinterlineskip\halign{\hfil$\scriptscriptstyle##$\hfil\cr
<\cr\sim\cr}}}}}
\def\ga{\mathrel{\mathchoice {\vcenter{\offinterlineskip\halign{\hfil
$\displaystyle##$\hfil\cr>\cr\sim\cr}}}
{\vcenter{\offinterlineskip\halign{\hfil$\textstyle##$\hfil\cr
>\cr\sim\cr}}}
{\vcenter{\offinterlineskip\halign{\hfil$\scriptstyle##$\hfil\cr
>\cr\sim\cr}}}
{\vcenter{\offinterlineskip\halign{\hfil$\scriptscriptstyle##$\hfil\cr
>\cr\sim\cr}}}}}

\def\lrang#1{\left\langle #1 \right\rangle}

\def\ib#1#2#3{{\em ibid.}~\underline{#1} (#3) #2}
\def\np#1#2#3{{\em Nucl.Phys.}~\underline{B#1} (#3) #2}
\def\pl#1#2#3{{\em Phys.Lett.}~\underline{#1B} (#3) #2}
\def\pr#1#2#3{{\em Phys.Rev.}~\underline{D#1} (#3) #2} 
\def\prc#1#2#3{{\em Phys.Rev.}~\underline{C#1} (#3) #2}
\def\prep#1#2#3{{\em Phys.Rep.}~\underline{#1} (#3) #2}
\def\prl#1#2#3{{\em Phys.Rev.Lett.}~\underline{#1} (#3) #2}

\def\sjnp#1#2#3{{\em Sov.J.Nucl.Phys.}~\underline{#1} (#3) #2}
\def\spj#1#2#3{{\em Sov.Phys.JETP}\/~\underline{#1} (#3) #2}

\def\zp#1#2#3{{\em Zeit.Phys.}~\underline{C#1} (#3) #2}

\def\epj#1#2#3{{\em Eur. Phys.J.}\/ {\underline {C#1}} (#3) #2}
\def\jhep#1#2#3{{\em JHEP}~{\underline{#1}} (#3) #2}
 \def\cite#1{[\ref{#1}]}
 \def\citd#1#2{[\ref{#1},\ref{#2}]}
 \def\citt#1#2#3{[\ref{#1},\ref{#2},\ref{#3}]}
 
 \def\citm#1#2{[\ref{#1}--\ref{#2}]}

\begin{document}
\begin{titlepage}
\begin{flushright}
% `Explain all that,' said the Mock Turtle. \\
% `No, no! The adventures first,' said the Gryphon in an impatient tone: \\
 % `explanations take such a dreadful time.'
'No, no! The adventures first,  explanations take such a dreadful time' \cite{LC}
% \\ Lewis Carrol. ``Alice's Adventures in Wonderland"
\end{flushright}              
\vspace*{\fill}
\begin{center}
{\Large\bf\boldmath Hadron multiplicity fluctuations in perturbative QCD$^\dagger$}
\end{center}
\par \vskip 5mm
\begin{center}
        {\bf\boldmath Yu.L.\ Dokshitzer$^*$}\\  
        Riga Technical University,  Latvia
   %      \footnote{on leave from         St Petersbourg Nuclear Physics Institute,  Russia}
        \\
        \vskip 0.3 cm
        {\bf B.R.\ Webber}\\
        Cavendish Laboratory, University of Cambridge, UK
        \end{center}
\par \vskip 6mm
\begin{center} {\large \bf Abstract} \end{center}
\begin{quote}
  We examine hadron multiplicity fluctuations in hard processes and confront analytic QCD predictions with the pattern of multiplicity fluctuations observed in \ee annihilation and high-$p_t$ jets produced in $pp$ collisions at the LHC. 
  Special emphasis is placed on high-multiplicity fluctuations in jets.  
 Selecting events with hadronic multiplicity exceeding the average value by a factor of 3 or more in various processes has been a source of conundrums for many years.    \\
 We discuss two recent high-multiplicity puzzles and attempt to reveal their common origin. 
 
 \vfill
  
\end{quote}
\vspace*{\fill}

 \hrule
 \bigskip
 
$^\dagger$ We dedicate this paper to the memory of our late collaborator and friend Pino Marchesini, in the hope that he would have found it sufficiently amusing and provocative.
 
$^*$ on leave from St Petersburg Nuclear Physics Institute,  Russia

\end{titlepage}

\tableofcontents

\mysection{Introduction}
\begin{flushright}
`Begin at the beginning and go on till you come to the end: then stop'
\cite{LC}
\end{flushright}              

A consistent QCD description of the pattern of multiplicity fluctuations in hard processes is lacking.  
In this paper, we extend an improved perturbative QCD analysis of multiplicity fluctuations in a gluon jet \cite{D93} to arbitrary ensembles of jets with a commensurate hardness scale, such as \ee\ annihilation into hadrons (viewed as 2 quark jets), hadronic Higgs or $\Upsilon$ decay (2 or 3 gluon jets), multiplicity distribution in one hemisphere of \ee\ annihilation and in high-$p_t$
LHC jet studies (one quark or, one day, one gluon), {\em etc.}

In terms of QCD predictions being compared with experiment, we do not aim at fitting the data optimally.
 Instead, we demonstrate that the QCD dynamics of parton multiplication is compatible with observation and provides reasonable quantitative agreement.  
 
In Section \ref{Sec:MF} necessary theoretical formulae are presented.

Section \ref{Sec:Ph} is devoted to phenomenology. 

In Section \ref{Sec:CA} we discuss two strange phenomena in high-multiplicity fluctuations that may have a common origin, namely an unexpected flattening of the high-multiplicity tail of ATLAS jets \cite{ATLAS-2019} and a surprising behaviour of long-range ellipticity recently observed by CMS \cite{CMS}.

\mysection{Multiplicity fluctuations\label{Sec:MF}}
\begin{flushright}
     `It would be so nice if something made sense for a change' 
\cite{LC}
\end{flushright}

\subsection{Polyakov -- KNO}

In 1970 A.M.~Polyakov \cite{AMPolyakov} considered multiparticle production in $e^+e^-$ annihilation in the framework of an abstract QFT.  
By assuming {\em scale similarity}\/ of microscopic dynamics he arrived at what we know now as  KNO scaling~\cite{KNO}:
\begin{subequations}\label{PKNO}
\beq\label{PKNO1}
   P_n(Q) \equiv \frac{\sigma_n}{\stot} = [N(Q)]^{-1} \Psi\big(\nu\big),  \quad \nu \equiv \frac{n}{N(Q)} ,
\eeq
with $N(Q)$  the mean particle multiplicity
\beq
 N(Q) = \sum_{n=1}^\infty nP_n(Q)  \>\equiv \lrang{n}\!(Q) .
\eeq
\end{subequations}
In the QCD framework, the validity of the asymptotic scaling regime \eqref{PKNO} was envisaged by Bassetto, Ciafaloni and Marchesini as a consequence of the cascading nature of parton multiplication in \cite{BCM}.  
An analytic solution of the scaling distribution in the limit $\as\to 0$ was obtained later in \cite{DFK} (see \cite{Book} for a review).

Polyakov arrived at the scaling law by employing the general properties of unitarity and dispersion relations.   
More than that: he also predicted that the probability of high-multiplicity fluctuations should fall at $\nu\gg1$ faster than exponentially, 
namely as 
\begin{subequations}\label{eq:gammamu}
\beq\label{eq:mudef}
\Psi(\nu) \>\propto \> e^{-\nu^\mu}  , \qquad \mu \>=\> \frac1{1\!-\!\gamma},
\eeq   
where the exponent $\mu$ is related to the anomalous dimension of mean multiplicity energy growth 
\beq\label{eq:gammadef}
\gamma \>\equiv\> \frac{d}{d\ln Q} \ln N(Q).
\eeq
\end{subequations}
Polyakov's amusing prediction remained virtually unnoticed.  And understandably so: back in 1970  % -- before QCD  --  
hardly anyone believed in the possibility of applying QFT dynamics to the physics of hadrons. 

In Polyakov's scale-invariant model $\gamma$ was constant. 
In QCD, scale invariance is broken by the running of the coupling, and $\gamma=\gamma(\as(Q))$ implying logarithmic scaling violation.

\medskip
Hadron--hadron interactions fall into two categories: large-cross-section,  limited-transverse-momenta phenomena (total,  diffractive,  minimum bias inelastic ``soft'' processes) and small-cross-section,  large-transverse-momenta ``hard'' collisions. 

The physics of the two is essentially different.  
In particular,  the validity of phenomenological KNO scaling in hadron--hadron interactions was always a mystery.  
The tension only eased when it became clear that the ``soft" KNO scaling vanishes with increasing collision energy \cite{CMSsoft}. 

In hard interactions, the situation is directly opposite.  Here, the scaling law \eqref{PKNO} {\em has to hold},  at least at a sufficiently large hardness scale $Q$.

In order to stress the difference, and in tribute to Polyakov's precocious discovery,  we will hereafter refer to the multiplicity pattern \eqref{PKNO} {\em in hard processes}\/ as {\bf P-KNO scaling}.

\medskip
It is important to stress that the scale dependence of the perturbative QCD P-KNO distribution is contained in a single quantity, that is, the multiplicity anomalous dimension $\gamma(\as(Q))$.  

\subsection{Multiplicity anomalous dimension}
There are different ways to determine the value and the momentum dependence of $\gamma$. 
For example, one could read it directly from the data on the mean multiplicity of hadrons in \ee annihilation. 
We chose a more ``theoretical'' approach: to use the two-loop QCD expression \cite{BW84}
\beq\label{eq:2lcoup}
\gamma(\as) \>=\> \sqrt{2N_c\frac{\as}{\pi}} - \left( \frac{\beta_0}{4} + \frac{10n_f}{3N_c^2}\right) \frac{\as}{2\pi}, \quad  \beta_0 = \frac{11}3N_c-\frac23 n_f . %  = 11 - \frac23 n_f .
\eeq
We remind the reader that the appearance of $\sqrt\as$ as an expansion parameter is typical for QCD observables that are sensitive to soft-gluon radiation such as particle multiplicity.
Gluons with energy fractions $z\ll1$ bring about large corrections $\propto \as\ln^2z$.
Resummation of such terms in all orders results in transmutation of the perturbative series expansion; see, e.g., \citd{Book}{ESW} and references therein.

The coupling constant in \eqref{eq:2lcoup} refers to the ``physical" coupling scheme also known as the ``bremsstrahlung" or CMW scheme \cite{CMW}, see also \cite{DKTSpec} and references therein.  

\medskip

\noindent
\begin{minipage}{0.35\textwidth}
In order to make our endeavour more challenging and in the hope of being more convincing, we chose not to play with the value of $\as$.   

\bigskip

We accept the world average
\[
\al_{\overline{\mbox{\scriptsize\rm MS}}}(M_Z)=0.119
\]
which translates into 
\[
\as(M_Z)\simeq 0.127
\]
for the CMW coupling ($n_f=5$). 

\medskip

Details of our coupling model are discussed in Appendix \ref{AppCoupl}.
\end{minipage}
\hfill
\begin{minipage}{0.65\textwidth}
  \centering%
 \includegraphics[width=0.95\textwidth]{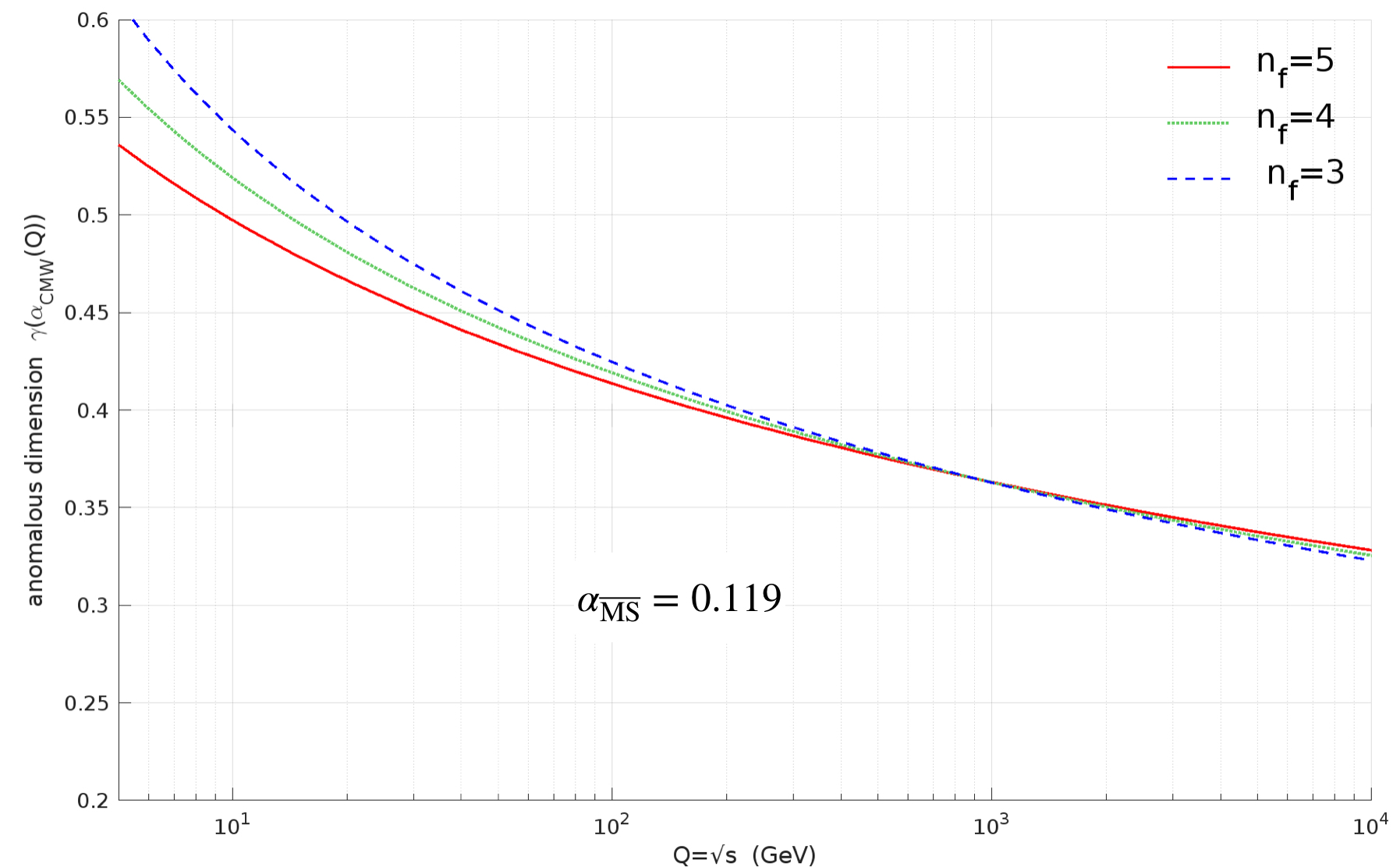} 
\stepcounter{figure}
Figure \thefigure: Anomalous dimension of mean multiplicity 
\end{minipage}

\subsection{DLA}
The asymptotic P-KNO function \cite{DFK} was derived in perturbative QCD in the leading double-logarithmic approximation (DLA).  The
normalized moments of the multiplicity distribution 
\beq\label{eq:gkDLA}
 g_{k}(Q) \equiv \frac{\lrang{n^k}(Q)}{[N(Q)]^k} = \int_0^\infty d\nu\, \nu^k\,  \Psi(\nu, Q)
\eeq
satisfy a recurrence relation following from the evolution equation for cascading multiplication of gluons and can be calculated recursively.  
In the $Q\to\infty$ limit $(\as,\,\gamma\to 0)$ they cease to depend on the hardness scale $Q$, thus satisfying P-KNO scaling.  
For large rank $k$ the moments can be found analytically,  
\begin{subequations}\label{eq:ek}
\beq\label{eq:ekas}
   g_{k}^{\mbox{\scriptsize DLA}} \simeq \frac{2 k!}{C^k} \left( k + \frac1{3k}\right) \qquad (k\gg 1).
\eeq
Here $C$ is a number determined by the equation \citd{DFK}{Book}
\beq\label{eq:Cdef}
\ln C = \int_1^\infty \frac{dx}{x} \left( \frac1{\sqrt{2(x-1-\ln x)}} - \frac{1}{x-1} \right) \simeq 0.937, \quad C\simeq 2.552.
\eeq 
\end{subequations}
Properly processed,  \eqref{eq:ek} results in exponential falloff of the P-KNO distribution
\beq\label{eq:DLAtail}
  \Psi(\nu) \simeq 2C\left(C\nu-1+\frac{1}{3C\nu}\right) \cdot e^{-C\nu} ,\quad \nu \gg 1.
\eeq
The accuracy of  \eqref{eq:ek} and of the corresponding expression for the tail 
\eqref{eq:DLAtail} was estimated in \cite{DFK} as
\beq\label{eq:DLAacc}
\left[ 1 +\cO{\frac{\ln(C\nu)}{(C\nu)^4}} \right].
\eeq
Surprisingly, the asymptotic formula for multiplicity moments \eqref{eq:ek} also proves to agree quite well with the exact values
of low rank moments (see Appendix \ref{AppDLA}).
\medskip

The DLA solution proved to be too broad and fatally unrealistic.  
QCD corrections to the multiplicity moments were found to be ``{\em so large that terms of yet higher order are unlikely to be negligible}\/''~\cite{MW84}.  

In next-to-leading order, the lowest few moments were found to be similar to those of a negative binomial distribution~\cite{MW86}.

\subsection{Improved QCD treatment of the P-KNO phenomenon (MDLA)}

\def\Ct{\tilde{C}}

Observables that are driven by soft-gluon multiplication have perturbative series running in powers of the multiplicity anomalous dimension $\gamma=\cO{\sqrt{\as}}$.   

In \cite{CT} the observation was made that specific {\em next-to-next-to-leading order}\/  terms of the series in $\sqrt{\as}$ {\em explode}\/: their magnitude increases linearly with the rank $k$ of the multiplicity moment.  

Expressed in terms of $\gamma$, the perturbative expansion for the moments of the P-KNO distribution $g_k$ can be cast as 
\beq
g_k =g_k ^{\big[\mbox{\scriptsize DLA}\big]}  \>+  g_k^ {\big[\mbox{\scriptsize MLLA}\big]}\cdot  \gamma(\as) \> +  g_k^{\big[\mbox{\scriptsize NMLLA}\big]} \cdot  \gamma^2(\as) \>+  \ldots
\eeq
The {\em second}\/ subleading term in this series possesses an explosive piece,
\beq
    g_k^{\big[\mbox{\scriptsize NMLLA}\big]} \>=\> g_k ^{\big[\mbox{\scriptsize DLA}\big]} \left[c\,k^2 \>+\> \cO{k}\right].  
\eeq
Such ``explosiveness'' systematically reproduces itself at higher orders in $\gamma$ 
in the form of a series in $k\gamma$.

In \cite{D93} the origin of such corrections was understood to be due to energy conservation in parton cascading, which is deliberately ignored in the DLA.   
A simplified theoretical model was constructed that permitted one to single out these specific contributions and resum them to all orders in $k\gamma$ for a gluon jet. 

This approximation one might label as the {\em Modified Double Log Approximation}\/ (MDLA) in analogy with MLLA --- the Modified Leading Log Approximation of \citt{AHM}{EAO}{Book}.

In this paper, we generalise MDLA to the case of a quark jet or an arbitrary jet 
ensemble and confront the resulting theoretical formulae with experiment.

\smallskip

In the MDLA treatment, high-rank multiplicity moments $k\gg 1$ acquire a multiplier
\beq\label{eq:gkdef}
   g_k \simeq g_{k}^{\mbox{\scriptsize DLA}}\cdot \frac{[\Gamma(1+\gamma)]^k}{\Gamma(1+k\gamma)}
   = \frac{2}{\Ct^k}\left( k + \frac1{3k}\right) \cdot \frac{\Gamma(1+k)}{\Gamma(1+k\gamma)} ,
      \qquad \Ct \equiv \frac{C}{\Gamma(1+\gamma)}.
\eeq
MDLA is likely to have imperfections. 
In particular, expansion of the MDLA factor in \eqref{eq:gkdef} produces powers of $\pi$ that are not apparent
in the exact NNLO calculation of the low-rank multiplicity moments \cite{Malaza}. 
Nevertheless, we consider it trustworthy in that it takes proper care of an important part of the physics of parton cascades: the energy balance. Since the MDLA resums terms that are enhanced at large $k$, corresponding to large $\nu$, we expect it to apply best to the tail of the P-KNO distribution, {\it i.e.}\ the region beyond the peak.
\smallskip

The ratio of $\Gamma$-functions is analysed in Appendix \ref{AppGam}.
Introducing  
\beeq   \label{eq:Ddef}
   D &\equiv&
   \Ct\, \gamma^\gamma (1\!-\!\gamma)^{1\!-\!\gamma} \>= \>\>
   C \frac{\gamma^\gamma (1\!-\!\gamma)^{1\!-\!\gamma} }{\Gamma(1+\gamma)},
\eeeq
it can be cast in the form
\begin{subequations}\label{eq:chiform}
\beeq\label{eq:gkimp}
   g_k &=& \frac{2}{D^k} \left( k + \frac1{3k}\right)\,\Gamma(1+k(1\!-\!\gamma))\cdot \chi(k) .
\eeeq
% \end{subequations}
The factor $\chi$ in \eqref{eq:gkimp} can be approximated as (see Appendix \ref{AppGam})
\beq\label{eq:chidef}
\chi(k) \approx \sqrt{\frac{1+k}{2\pi(1+\gamma k)(1+(1\!-\!\gamma)k)}}\cdot e \left( 1+\frac1{\gamma k}\right)^{-\gamma k}. 
\eeq
 \end{subequations}
From \eqref{eq:chidef} it is clear that $\chi(k)$ contains no factorial nor exponential growth with $k$. \\
It starts from $\chi(0)\!\approx\!1$ and gradually decreases as $1/\sqrt{k}$
% $({2\pi(1\!-\!\gamma)(1\!+\!\gamma k)})^{-1/2}$ 
when $k\to \infty$.  
\medskip

An alternative to \eqref{eq:chiform} covers factorials, exponents {\em and} powers of $k$:
\begin{subequations}\label{eq:cFform}
\beeq\label{eq:gkcF}
g_k &=& \frac{2}{D^k} \left( k + \frac1{3k}\right)\,{\Gamma (\half + (1\!-\!\gamma) k)} \cdot \cF(k).
\eeeq
Here $\cF$ is another slowly varying prefactor analogous to $\chi$ \eqref{eq:chidef}. Approximately, (see Appendix \ref{AppPsi} for a more accurate formula)
\beeq\label{eq:cFdef}
\cF (k,\gamma) &\simeq &   \frac1{\sqrt{2\pi}}   \sqrt{\frac{1+k}{1 +\gamma k} } .
\eeeq
\end{subequations}
This one starts off from $(2\pi)^{-\half}$ and freezes when $\gamma k\gg 1$.
The transition between the two regimes for each prefactor is governed by the value of the product $\gamma k$. 
It is that very key parameter that formed the basis for the MDLA approach.
%\beq
%\eeq

    \def\ksd{k_{\mbox{\scriptsize sd}}}

The two approximations are identical at large $k$ and serve different purposes. 
\smallskip

The first is better suited for analysis of the interval of moment ranks $\gamma k <1$ and, in particular, for verification of the formal asymptotic limit $\gamma\to 0$
(which we hereafter refer to as the DLA limit).
\smallskip

Given that $\gamma$ in reality is not so small, the second form \eqref{eq:cFform} is more practical.

 \medskip
 
The moments \eqref{eq:gkimp}, \eqref{eq:gkcF} are characteristic of a {\em generalised Gamma distribution}\/ and 
result in the following alternative expressions for the P-KNO tail (see Appendix~\ref{AppPsi} for the derivation):
% The reconstructed spectra look as follows:
%
\begin{subequations}\label{eq:Psi_fin}
\beeq\label{eq:Psi_chi}
 \Psi(\nu) &=&  \frac{2\mu^2}{\nu } \, \left( [D\nu]^{2\mu} -[D\nu]^{\mu} +\frac1{3\mu^2}\right)  e^{-[D\nu]^\mu}   \cdot  \chi(\ksd) , \\
 \label{eq:Psi_cF}
\Psi(\nu) &=&   \frac{ 2\mu^2}{\nu } \left( [D\nu]^{\threehalf\mu} 
- \frac12[D\nu]^{\half\mu} + \frac{1}{3\mu} [D\nu]^{-\half\mu}  \right) e^{-[D\nu]^\mu} \cdot \cF(\ksd) ,
 \eeeq 
\end{subequations}
where $\mu$ and $D$ are given in terms of $\gamma$ by Eqs.~\eqref{eq:mudef} and \eqref{eq:Ddef}, respectively.
In the  limit $\gamma\to 0$ we have 
$\mu\!=\!1$, $D\!=\!C$, $\ksd = C\nu\!-\!2$,  $\chi(k)\!=\!1$, and \eqref{eq:Psi_chi} turns into 
the asymptotic DLA expression \eqref{eq:DLAtail}.

\smallskip

\stepcounter{figure}
\noindent
\begin{minipage}{0.6\textwidth}
  \centering%
  \includegraphics[width=0.95\textwidth]{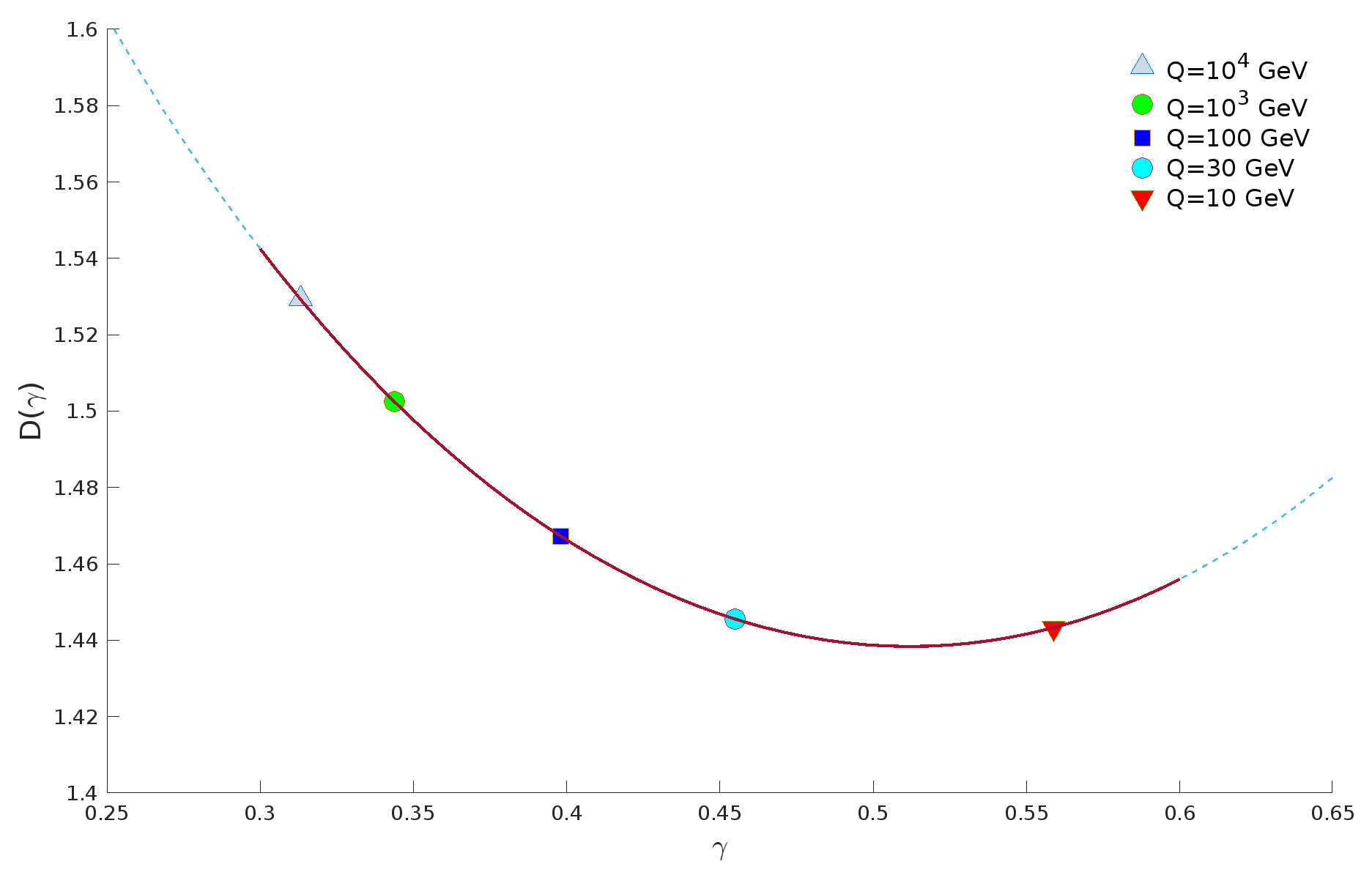} 

Figure \thefigure: $D(\gamma)$ with typical hardness scales marked
\end{minipage}
\hfill
\begin{minipage}{0.4\textwidth}
Fig. \thefigure\ displays the function $D(\gamma)$ defined by \eqref{eq:Ddef}.
It shows that on realistic hardness scales $D$ stays near 1.5.

\smallskip
In the course of reconstruction of the distribution $\Psi(\nu)$ % \eqref{eq:Psi_fin}
from its moments, we evaluate the prefactors $\chi(k), \, \cF$ at a characteristic point $k=\ksd(\nu)$. 
The value of 
$\ksd(\nu)$ that corresponds to a given multiplicity ratio $\nu$ in a gluon jet is derived in Appendix \ref{AppPsi} 
with $1/k^2$ accuracy as
\beq\label{eq:ksddef}
 \ksd(\nu) \equiv \mu\big( [D(\gamma)\nu]^{\mu} -2 \left[\threehalf\right]\big).
\eeq
The different numbers to be subtracted depend on the choice between Eqs.~\eqref{eq:Psi_chi} and \eqref{eq:Psi_cF}, respectively.

\end{minipage}

\medskip

As discussed in detail in Appendix \ref{App:charmom}, for practical purposes we prefer to use the simplified version 
\beq
\ksd(\nu,\gamma) \>=\> \mu(\gamma) \big[D(\gamma)\,\nu\big]^{\mu(\gamma)},
\eeq
which adequately describes the tail of the multiplicity distribution and, at the same time, permits us to descend to the maximum at $\nu\sim 1$.  All the phenomenology in the following Sections was done using Eq.~\eqref{eq:Psi_cF} with this simplification.

\medskip

\medskip

Experimental information on multiplicity fluctuations inside identified {\em gluon}\/ jets is quite poor. 
In order to apply MDLA wisdom to the data we first need to generalise \eqref{eq:Psi_fin} to describe a quark jet,  2-quark systems (\ee),  etc.

 \subsection{Quark jet and multi-jet ensembles}
To compute the P-KNO distribution for emission from an assembly of $m_q$ quarks and/or antiquarks and $m_g$ gluons, with comparable hardness, from that of a single hard gluon, (2.11), one must introduce a  ``source strength" parameter
\beq\label{eq:rhodef}
\rho = m_q \rho_q + m_g
\eeq
where $\rho_q$ is the mean multiplicity for a quark jet, relative to that of a gluon jet.

The derivation of $\Psi^{(\rho)}$ for $\rho\neq 1$ then goes through three steps:

\begin{subequations}
1. Laplace transform the gluon P-KNO function \eqref{eq:Psi_fin} onto the complex $\beta$-plane
\beq\label{eq:Mto}
 \Phi(\beta) = \int_0^\infty d\nu\, \Psi(\nu)\, e^{-\beta \nu} .
\eeq

2.  Construct the  Laplace image% of the general 
 \beq\label{eq:PhiR}
   \Phi^{(\rho)} (\beta) \>\simeq\>  \left[ \Phi\left(\frac{\beta}{\rho}\right) \right]^{\rho} . 
 \eeq
 
 3. Evaluate the inverse Laplace transform 
\beq\label{eq:Mfrom}
   \Psi^{(\rho)}(\nu)  = \int_{\cal{C}} \frac{d\beta}{2\pi i} \,   \Phi^{(\rho)} (\beta)\, e^{\beta \nu} . 
\eeq 
 \end{subequations}
 As far as the tail of the multiplicity distribution is concerned, {\em i.e.}\ for `large' $\nu$, the first and third steps  \eqref{eq:Mto} and  \eqref{eq:Mfrom} can be carried out with the help of the steepest descent and stationary phase methods, see Appendix~\ref{AppRho} for details.

The final result reads
\beeq
\label{eq:PsiRHO}    
  \Psi^{(\rho)}(\nu)  &=&  \,\frac{{\sqrt{ \rho}}} {\nu}\,   
\bigg[\> \nu \Psi(\nu) \> \bigg]^\rho
      \cdot  \left( \frac{ \sqrt{2\pi} \, (1-\gamma)} {\sqrt{\gamma[D\nu]^\mu + 2\left[\threehalf\right](1\!-\!\gamma) }}
        \right)^{\rho-1}  .
 \eeeq
When $\rho=1$ this expression coincides with the original {\em gluon jet}\/ distribution \eqref{eq:Psi_fin}. 

For $\nu\Psi$ in square brackets one can substitute either of representations \eqref{eq:chiform}, \eqref{eq:cFform}, with the corresponding change in the denominator of the Gaussian factor.

In practice, we shall use the latter and compare with data in the range $\nu>1$.

\mysection{P-KNO phenomenology\label{Sec:Ph}}
\begin{flushright}
    `Oh, don't bother me,' said the Duchess; `I never could abide figures!' \cite{LC}
\end{flushright}

Now that we have all the necessary formulas prepared, we can begin to test them phenomenologically. Recall that all aspects of the general MDLA P-KNO distribution depend only on the two parameters $\gamma$ and $\rho$, which can in principle be extracted directly from data on average multiplicities alone.

\medskip

The bulk of high-quality data on multiplicity fluctuations come from \ee annihilation.

 \subsection{Quark jet vs.\ Gluon jet}
 
One routinely identifies the \ee annihilation final state with a \qq ensemble.  

\bigskip

\stepcounter{figure}
\noindent
\begin{minipage}{0.65\textwidth}
   \centering%
  \includegraphics[width=0.95\textwidth]{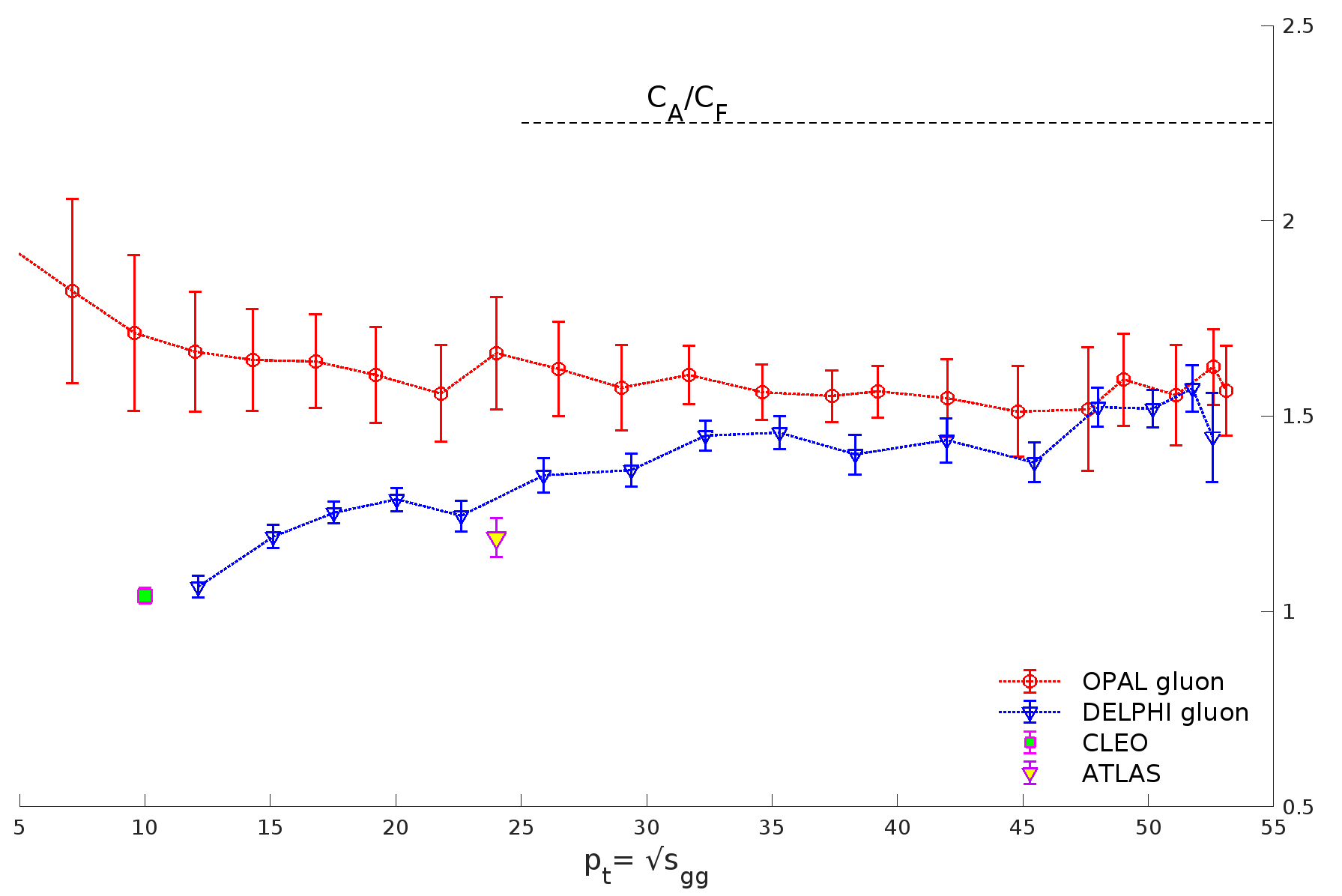}
Figure \thefigure:  Ratio of charged hadron multiplicities in g- and q-initiated jets, \citm{DELPHI2005}{ALEPHgq}
%{OPAL2002}{CLEO}
\end{minipage}
\hfill
\begin{minipage}{0.35\textwidth}
This is a reasonable thing to do with a few percent accuracy.   
\smallskip

Within this logic, the parameter $\rho$ should be fixed in \eqref{eq:PsiRHO} as $\rho=2\rho_q$.

For gluon emission one would expect asymptotically 
\[
\rho_q = \frac{C_F}{C_A} = \frac49 .
\]
However, for charged hadron emission at present energies the gluon-to-quark multiplicity ratio stays stubbornly close to 1.5; see Fig. \thefigure. 
\medskip

Therefore for phenomenology we adopt the value $\rho_q=2/3$.
\end{minipage}

\subsection{Sensitivity to {\protect $\rho$} and scaling violation}

\stepcounter{figure}
Fig. \thefigure\ demonstrates the sensitivity of the distribution \eqref{eq:PsiRHO} to the value of $\rho$ and to the jet hardness scale.

\noindent
\begin{minipage}{0.5\textwidth}
  \centering% 
  \includegraphics[width=0.95\textwidth]{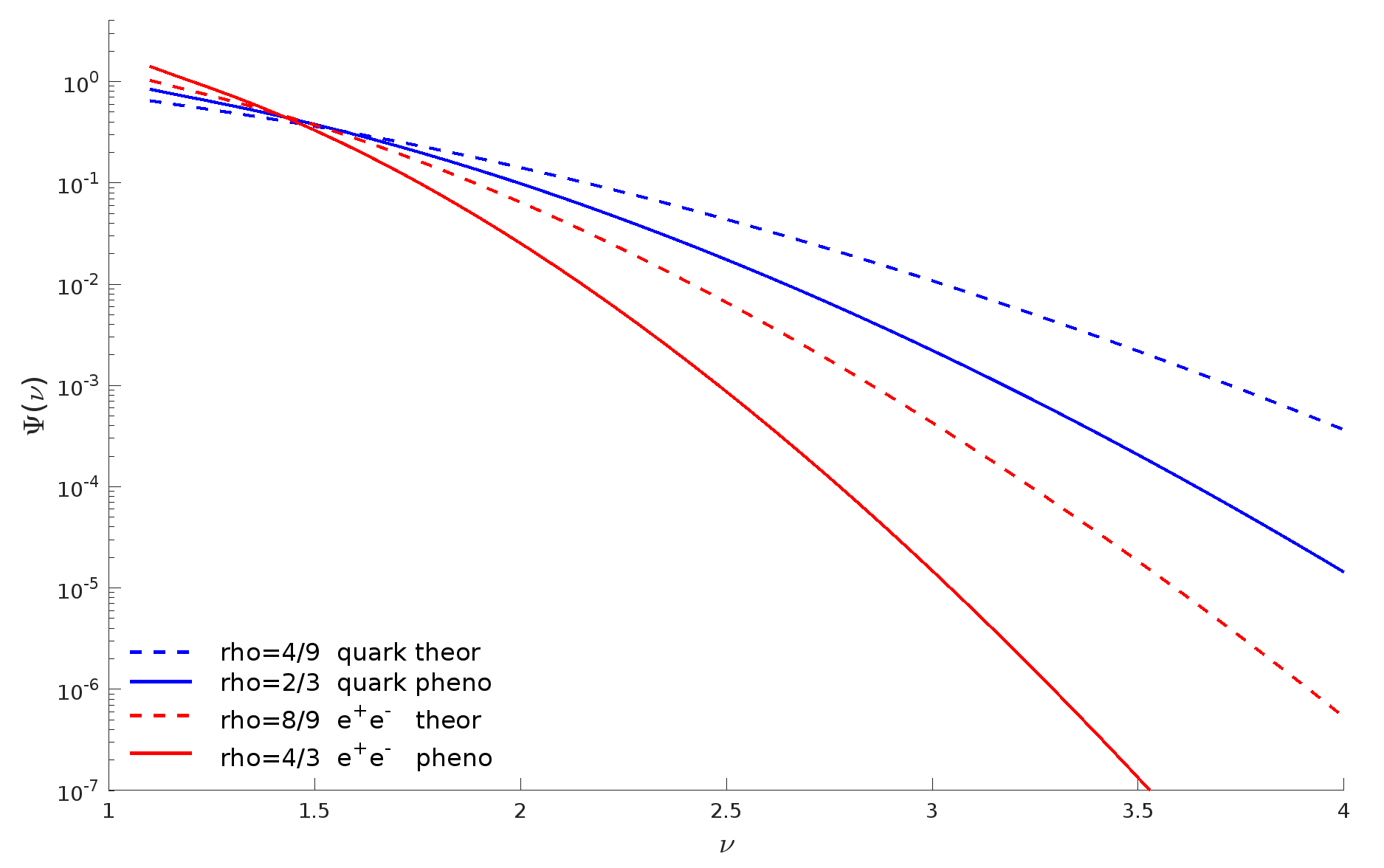}
Figure \thefigure a: Quark jet % P-KNO distribution 
variation with $\rho$.
\end{minipage}
\hfill
\begin{minipage}{0.5\textwidth}
  \centering% 
  \includegraphics[width=0.95\textwidth]{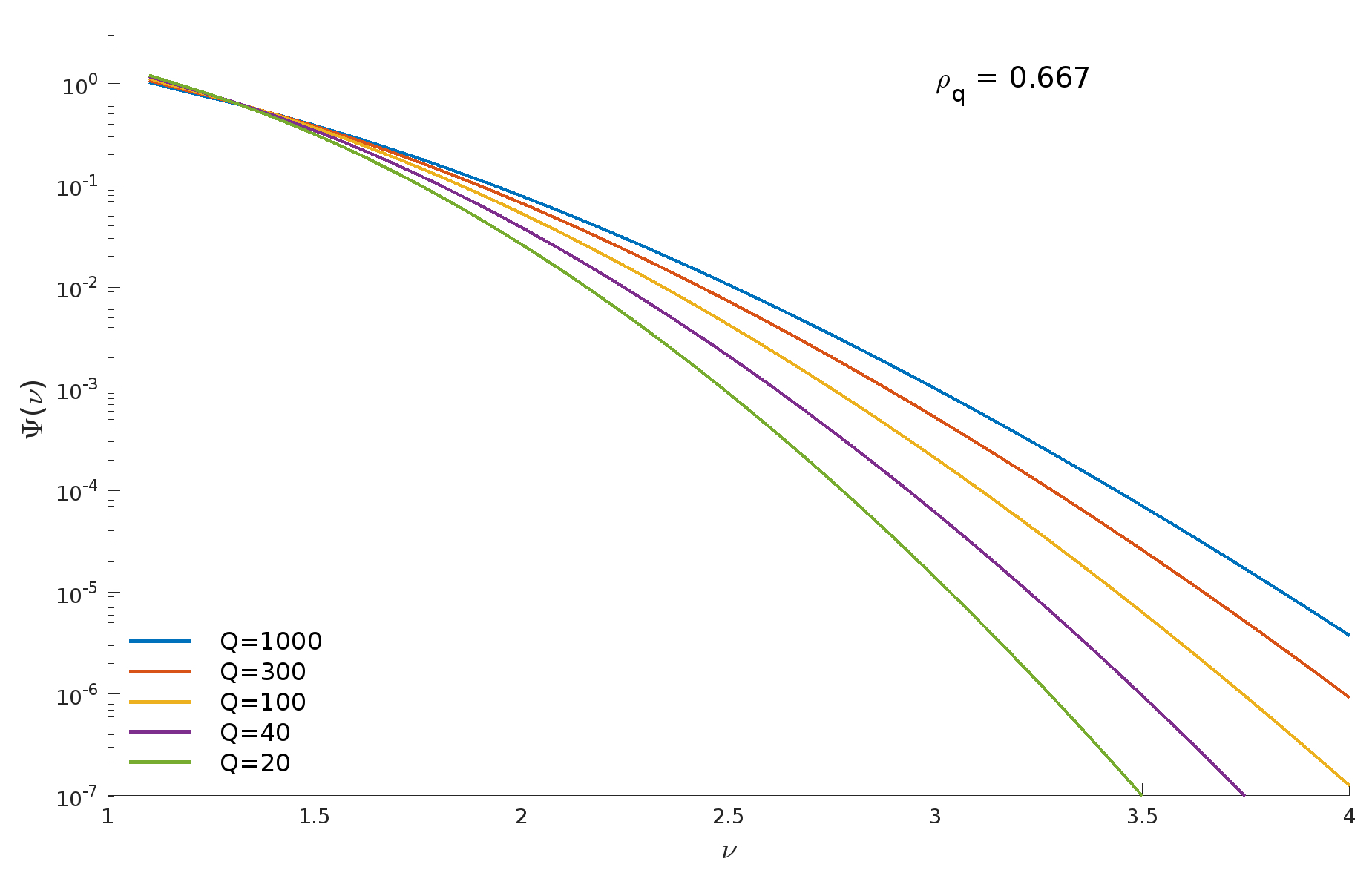}
Figure \thefigure b: Gluon jet variation with hardness $Q$.
\end{minipage}

\vspace{1 cm}
\subsection{P-KNO $e^+e^-$ tail}

\begin{flushright}              
`Well, now that we have seen each other,'
said the Unicorn,\\
`if you'll believe in me, I'll believe in you.' 
\cite{LC}
\end{flushright}              

\begin{figure}[htb]
   \centering%
   \includegraphics[width=0.8\textwidth]{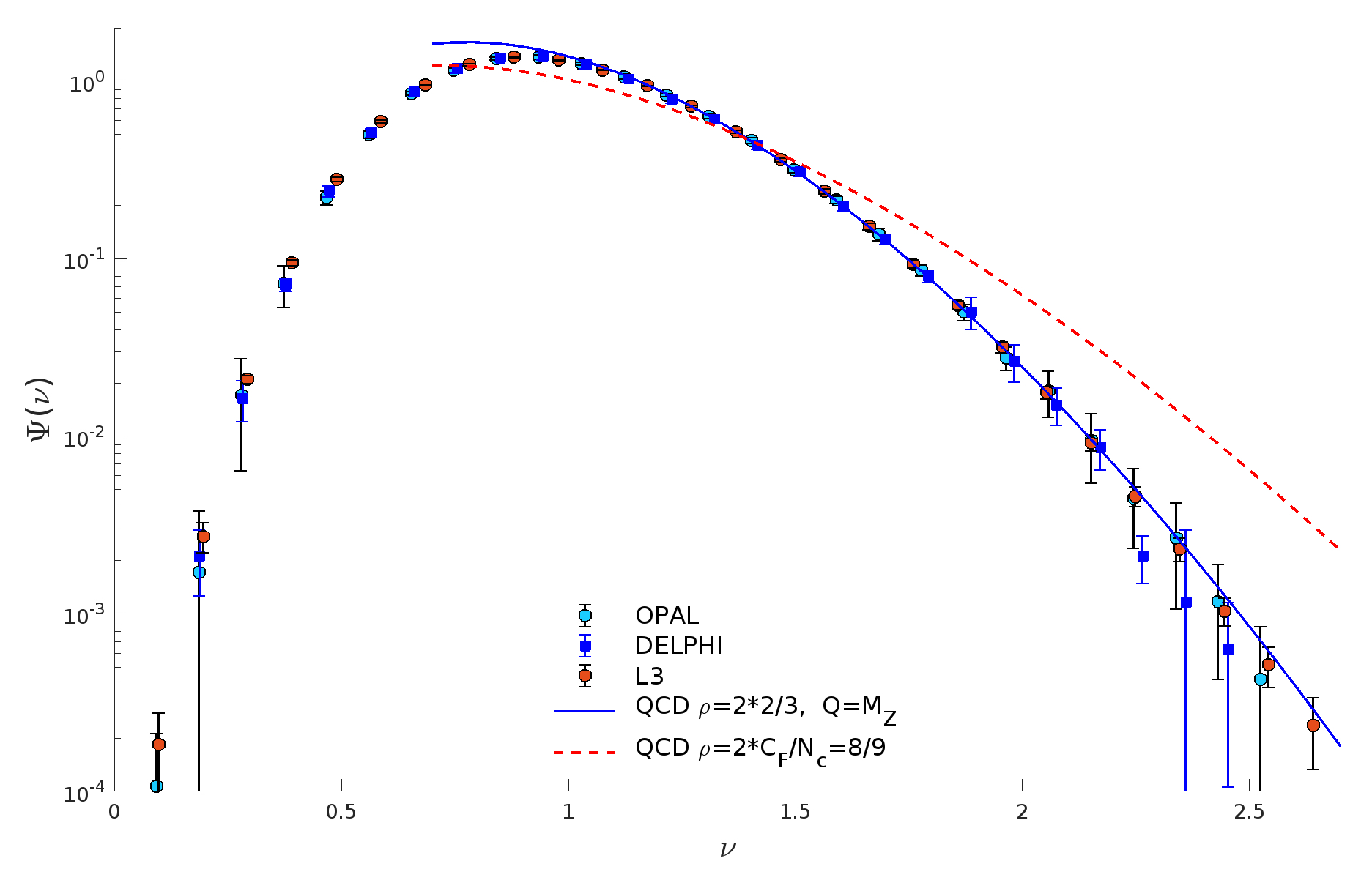}
   \caption{\label{fig:g2qrattest} Multiplicity fluctuations in  \ee annihilation: checking $\rho_q$ }
\end{figure}

In Fig.~\ref{fig:g2qrattest} we compare experimental data from \ee annihilation 
\citt{OPAL1992}{DELPHI}{L3} with theoretical curves corresponding to two prescriptions for the parameter $\rho$. 
This shows what happens when we substitute the phenomenologically motivated $\rho_q=\twothird$ ratio for the $\rho$ parameter in \eqref{eq:PsiRHO}.  
Now, the QCD curve agrees significantly better with the data.

\subsection{Scaling violation (?)}

\begin{figure}[ht]
   \centering%
  \includegraphics[width=0.75\textwidth]{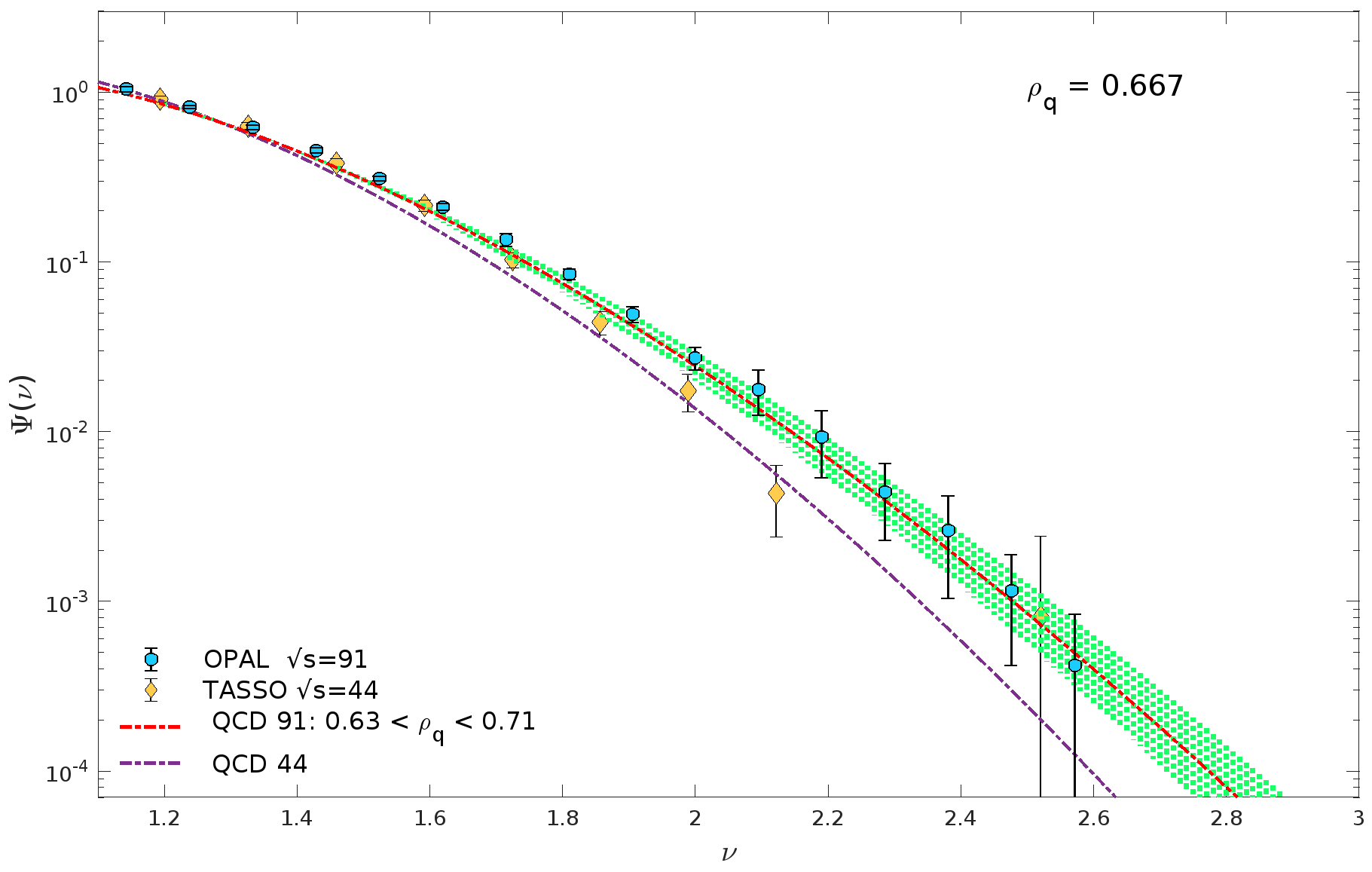} 
\caption{\label{fig:ODT} Hint at scaling violation in the P-KNO tail}
\end{figure}

Being an optimist, one could even see a hint of perturbatively
controlled {\em scaling violation}\/ by comparing the P-KNO tails of
the TASSO (44 \GeV) \cite{TASSO} and OPAL (91 \GeV)  \cite{OPAL1992}
distributions shown in Fig.~\ref{fig:ODT}.

\subsection{Jet hardness }
Meanwhile one subtle point should be mentioned here.  
There is a problem of how to determine the hardness scale for a jet stemming from a hadron--hadron collision.    
This is not as straightforward as for \ee annihilation.  

Hardness is a parameter that measures the state of development of the parton cascade and eventually the multiplicity of final hadrons.  

In terms of quantum physics, one cannot state that a given particle {\em belongs}\/ to a given jet.  
We somewhat arbitrarily assemble the jets by {\em assigning}\/ particles to them.  In doing so we follow the QFT-dictated pattern of collinear enhancements.  
The property of Angular Ordering in time-like parton cascades allows one to embody the total yield of final particles into a sum of a small number of jets with definite (individual geometry-dependent) hardness. 

For an individual jet, the hardness parameter is determined by  {\em the maximal transverse momentum}\/ of particles assigned to this jet:
\beq\label{eq:hardness}
  P_{t,\max} = 2E_{\mbox{\scriptsize jet}} \sin\frac\Theta 2 .
\eeq
Here $E_{\mbox{\scriptsize jet}}$ is the energy of the jet and $\Theta$ its opening angle.  
This definition implies that studying the jet content (particle energy spectra,  multiplicity,  etc.) one has count only particles that fit inside a given angular cone.  

For example, in $\ee\to\qq$ one sets in \eqref{eq:hardness} the opening angle $\Theta\!=\!\pi$ for each of the two jets to obtain 
\beq{\label{eq:Nscale}}
   N_{\ee}=2\,n_q(P_{t,\max}^2={s}).
\eeq
This means one treats the multiplicity of \ee annihilation events as the sum of two independent quark jets.  
In so doing, hard-collinear as well as soft-large-angle emissions are taken care of, which contribute at the level of $\cO{\gamma} \propto\sqrt{\as}$ relative to the leading DLA answer.   
The next-to-next-to leading correction $\cO{\gamma^2}\propto \as $ would come from large-angle hard-gluon radiation (3 jets) and from correlated large-angle radiation of two energy-ordered soft gluons.    

For 3-jet \ee annihilation events, the full secondary particle multiplicity (including soft large-angle radiation) can be cast as  \cite{LuLe}
\beq\label{eq:ghardness}
 N_{\ee\to \qq g}(s)  =  N_{\ee\to\qq}(s_{\qq}) + \half N_{gg}(K_t^2), \quad s_{\qq} = 2(p_q p_{\bar{q}}), \quad K_t^2 \equiv \frac{2(p_g p_q ) (p_g p_{\bar{q}})}{(p_q p_{\bar{q}})} ,
\eeq
where $N_{gg}$ stands for the multiplicity in an event of two-gluon production by a colourless source, and 
$K_t$ is the transverse momentum of the gluon in the \qq center of mass frame.  

Eq.~\eqref{eq:ghardness} was tested experimentally and resulted in a measurement of the QCD $C_F/C_A$ ratio \cite{DELPHI2005}.  

\smallskip

 \subsection{One hemisphere}

\begin{figure}[ht]
   \centering%
  \includegraphics[width=0.8\textwidth]{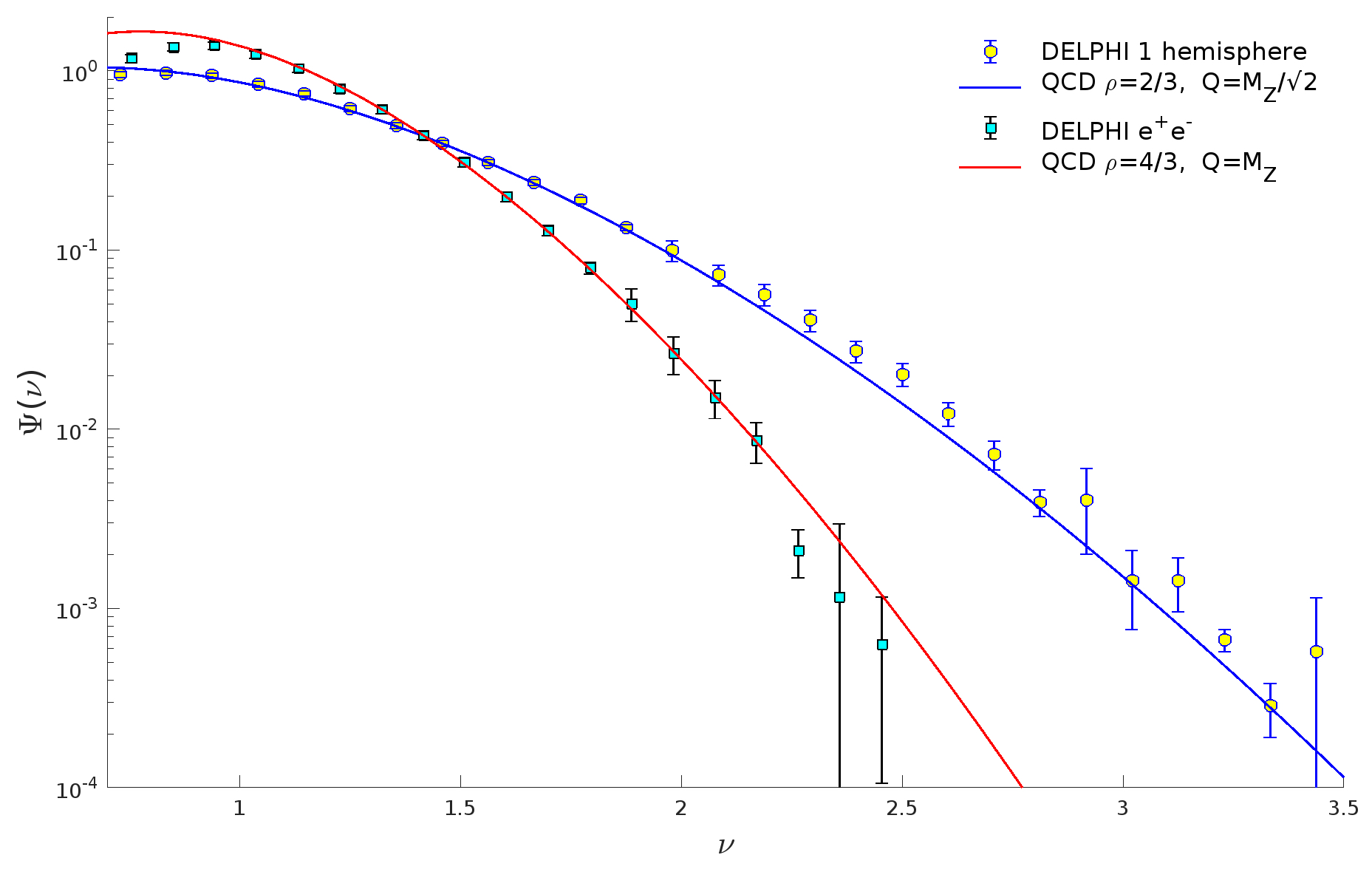} 
  \vspace{-0.5cm}
\caption{\label{fig:1hem}Comparison of multiplicity fluctuations in one hemisphere with full \ee event \cite{DELPHI}.}
\end{figure}

A hard test is provided by the DELPHI measurement of multiplicity fluctuations in a single quark jet (one hemisphere of \ee annihilation) \cite{DELPHI}.
Here, instead of $\rho=2\rho_q=\frac43$, we have to use $\rho=\rho_q=\frac23$.  \\
The result is displayed in Fig.~\ref{fig:1hem}.

We do not have a solid explanation for why a 
``tail formula" permits extension all the way down almost to the peak of the P-KNO distribution.

The agreement is not perfect, but still impressive given that there are no tuning parameters involved. 

\smallskip
There is an interesting point here in relation to the hardness scale.

Measurement in one hemisphere means that the maximum allowed radiation angle equals $\frac{\pi}{2}$. 
According to \eqref{eq:hardness}, this implies the hardness scale
\beq\label{eq:Q1}
  Q_1 = 2E_{\mbox{\scriptsize jet}}\sin\frac{\Theta}{2}  = 2E_{\mbox{\scriptsize jet}}\sin\frac{\pi}{4}  = \sqrt{s} \cdot \frac{1}{\sqrt2}.
\eeq
For full \ee annihilation events, we have $Q=\sqrt{s}$ ($\Theta=\pi$). 
This may seem confusing: \ee\ annihilation is a sum of two hemispheres, is it not? 
It is when it concerns mean multiplicity, but not so for multiplicity correlators!

A quark radiates in all directions, producing the bulk of offspring
\[
  N_{own} = n\left(\frac{Q}{\sqrt2}\right)
\]
 in its own hemisphere, and
\[
  N_{opp} = n\big(Q\big) - n\left(\frac{Q}{\sqrt2}\right) \>\simeq\> \ln\sqrt2\cdot n'\big(Q\big) =\cO{\sqrt{\as}\, n\big(Q\big)}
\]
particles backward. This does not affect the total multiplicity of an event, since the missing piece in the quark's hemisphere is covered by an identical contribution from antiquark backward radiation.

However, these two parts, while flying in the same direction, are independent by origin. They do not participate in a single cascade and therefore do not contribute to the P-KNO phenomenon.

 \subsection{ATLAS jets}

The transverse momenta of the high-$p_t$ jets studied by ATLAS at the
LHC at $\sqrt{s}=13\,\TeV$ span a wide range of $p_t$, from 0.1 to 2.5
\TeV~\cite{ATLAS-2019}.  Among other things, the multiplicity
distributions of charged hadrons were reported.

\subsubsection{Separated quark and gluon jets}
Hard hadron-hadron collisions are expected to produce a mixture of
quark- and gluon-initiated jets, with a quark fraction that
increases with $p_t$ and rapidity $|\eta|$.  In the ATLAS analysis,
the quark and gluon fractions were estimated using simulated jets from
Pythia 8, the jet flavour being defined as that of the hardest parton
assigned to the jet before hadronization.   There are problems with
this definition, which we discuss in the following.  However, by accepting it
one can use the variation of the quark fraction with
$|\eta|$ to solve for the multiplicity distributions of pure quark and
gluon jets at a given $p_t$.

\begin{figure}[ht]
   \centering%
  \includegraphics[width=0.75\textwidth]{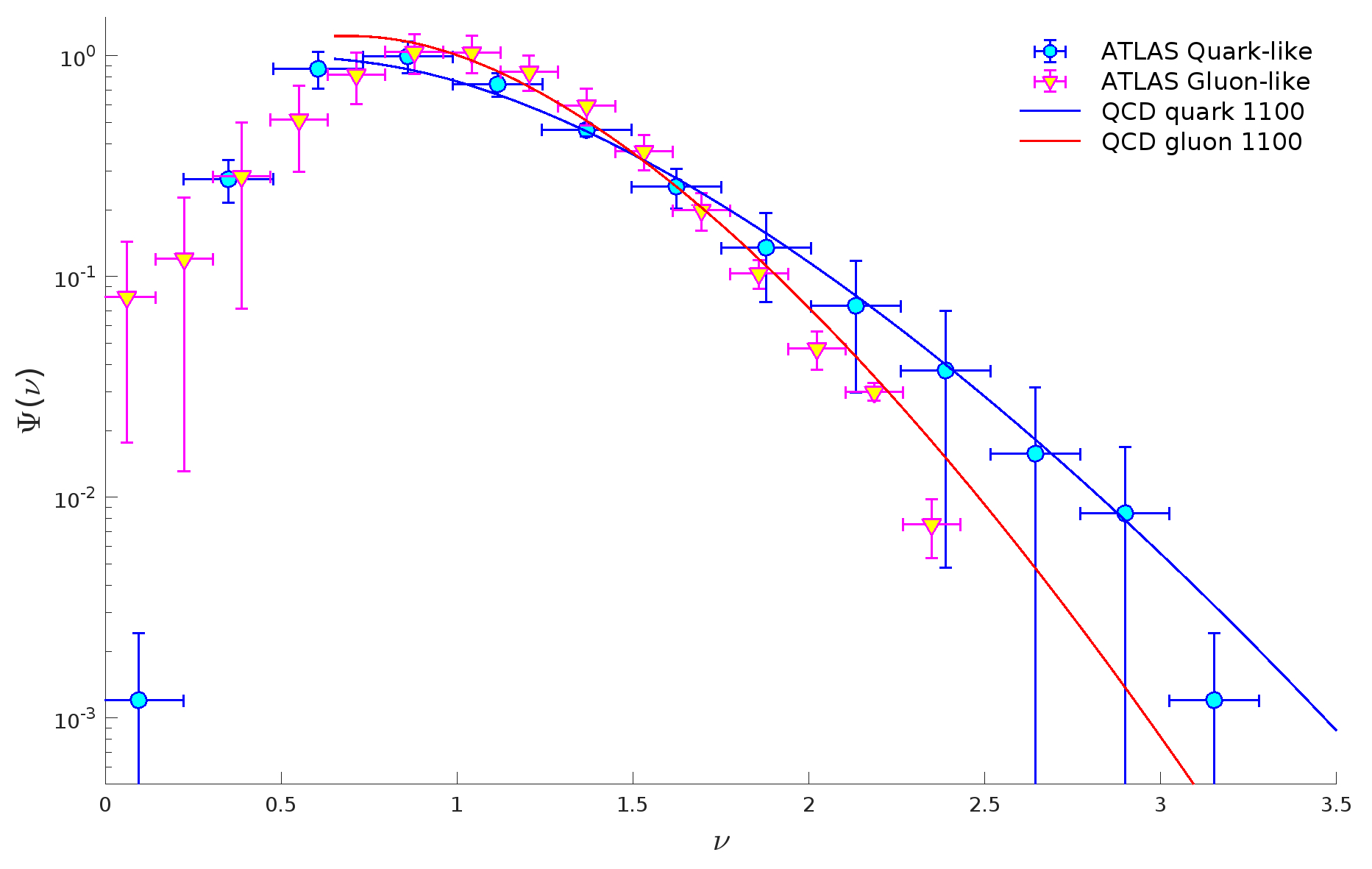} 
\caption{\label{fig:AtlasQGsep} P-KNO distributions of separated quark and
  gluon jets at $1000<p_t<1200$ GeV. }
\end{figure}

Results of this procedure were shown in~\cite{ATLAS-2019} for jets
with $1000<p_t<1200$ GeV.  Translated into the P-KNO form, as shown in
Fig.~\ref{fig:AtlasQGsep}, the tails of the distributions are quite
well matched by the QCD formula \eqref{eq:PsiRHO}.
We stress that this ``fit'' does not contain any fitting parameter.
The only input is the value of $\rho_q=\frac23$ for the quark jets.
For the mean multiplicity ratio the data give $N_g/N_q=1.56\pm 0.08$,
consistent with the value obtained at lower scales from \ee\ data in
Fig.~3, corresponding to $\rho_q=0.64\pm 0.03$.

For the hardness scale $Q$ we use
\beq\label{eq:ATLAS-Q}
  Q = 2\, \overline{P_t} \, \sin \frac{R}{2},  \quad R=0.4 ,
\eeq
with $\overline{P_t}$ the middle point of the $p_t$ interval
and $R$ the ``angular radius" parameter of the anti-kt jet finder employed by ATLAS.  
Varying the scale in \eqref{eq:ATLAS-Q} by a factor of 2 or more has a
negligible effect on the predicted P-KNO distribution, because at such
large $p_t$ the anomalous dimension $\gamma(\as(Q))$ varies very little.

\subsubsection{Quark and gluon jet mixture}

For a mixture of a fraction $f_q$ of quark jets and $f_g=1-f_q$ gluon
jets, the multiplicity distribution is
\beq
P(n) = f_q P_q(n) + f_g P_g(n) \equiv N\Psi\left(\frac nN \right)
\eeq
where the overall mean multiplicity is $N = f_qN_q + f_gN_g$.  Hence
\beq
\Psi(\nu) = \frac 1N\left[ f_qN_q \Psi_q\left(\nu \frac{N}{N_q}\right) + f_gN_g
  \Psi_g \left(\nu \frac{N}{N_g}\right)\right].
\eeq
We are using $N_q/N_g \equiv \rho_q= 2/3$, so $N = (2+ f_g)N_g/3$ and
\beq
\Psi(\nu) = \frac 1{2+f_g}\left[2(1-f_g)\Psi_q\left(\nu\frac{2+f_g}{2}\right) + 3f_g
  \Psi_g\left(\nu\frac{2+f_g}{3}\right)\right].
\eeq

\begin{figure}[ht]
   \centering%
  \includegraphics[width=0.75\textwidth]{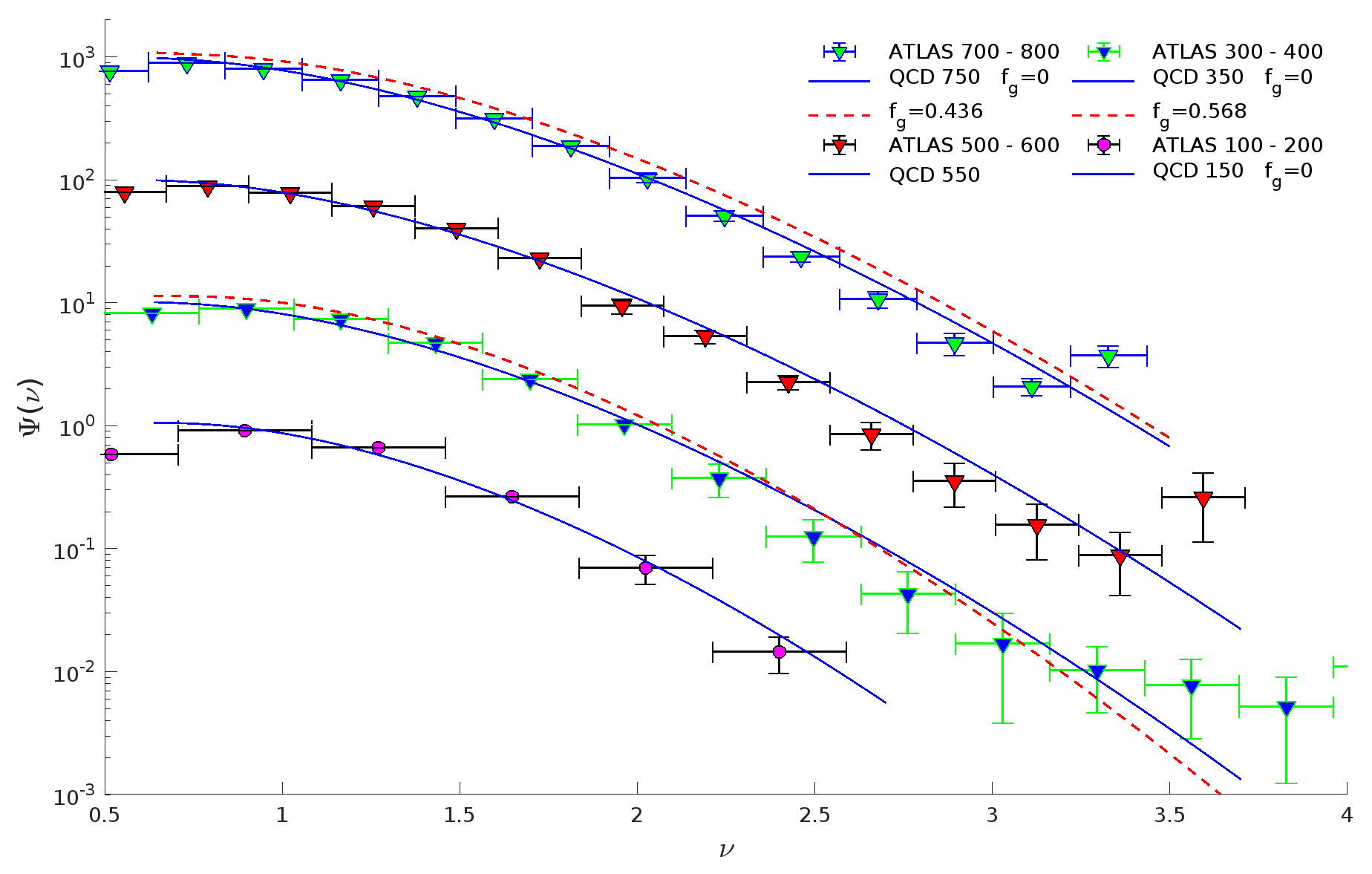} 
\caption{\label{fig:ATLAS_small} P-KNO distributions of unseparated jets at $100<p_t<800$ GeV. }
\end{figure}

For the ATLAS jet sample, the gluon fraction $f_g$, defined as explained
above, varies from 68\% for $p_t=100-200$ GeV to 44\% at 700--800 GeV.
Somewhat surprisingly, as shown in Fig.~\ref{fig:ATLAS_small},  for
fractions around 50\% the predicted tail of the unseparated P-KNO
distribution is quite similar to that of a pure quark jet sample, both of which give reasonable agreement with the data.  The reason is that for such a mixture the higher average multiplicity of
the gluon jets happens to compensate for their narrower
P-KNO distribution.

\begin{figure}[ht]
   \centering%
  \includegraphics[width=0.75\textwidth]{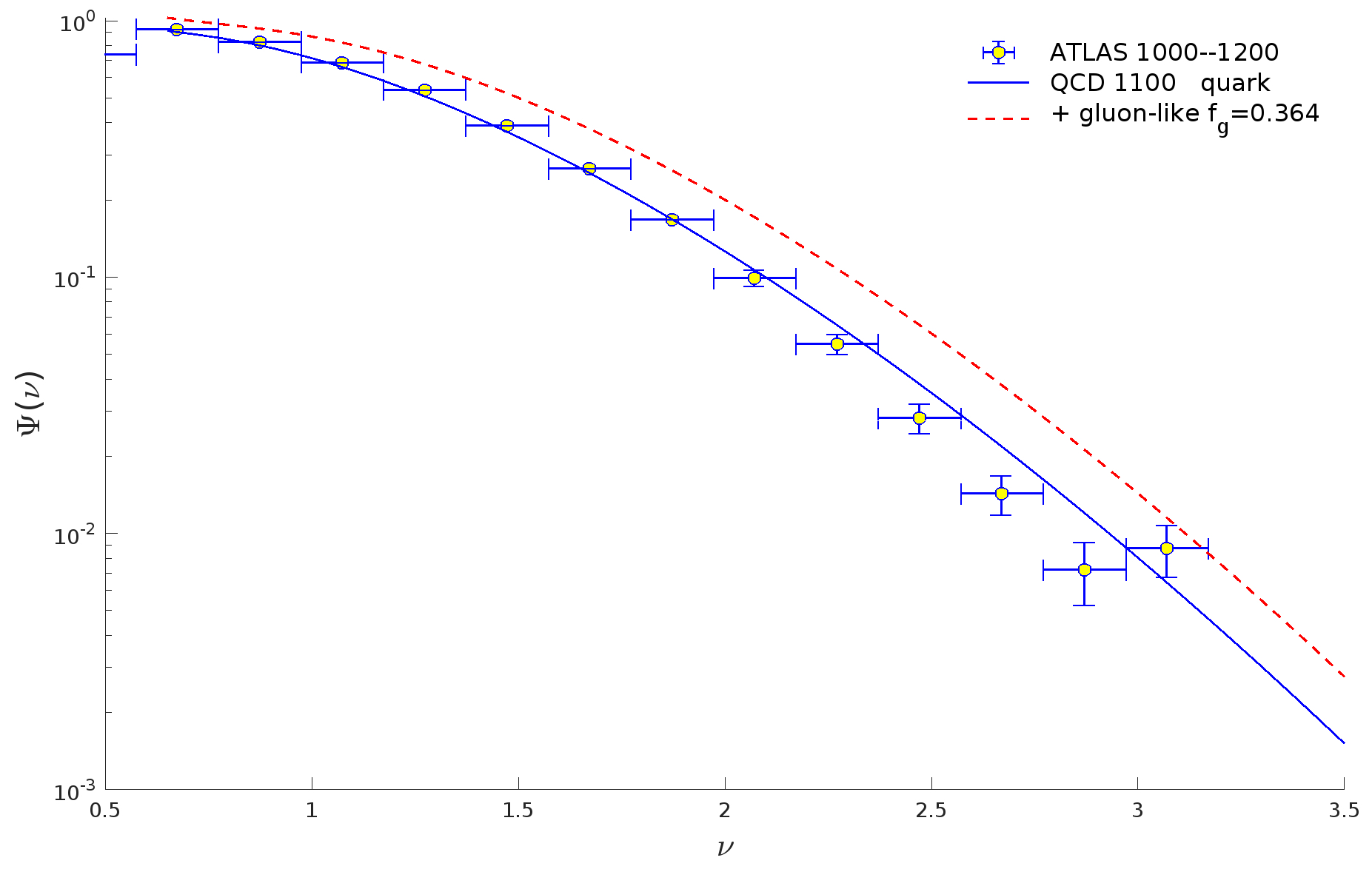} 
\caption{\label{fig:ATLAS_1100} P-KNO distributions of unseparated jets at $1000<p_t<1200$ GeV. }
\end{figure}

For lower gluon jet fractions, the predicted tail of the unseparated P-KNO 
distribution actually moves above the pure quark jet curve, only
approaching it from above at very small fractions.  This is because
gluon jets, although rare, are preferentially selected at high multiplicity.
For example, at 1100--1200 GeV the ATLAS estimate of the gluon jet
fraction is 36\%.  As shown in Fig.~\ref{fig:ATLAS_1100}, this would
imply a tail well above that for pure quark jets, whereas the latter 
agrees better with the data.  This is paradoxical because we saw in
Fig.~\ref{fig:AtlasQGsep} that the same data, separated into quark and
gluon components according to the ATLAS procedure, agrees with the
predictions for each separate component.

Without proposing a resolution of this paradox, we remark in this
context that the ATLAS procedure for estimate the quark and gluon jet
fraction from parton shower Monte Carlo data could be problematic.  A leading quark
can emit a collinear gluon carrying most of its energy, which would lead to a quark jet
being classified as a gluon jet, and vice-versa for a leading gluon splitting
to $q\bar q$. A safer procedure would be to associate jets with a
primary parton from the hard subprocess, before showering.

\subsubsection{Tails up!}
\begin{flushright}
`It is a long tail, certainly,  but why do you call it sad?' 
\cite{LC}
\end{flushright}              

The multiplicity distributions presented in \cite{ATLAS-2019} were plotted on a linear $y$-scale and looked uneventful.  However, focusing on the tails of the log-plots of the same data in Fig.\ref{fig:ATLAS_small} hints at a curious feature:
very high-multiplicity fluctuations seem to occur more often than would be expected.
Apparently something strange is going on on the high-multiplicity end when the multiplicity ratio hits $\nu=3$ \ldots

\begin{figure}[ht]
   \centering%
  \includegraphics[width=0.9\textwidth]{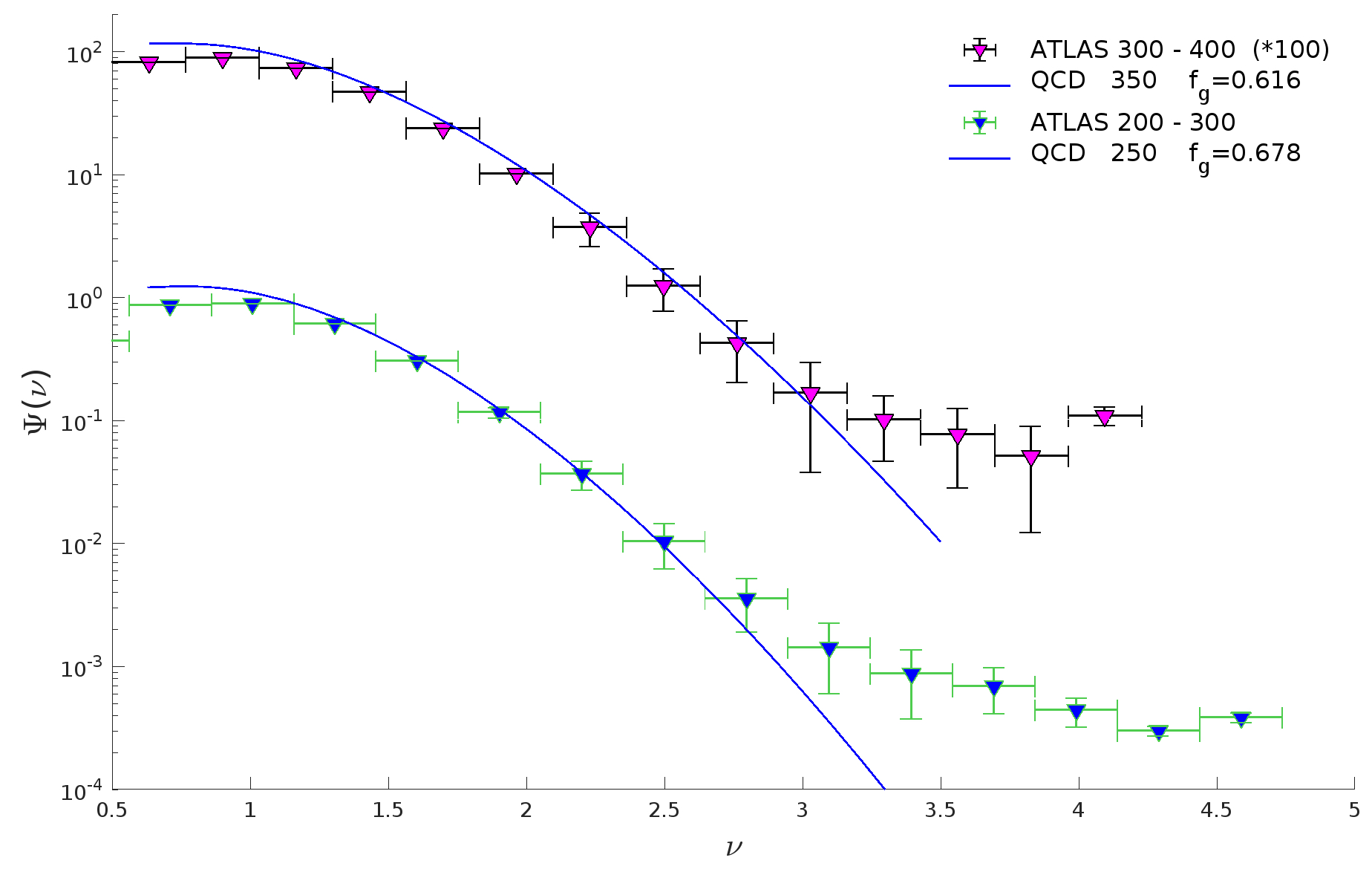} 
\caption{\label{fig:3ATLAS} Charged hadrons  in ATLAS jets:  $200\,\GeV\le p_{t, \mbox{\scriptsize{jet}}} \le 400\, \GeV$ }
\end{figure}

ATLAS multiplicity distributions were discussed in a recent review
\cite{KNO ATLAS} devoted to the KNO phenomenon.   However, the authors
skipped the intervals $p_t =$ 200--300 and 300--400 \GeV, the ones
where this misbehaviour is most prominent, as shown in more detail in Fig.~\ref{fig:3ATLAS}.  In our opinion,  this ``{\em tails up}\/'' feature that can be seen
in almost every data set is not an unfortunate glitch in the estimate
of systematics, but rather some new unexpected phenomenon worth exploring.

We therefore turn to a discussion of the possible origin --- and the possible impact --- of {\em flattening}\/ of the rate of high multiplicity fluctuations.

\mysection{ATLAS tail and CMS ellipticity\label{Sec:CA}}
\begin{flushright}
`Visit either you like: they're both mad' \cite{LC}
\end{flushright}              

\subsection{ATLAS high-$p_t$ jets}

\subsubsection{Large-multiplicity bias: angular push}
%\begin{flushright}
%`What’s the answer?'\\`I haven’t the slightest idea' \\
% \cite{LC}
%\end{flushright}              

\stepcounter{figure}

\noindent
\begin{minipage}{0.25\textwidth}
   \centering%
  \includegraphics[width=0.9\textwidth]{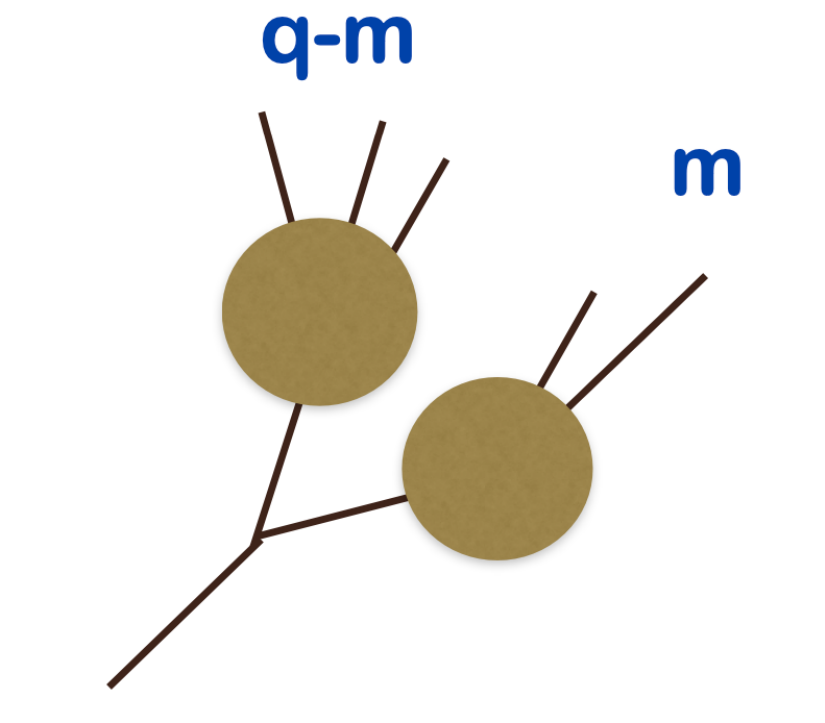}
    \centering%
Figure \thefigure: Topology of large-$\nu$ 
measurement
\end{minipage}
\hfill
\begin{minipage}{0.7\textwidth}
Selecting events with an abnormally large multiplicity creates a bias.  

In order to see that, let us look into the angular and energy pattern of parton splittings that form such events.  

\smallskip
The characteristic rank of multiplicity moment that governs the  yield of events at a given $\nu$ was estimated above as
\beq
     \ksd \sim \mu\big[D(\gamma) \nu\big]^\mu
\eeq
where $\mu=1/(1-\gamma)\sim 1.7$. With $\nu$ increasing we start to probe higher and higher multiplicity moments $q\sim  \ksd $. 
\end{minipage}

\medskip
Now consider the evolution equation for the multiplicity moments. 

It has the following structure \cite{D93}:
\beq\label{eq:EvEq}
  g_q^{(\rho)} = \rho\int^\Theta \frac{d\theta'}{\theta'}
   \sum_{m=1}^{q} C^m_q g_m g_{q-m}^{(\rho)}  \int^{1-\delta'}_{\delta'} \frac{dz}{z} \gamma^2(k_\perp)\,
  \frac{N^m(zE\theta')N_{(\rho)}^{q-m}((1\!-\!z)E\theta')}{N _{(\rho)}^q(E\Theta)} , \quad \left( \delta' = \frac{Q_0}{E\theta'} \right)
\eeq
where scaling was envisaged,
$g_m \cdot [N(Q)]^m = g_m \cdot [N _{(\rho)}(Q)/\rho]^m$ being the $m^{\rm th}$
multiplicity moment of the cascade from the gluon radiated with momentum fraction $z$ off the parent parton with source strength $\rho$ (quark or gluon).\footnote{In \cite{D93} the evolution equation was derived for factorial moments $\lrang{n(n-1)\ldots(n-m+1)}$, but the same form of equation holds for the normal moments $\lrang{n^m}$.}

The evolution equation is written {\em as if}\/ this gluon were the first parton emitted in the evolution of the  jet. This is not necessarily so.  Gluon \br\ may be also present between $\theta'$ and the opening angle $\Theta$. These are inclusive gluons that do not participate in formation of the observable. They are capable of taking away some  fraction $(1\!-\!y)$ of the parent parton energy. However, since 
the multiplicity factors $N$ in \eqref{eq:EvEq} are growing functions of the argument proportional to $\theta'$, energy loss would diminish the yield of final particles, and therefore this initial stage collapses into $D_q^{q'}(y, \as\ln(\Theta/\theta')) \to \delta(1-y)$.

As a result, integration over the angle $\theta'$ of the first
splitting that forms the cascade converges, and the collinear cutoff parameter $\delta'$
can be dropped. Since in the essential integration region $\ln(\Theta/\theta')$ is finite, we can approximate the multiplicity factors as
\beeq
  \ln N(\theta' E)  &=&  \ln N(E) + \ln \frac{\theta'}{\Theta}\cdot \frac{dN(E\Theta)}{d\ln E} +\cO{\gamma^2} , \\
  N(\theta' E) &\simeq & \left(\frac{\theta'}{\Theta}\right)^\gamma \cdot N(E).
\eeeq
Substituting into \eqref{eq:EvEq} results in
\beq\label{eq:squeeze}
g_q^{(\rho)} = \int^\Theta \frac{d\theta'}{\theta'} \left(\frac{\Theta}{\theta'}\right)^{-q\gamma} \times  
   \sum_{m=1}^{q} C^m_q g_m g^{(\rho)}_{q-m} \rho^{1-m}
    \int^{1-\delta}_{\delta} \frac{dz}{z} \gamma^2\, 
  \frac{N^m(zE\Theta)N^{q-m}((1\!-\!z)E\Theta)}{N^q(E\Theta)} ,  
\eeq
The entire dependence on $\theta'$ is now localised in a simple integral:
\beq
I(\gamma q) =\int^\Theta \frac{d\theta'}{\theta'} \left(\frac{\Theta}{\theta'}\right)^{-q\gamma} = \frac1{\gamma q}.
\eeq
We may characterise the typical ``angular distance'' between the % first 
splitting and the jet opening angle by an average pseudorapidity distance\footnote{$\eta^*$ stands for pseudorapidity with respect to the parent parton, measured along the jet axis.}
\beq\label{eq:angpush}
  \Delta \eta^*\sim \lrang{\ln \frac{\Theta}{\theta_1}} = I^{-1}\int^\Theta \frac{d\theta'}{\theta'}  
  \ln \frac{\Theta}{\theta'}  \left(\frac{\Theta}{\theta'}\right)^{-q\gamma}
  = \left. -\frac{1}{I} \frac{d\,I(\alpha)}{d\alpha}\right|_{\alpha=\gamma q}    \> =\> \frac{1}{q\,\gamma}.
\eeq
It is supposed to be large, since formally $\gamma\propto {\sqrt{\as}}$ is a small PT-expansion parameter.  
As we know,  the anomalous dimension floats between $0.3$ (corresponding to the maximal foreseeable hardness, of the order of 10 \TeV) 
up to $\gamma\la 0.6$ (for $Q\sim 10\,\GeV$),  see Fig.~3.   %%% \ref{}.  
In real life $\gamma \in \frac 13 \div \frac12$. 
When studying jet multiplicity, $q\!=\!1$, the pseudo-rapidity difference is still significant,  $\Delta \eta^* >2$ 
(for smaller hardness scales $Q$) or even $\Delta \eta^* >3$ (the largest $Q$) .  

However,  when one turns to multiplicity moments, the situation changes drastically:
already for multiplicity ranks $q\ge  2\div 3$ one has $\Delta \eta^* <1$. 
This means that the first parton splitting gets pushed out more the larger the moment rank one considers.

\subsubsection{Large-multiplicity bias: energy share \label{sec:share}}

A similar bias occurs in the pattern of parton energy sharing.
  
  \medskip
  
  \stepcounter{figure}
\noindent
\begin{minipage}{0.58\textwidth}
  \includegraphics[width=0.95\textwidth]{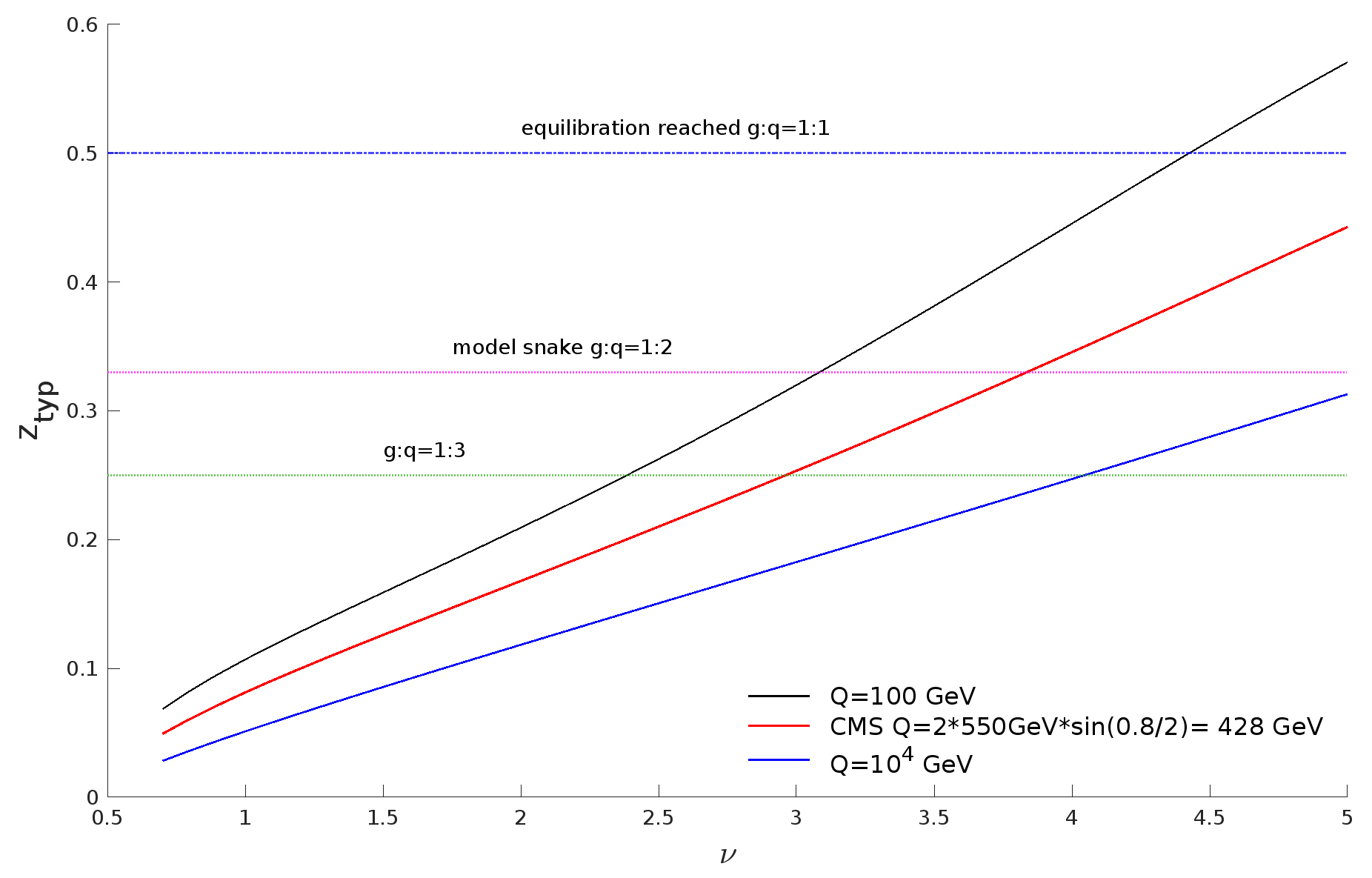} 
  
\centering{Figure \thefigure: Typical gluon energy fraction as a function of $\nu$}
\end{minipage}
\begin{minipage}{0.42\textwidth}
From the structure of the $z$ integral in \eqref{eq:squeeze},   
\[
  \sum_m^q C^m_q g_m g^{(\rho)}_{q-m} \, \rho^{1-m} \cdot \int \frac{dz}{z}\,  z^{m\gamma}(1-z)^{(q-m)\gamma},
\]
it is straightforward to deduce (using the trick of partial differentiation w.r.t.\ $m\gamma$ while keeping $(q\,-\,m)\gamma$ fixed)
\beq\label{eq:meanlnz}
   \lrang{\ln\frac1{z}} \simeq  \psi \big(1+q\gamma\big) - \lrang{\psi\big({m} \gamma\big)}.
\eeq
In order to calculate the average \eqref{eq:meanlnz} we need to recall the gluon multiplicity moments $g_m$ \eqref{eq:gkdef} as well as $g_{q-m}^{(\rho)}$ describing quark fluctuations.
The $m$-dependence of the latter, which is essential for our discussion, is analysed in Appendix \ref{App:rhomom}.
\end{minipage}

\medskip

In \eqref{eq:gqmm} important $m$-dependent factors are singled out:
\beq\label{eq:prop}
g_{q-m}^{(\rho)} \> \propto\>  \frac{\rho^{m-q} }{D^{q-m}} \cdot \rho^{\gamma (q-m)} .
\eeq
The first factor, assembled  with the gluon moment and the power of $\rho$ from  \eqref{eq:squeeze}
form the $q$-moment $g_{q}^{(\rho)}$ on the l.h.s.\ of the Evolution Equation:
\[
g_m\cdot g_{q-m}^{(\rho)} \propto \left[\frac1{D}\right]^m \cdot \left[\frac1{\rho D}\right] ^{q-m} \times \rho^{-m}
\>\Longrightarrow\> g_{q}^{(\rho)}.
\]
The second coupling-dependent factor from \eqref{eq:prop} is responsible for "squeezing". 
When $\rho<1$, it assigns a larger weight to the configurations with larger $m$, that is, with more offspring stemming from the gluon cascade. 

This effect is displayed in Fig.~\thefigure\ where the typical \br\ gluon energy fraction $z_{\mbox{\scriptsize typ}}$ is shown for different jet hardness scales. Producing this plot, we substituted the full quark$\to$gluon splitting function for the soft gluon spectrum,
\[
\frac1z \>\to\> \frac1z - 1 + \frac{z}{2},
\]
to make the estimate more realistic. Having calculated the average $\ln z$ at a given $q$, we used the value $q_{\mbox{\scriptsize char}}(\nu)$ of the characteristic rank that corresponds to a given $\nu$ and  can be estimated from the position of the steepest-descent point in the defining integral 
\beq\label{eq:qchardef}
  g_q = \int_0^\infty d\nu\,  \nu^q \, \Psi^{(\rho)}(\nu,\gamma), \qquad 
  q_{\mbox{\scriptsize char}}(\nu) \simeq \mu \rho [D\nu]^\mu .
\eeq

For finite ranks and small coupling, the gluon stays soft and free to provide a logarithmic enhancement.
As the rank $q$ increases, the logarithmic $z$-integration gets squeezed. 
Soft gluons are forced to carry an energy fraction larger than they were used to in average unbiased jets.
For example,  $z_{\mbox{\scriptsize typ}}=\third$ around $\nu\sim 3.5\div 4$.  
In the asymptotic limit $q\to\infty$ the gluon "takes it all".

\medskip

 \subsubsection{Tail flattening}
At this point, nothing bad happens to the QCD expression \eqref{eq:PsiRHO} for the tail.  
It still describes multiplicity fluctuations {\em in a  quark jet}\/ dressed up with secondary collinear/soft partons.   

What really changes above $\nu\!=\!2.5$ is the {\em  nature of the object under study}: it is a single jet no longer.  

As can be seen in Fig.~\ref{fig:3ATLAS} the tail shape changes above $\nu\simeq3$.   
Apparently, the interval between $\nu\!=\!2.5$ and $\nu\!=\!3$ is the transition region.  

For still larger $\nu$ one should look upon the jet 
as a two-jet system ($q+g$) as a source of new particles.  
Its ``source strength'' is larger than that of the original quark,  $\rho\to \rho+1$,  and the tail flattens.  

This conclusion might look surprising, since according to \eqref{eq:PsiRHO} 
the fluctuation distribution should get {\em narrower}\/ with an increase of $\rho$.  
True, {\em provided}\/ for each jet or system of jets we calculate $\nu$ by normalizing the number of particles by the proper mean,  
\begin{subequations}
    
\beq\label{eq:nRatio}
\nu= \frac{n}{\lrang{n^{(\rho)}}}.   
\eeq
What we are doing instead is counting particles in units of the original quark jet multiplicity in place of a larger split-tongue multiplicity,
\beq
\nu= \frac{n}{\lrang{n^{(\rho_q)}}}.   
\eeq
\end{subequations}
As a result the characteristic exponent gets {\em smaller}\/ when $\rho_q$ is replaced by a larger $\rho=\rho_q+1$: 
\beq
 -\rho\cdot \big( \nu \big)^\mu = -\rho\cdot \left( \frac{n}{\lrang{n^{(\rho)}}} \right)^\mu = -\rho^{1-\mu}\cdot \left( \frac{n}{\lrang{n_{g}}} \right)^\mu 
 \> \propto \> -\rho^{-\gamma\mu} \simeq -\rho^{-\frac23} \quad ({\mbox{for}}\>\> \gamma=0.4).
\eeq
The change in slope does not occur abruptly. 
At the moment, it is not clear to us how to treat the tail-up transition quantitatively.  

Meanwhile, it seems plausible that the resolution of the ATLAS puzzle may be related to the ``rattlesnake effect'' (RSE) at which we are pointing (split tongue causing a raised tail).

\stepcounter{figure}
\noindent
\begin{minipage}{0.7\textwidth}
\centering%
  \includegraphics[width=0.9\textwidth]{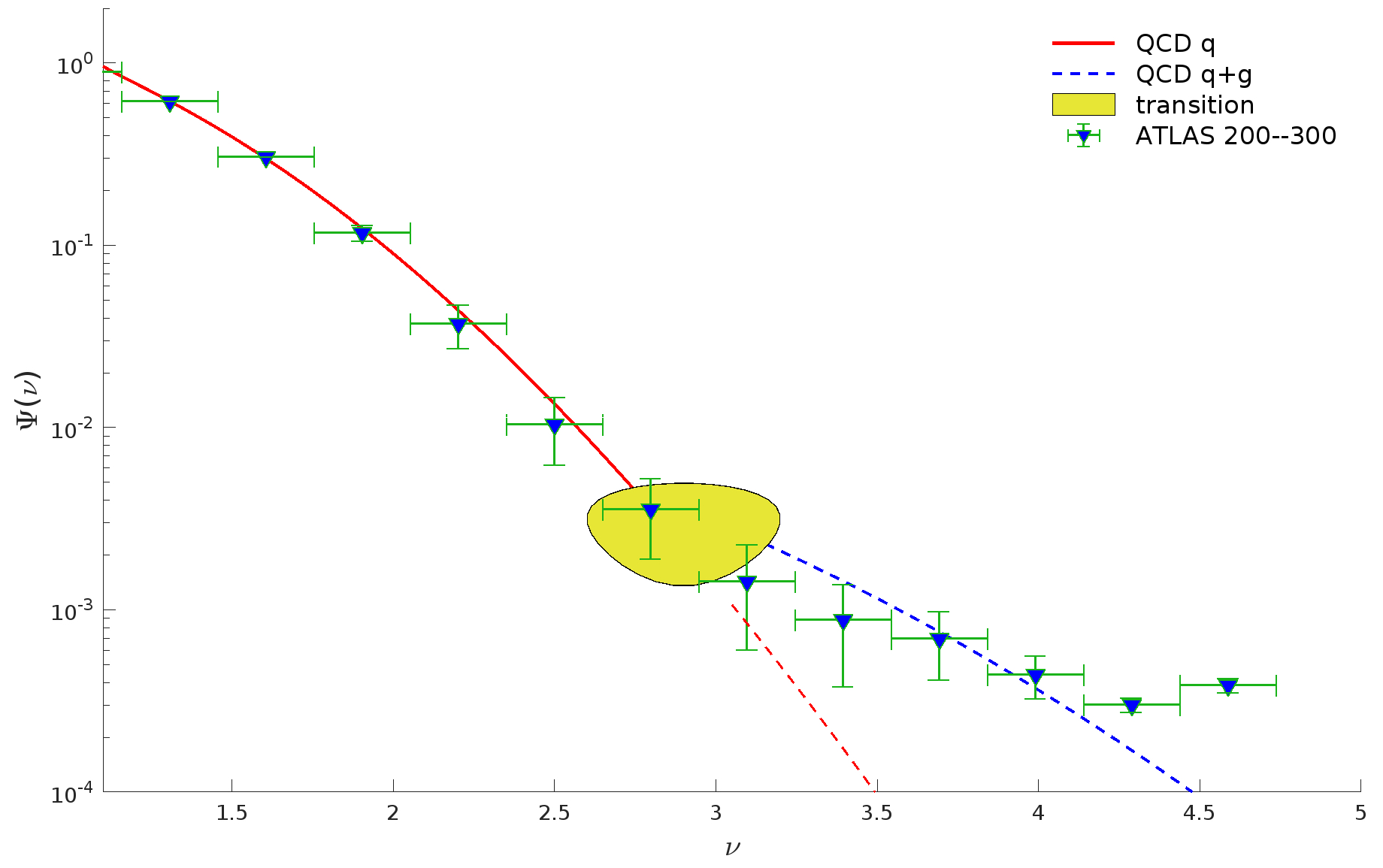} 
    \centering%
 { Figure \thefigure: Tail-up transition around $\nu=3$}
\end{minipage}
\begin{minipage}{0.3\textwidth}
To support this hypothesis, Fig. \thefigure\ shows what is about to happen when a quark jet is replaced by a pair $q+g$ with comparable hardness.

\medskip

The standard one-quark curve (red) is $\Psi^{(\rho_q)}(\nu)$, and the quark+gluon one (dashed blue) corresponds to 
$$
\Psi^{(\rho_q+1)}(x) \times \Psi^{(\rho_q)}(3)
$$
evaluated at  
$$
x=\frac{\nu}{\rho_q+1}. 
$$
\end{minipage}

%\clearpage

\subsection{Snake-tongue transition and CMS high-multiplicity jets}
\begin{flushright}
% 'There's no use trying. One can't believe impossible things.\\
'Why, sometimes I've believed as many as six impossible things before breakfast' 
\cite{LC}
\end{flushright} 

There is another puzzle for which RSE may assume responsibility.

\smallskip
\stepcounter{figure}

\noindent
\begin{minipage}{0.4\textwidth}
  \includegraphics[width=\textwidth]{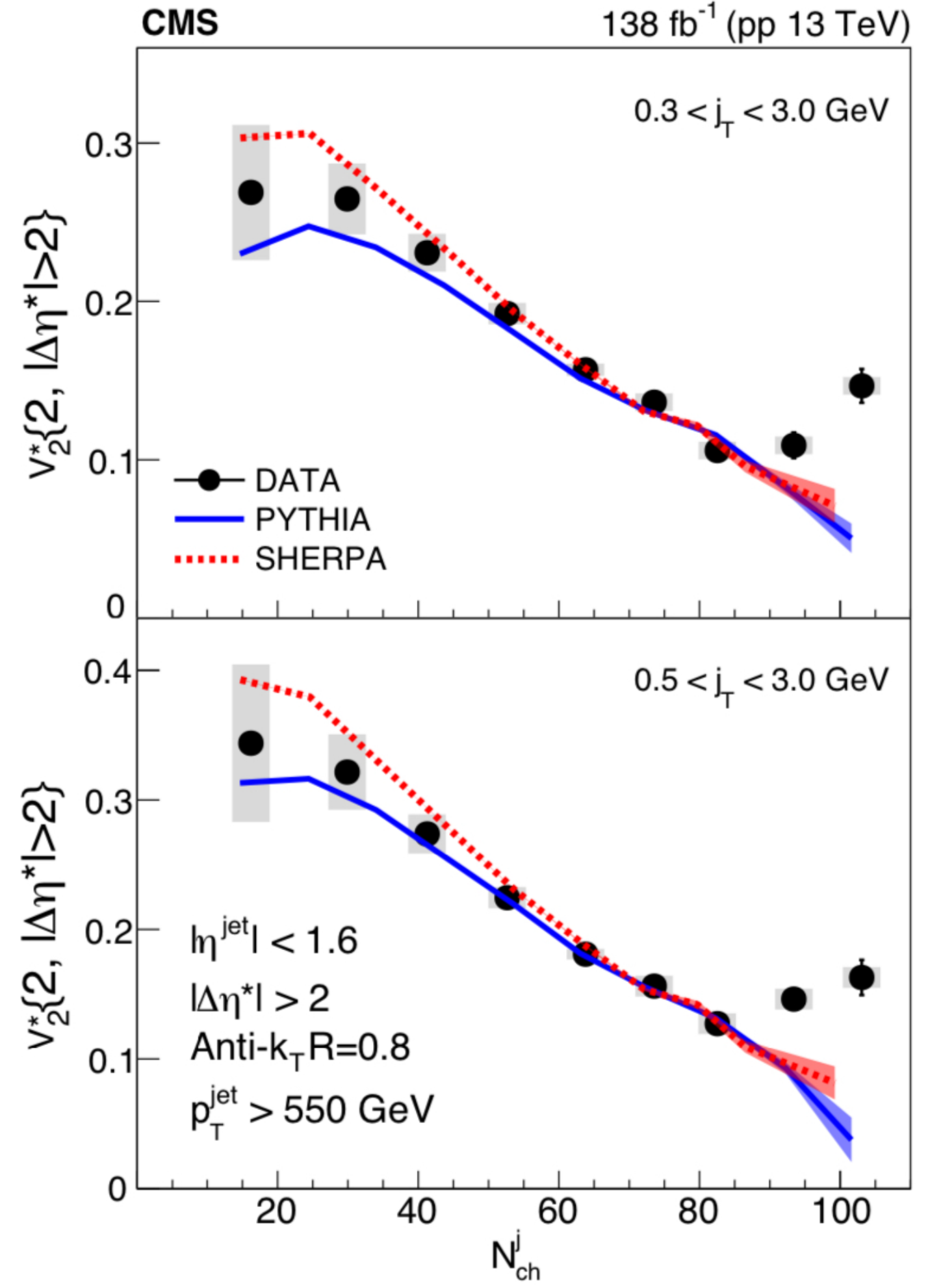} %  \epsfig{file=DW CMS.pdf, width= \textwidth, angle=0} 
    \centering%
   Figure \thefigure:  $v_2^*$ versus jet multiplicity \cite{CMS}
   \label{Fig:cms}
\end{minipage}
\hfill
\begin{minipage}{0.55\textwidth}
The CMS Collaboration recently found a spectacular change in the trend of the behaviour of the second azimuthal harmonic $v^*_2$ of long-range ($\abs{\Delta \eta^*}>2$) 
two-particle correlations \cite{CMS}. 
\bigskip

On our side we make the following observation:
\smallskip

Divided by the mean multiplicity $ %\lrang
{N_{\mbox{\scriptsize ch}}^j}\simeq 26$ reported by CMS,  
the turning point in Fig.~\thefigure\ translates into 
\[
\nu = 80/26\simeq3. 
\]

\medskip

Given that the hardness scales of CMS $p_t \!>\! 550\> \GeV$ jets  and the snaky ATLAS jets discussed above are close,
it is worth looking into the possibility that the two -- equally crazy -- phenomena have a common origin: the RSE scenario.
\medskip

As far as a split tongue goes together with a raised tail, it seems natural to refer to this picture as RSE. 

The name is flexible, as it can be read as ``rattlesnake event", or ``rattlesnake effect", depending on context.
\end{minipage}
\medskip

\subsubsection{Jet finder at work} 

Imagine we had two hard partons with comparable energies (1:2) flying apart at a $45^{\mbox{\scriptsize o} }$ angle. 

What will the final state look like?
\stepcounter{figure}

\noindent
\begin{minipage}{0.5\textwidth}
The anti-kt machinery first collects secondary particles in the 30+30 degree cone around the hardest primary parton (quark), labeled ``anti-kt borderline'' in Fig.~\thefigure.   
\smallskip

Then it assembles the second bunch around the less energetic parton from inside the 15+15 degree cone. 
\medskip

 Having done so,  the algorithm will combine the two freshly cooked subjets into one, as long as their angular distance is less than the chosen ``radius'' $R\!=\!0.8$.  
 \end{minipage}
 \hfill
\begin{minipage}{0.5\textwidth}
   \centering%
  \includegraphics[width=0.90\textwidth]{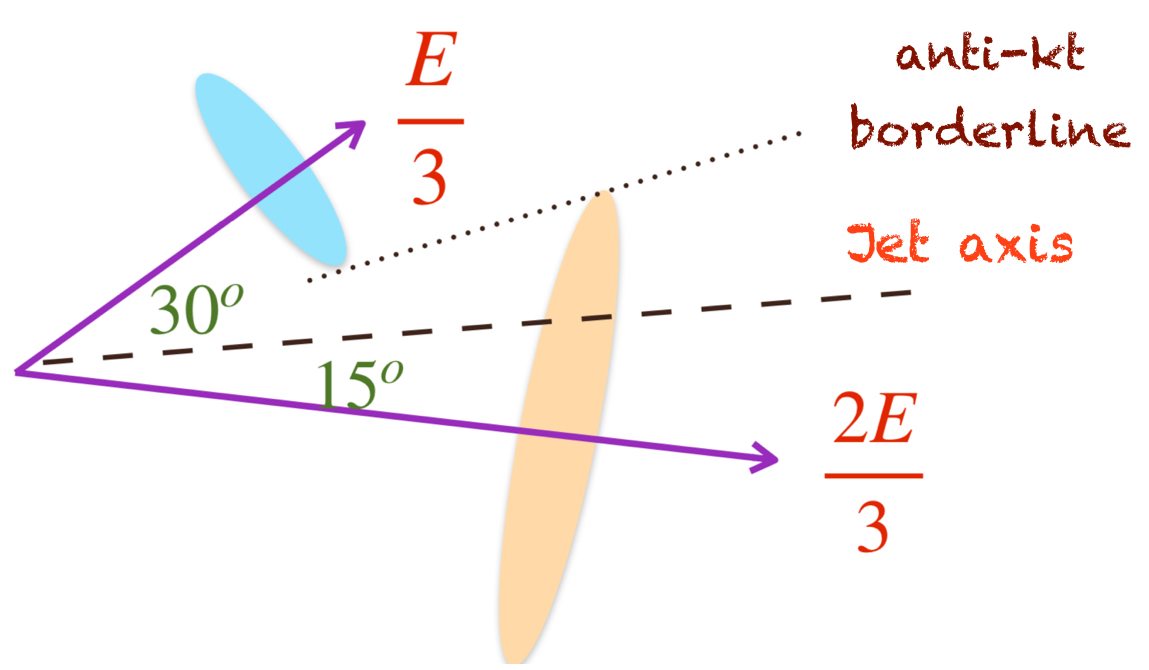} %  \epsfig{file=2 tongue.png,width= 0.9\textwidth, angle=0} 
Figure \thefigure: Snake-tongue jet substructure
\end{minipage}

\medskip

  The jet axis is determined with respect to which the rapidity $\eta^*$ and the azimuthal angle $\phi^*$ are measured.    
  {\em Note:}\/ here and in what follows, variables marked with a star are defined with respect to the jet axis, and $j_t$ refers to the transverse momentum of a hadron with respect to it \citd{Texas}{CMS}

  % \smallskip
  \subsubsection{Jet axis misused}
\begin{flushright}
% 'There's no use trying. One can't believe impossible things.\\
% 'You see the earth takes twenty-four hours to turn round on its axis--' \\
%`Talking of axes,' said the Duchess, `chop off her head!'
``When I use a word,
%’ Humpty Dumpty said in rather a scornful tone, ‘
it means just what I choose it to mean --- neither more nor less"
\cite{LC}
\end{flushright}

The ``jet axis" is purely formal: It points in the direction of an aggregate jet momentum, but it has little to do with the physics of radiation.  

\bigskip

Let us take the two-parton configuration of Fig.~\thefigure\/ as a toy model to see what kind of effects one should expect in the snake-tongue event scenario. 
\ben
\item[Expectation 1:]
In usual unbiased events the jet axis practically coincides with the direction of the quark momentum.
In this situation, one should see a ``Feynman plateau" in the inclusive distribution in $\eta^*$.

\item[Expectation 2:]
As a consequence of pQCD cascades, the height of this plateau should slowly but steadily increase towards smaller $\eta^*$ as
\beq
\frac{dn}{d\eta^*} \simeq \frac{d}{d\eta^*}\, N_q\left(E_{\mbox{\scriptsize jet}}\sin \frac{\theta}{2}\right) =\frac{d}{d\eta^*}\, 
N_q\left(\frac{E_{\mbox{\scriptsize jet}}}{\sqrt{1+\exp(2\eta^*)}}\right).
\eeq

\item[Expectation 3:]
In RSE instead, the bulk of offspring are radiated off the two hard partner partons that are away from the axis.  
Since radiation is predominantly collinear to their respective parents,  
the produced particles will peak at two $\eta^*$ values that correspond,  
in the particular kinematics of  Fig.~\thefigure, to 15 and 30 degree angles to the ``jet axis'', $\eta^*=2.1$ and $\eta^*=1.3$. 

So, a significant enhancement should be expected at $\eta^*\sim 1\div 2$ because it is the region covered by the snake tongue, where most secondary radiation flies.

\item[Expectation 4:]
For the same reason, the $\eta^*$-distribution should drop to $\exp(-2\eta^* )$ because the radiated particles can align with the axis only by accident (phase space).

\item[Expectation 5:]
Turning from inclusive distribution to particle correlations, the RSE is a planar event with its jet axis lying in the plane between the quark and the gluon. Therefore, among hadron pairs half will belong to subjets that are on opposite sides with respect to the jet axis. Massive $\phi^*=\pi$ correlations are expected.

\item[Expectation 6:]
Since RSEs coexist with normal events with one quark as the leader, the opposite-side correlations must grow when one cuts out ``small transverse momenta" hadrons, $j_t> [j_t]_{min}=$ few hundred \MeV. This would increase the portion of RSE in the event sample. Since "baby snakes" are omnipresent as rare fluctuations, imposing a $j_t$-cut will increase back-to-back correlations for any multiplicity, not necessarily extremely large.

\item[Expectation 7:]
Starting from the point where RSE becomes dominant, a $j_t$-cut should no longer matter.
 
\een

All these expectations are met by the CMS data.

\stepcounter{figure}

\noindent
\begin{minipage}{0.5\textwidth}
Expectations \# 1 to 4 find confirmation in the inclusive pseudorapidity distribution of charged hadrons 
% reported in 
\citd{CMS}{Gardner} shown in Fig. 17. %\ref{\thefigure}.
\medskip

A plateau is manifest in normal jets with multiplicity of hadrons close to the mean (red points).
\medskip

Its increase towards smaller $\eta^*$ values is seen. 
\medskip

In contrast, in the high-multiplicity event sample (blue points), the RSE is in action, and the jet axis no longer represents the direction of any hard parton, causing the $\eta^*$ distribution to fall exponentially, demonstrating that a small angle of a hadron with respect to the jet axis is an accident rather than the norm. 
\medskip

The particle density is by an order of magnitude higher at $\eta^*=1\div2$ than in unbiased events, the ones that are not forced to produce an exceptionally large multiplicity.

%\medskip
\end{minipage}
\hfill
\begin{minipage}{0.5\textwidth}
    \includegraphics[width=0.95\textwidth]{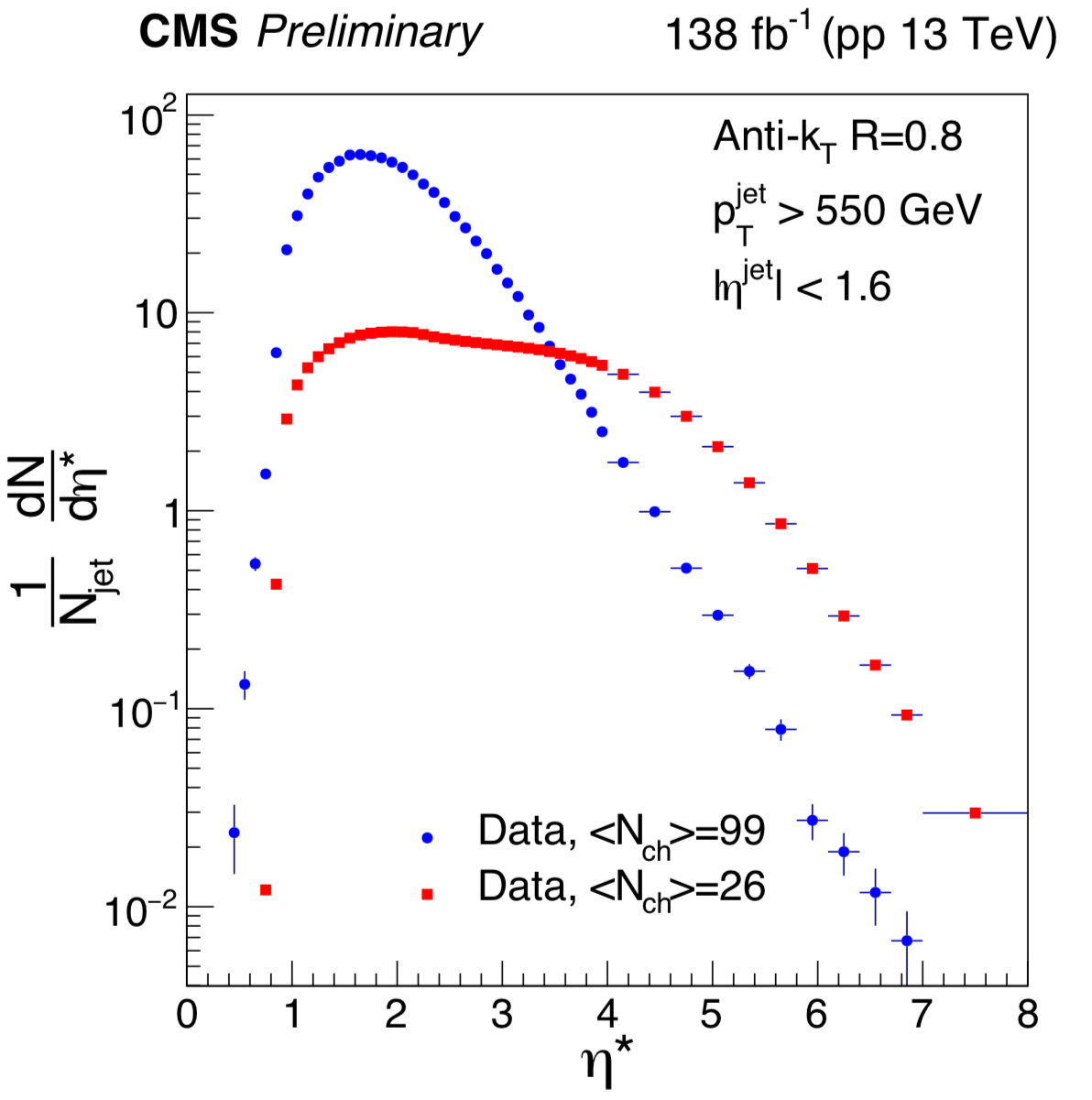} %    \epsfig{file=2GardnerETA.png,width= 0.95\textwidth, angle=0}
    \centering%
   Figure \thefigure:  Large-$\eta^*$ tail $\propto \exp(-2\eta^*)$.
\end{minipage}

\medskip

Now we move to correlations.
\medskip

Expectation \# 5 is what Fig. 16 %\ref{} 
is all about: increase of opposite-side azimuthal correlations with increase of registered multiplicity as a means of enriching the proportion of RSE. 
\medskip

\stepcounter{figure}
\noindent
\begin{minipage}{0.5\textwidth}
  \includegraphics[width=0.9\textwidth]{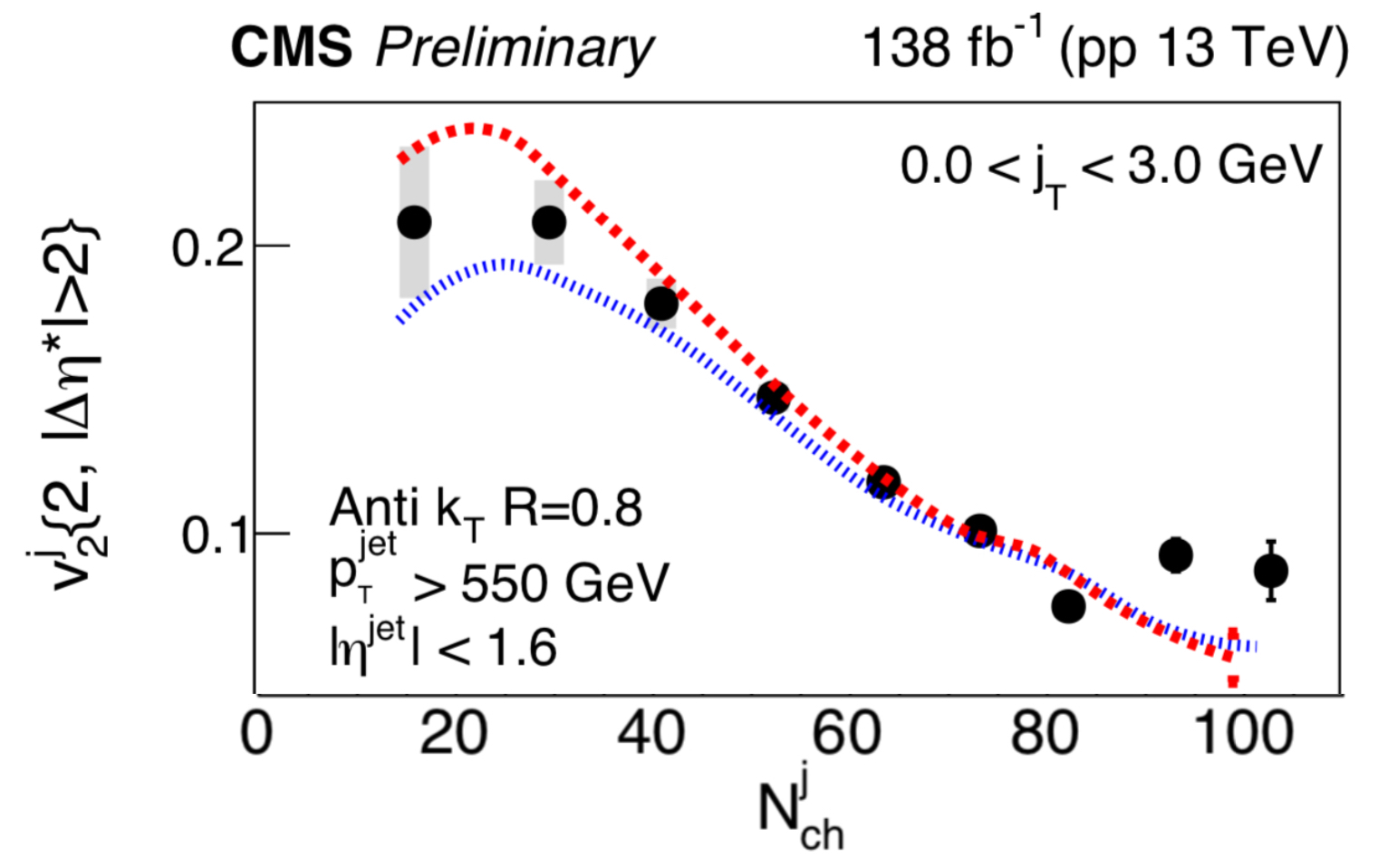} 
  \centering
   Figure \thefigure: No-$j_t$-cut data \cite{Gardner} 
\end{minipage}
\begin{minipage}{0.5\textwidth}
Expectations \# 6 and 7 find support in the comparison of the two panels of Fig.~15, published in the CMS article \cite{CMS}, with a third that was presented in \cite{Gardner}: Fig.~\thefigure\ taken from the latter shows the measurement of $v_2^*$ without imposing a $j_t$ cut.
\medskip

Compared with Fig.~15, the turnover at $N\sim 80$ is less pronounced and, more importantly, 
$v_2^*$ is systematically larger in Fig.~15 than in Fig.~\thefigure\ for {\em all}\/ multiplicities.
\end{minipage}

\medskip

This shows that "baby snakes" are an intrinsic part of parton cascades. 

Below $N_{ch}^j\sim 80$ ($\nu\simeq 3$), the cut-off value is important: Moving 
from $[j_t]_{min}=0$ to 0.3 and then to 0.5 $\GeV$ causes a steady increase in $v_2^*$.

Starting from $\nu\sim 3$, the value of $[j_t]_{min}$, 0.3 or 0.5, no longer makes any difference because a cut 
leaves snake-tongued configurations unaffected.\footnote{{\em Remark:}\/ Apparently, the upper bound on the transverse momenta of hadrons
% $j_t<3\GeV$ 
was inherited from an earlier study of collective effects where $p_t<3\,\GeV$ was chosen to minimize correlations from jets while studying minimum bias events \cite{CMS 2017}. In the present context, it is irrelevant.}

By then, the baby snakes have matured and started to take over. 

%%%%%%%%%%%%%
\subsubsection{Long-range away ridge}

The last point to pay attention to is whether the opposite-side correlations that the RSE generates are good enough to qualify as ``long-range" ($\abs{\Delta\eta^*}>2$).

\stepcounter{figure}

\noindent
\begin{minipage}{0.55\textwidth}
 %  \centering%
  \includegraphics[width=0.95\textwidth]{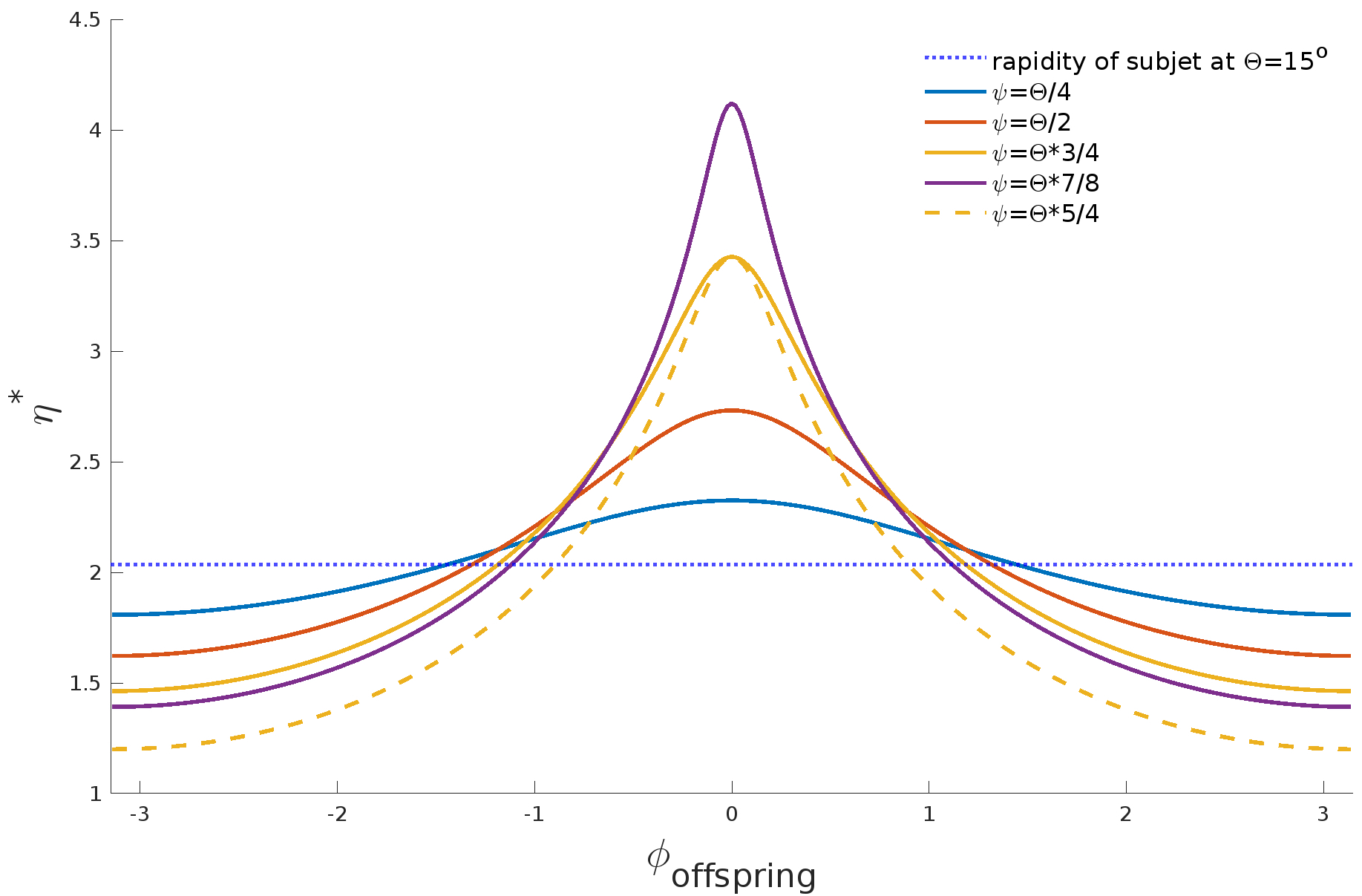} %  \epsfig{file=DW raptest.png,width= 0.95\textwidth, angle=0} 
  
\centering%
Figure \thefigure: Large-$\eta^*$ particles as a misunderstanding
\end{minipage}
\hfill
\begin{minipage}{0.43\textwidth}
\vfill

Recall that the RSE triggers a {\em plane}\/ in the jet fragmentation phase space and produces two peaks in $\eta^*$
that are always opposite in azimuth. 
Depending on whether the two triggered hadrons are picked from the same subjet or from opposite subjets, 
one obtains same-side or opposite-side correlations.

\medskip

If subjets were pencil-like, we would have short-range forward correlations in rapidity, $\abs{\Delta\eta^*} \!\simeq\! 0$ 
($\phi^* \!=\! 0$).  
The opposite side, in contrast, has a handicap: the varying opening angle of the emitters guarantees an initial ridge with a width of the order 1, $\abs{\Delta\eta^*} \!=\! 1.3-2.1 =0.8$.
This spread can be made larger by allowing the gluon to carry a somewhat smaller momentum fraction.
\end{minipage}

 \medskip
Another source of extending the rapidity span of the ridge is QCD \br. 
It is spread in angles, so that some part of the accompanying radiation may accidentally hit the ``axis'' direction and will be interpreted as having a large $\eta^*$.

Fig.~\thefigure\ demonstrates this effect.  
Here, $\Theta$ is the angle of the leading quark to the jet axis and $\psi$, $\phi_{\mbox{\scriptsize offspring}}$ are the polar and azimuthal angles of the secondary radiation with respect to its quark parent.  
The solid curves ($\psi<\Theta$) show the rapidity spread in back-to-back correlations (the dashed line ($\psi>\Theta$) gives an example of an additional contribution to the same-side correlation spread).  

The pseudorapidity of the harder parton (the quark with $2/3E$ in Fig.~\thefigure) is marked by the dotted line.
Picking up an offspring of the quark subjet and correlating it with any of an offspring of the second subjet (with smaller $\eta^*$) will generate a $\phi^*\sim\pi$ effect.  
Given the relative proximity of the quark to the jet axis, such an induced long-range correlation can be quite significant and spread in $\abs{\Delta \eta^*}$.  

A simple phase-space probability estimate leads to an exponential drop-off at large relative rapidity $dw\propto \exp(-2\abs{\Delta \eta^*})$.  

\medskip

\bigskip

Should the RSE picture prove viable, one may expect the fork structure of the leading partons to progress further with increasing $\nu$, from trident to ``used shaving brush".
The away ridge will first become shorter in $\eta^*$ and then recede, causing $v_2^*$ to decrease again, leaving room for higher harmonics.

\mysection{Conclusions}
%\bigskip
\begin{flushright}
Alice: Would you tell me, please, which way I ought to go from here? \\
The Cheshire Cat: That depends a good deal on where you want to get to. \\
Alice: I don't much care where... \\
The Cheshire Cat: Then it doesn't much matter which way you go. \\
Alice: ...So long as I get somewhere. \\
The Cheshire Cat: Oh, you're sure to do that, if only you walk long enough
\cite{LC}
\end{flushright}

We have presented arguments in favour of the QCD picture of particle multiplication in hard interactions,
by confronting with experiment the analytic perturbative QCD formulas describing multiplicity fluctuations (the Polyakov--Koba--Nielsen--Olesen scaling phenomenon).

An unconventional ingredient has been introduced into our approach:
\begin{quote}
Using an expression for the P-KNO distribution, we replaced the ratio of quark and gluon {\em colour charges} by the experimental ratio of quark and gluon jet {\em hadron multiplicities}. 
\end{quote}
This move allowed us to describe high-end multiplicity fluctuations both in \ee\ (full event and one-hemisphere distributions) and in LHC high-$p_t$ jets.  
The analytic prediction of the height and shape of the P-KNO tail agrees reasonably well with the data from \ee annihilation and LHC high-$p_t$ jets, despite having no tunable parameters.

Analysing the tails of multiplicity distributions measured by the ATLAS Collaboration, we dwelt on a certain flattening of the falloff above $\nu\simeq 3$, and proposed 
a possible qualitative explanation for this phenomenon.

We also linked it with another puzzling observation recently reported by the CMS Collaboration: the growth of long-range back-to-back correlations with increase of hadron multiplicity inside high-$p_t$ jets.  

 The CMS discovery was triggered by a theoretical suggestion to look for signs of collective effects inside jets as a result of final-state interaction between partons. % \cite{Texas}.  
 Quote \cite{CMS}:
 \begin{quote}
{\em
``The non-monotonic dependence of $v_2^*$ versus $N_{ch}^j$ \,{\em [ $\cdots$ ]}
may indicate an onset of novel QCD phenomena related to nonperturbative dynamics of a parton fragmenting in the vacuum. These phenomena could include the emergence of collective effects possibly driven by final-state rescatterings, as suggested in Ref.~{\em \cite{Texas}}" }
\end{quote}

However, hunting for collective ``QGP-like" phenomena inside a QCD jet propagating in the vacuum is doomed to fail. 
This follows from an old analysis of the space-time picture of perturbative multiplication of QCD partons. 
This conclusion motivated the hypothesis of local parton--hadron duality (LPHD \citd{EAO}{ADKT85}), which has been experimentally verified since then and continues to march with flying colours today.  
\smallskip

In simple terms, reinteraction between partons inside the jet would contradict causality. 
Two partons originating from a QCD cascade, either having a common parent or coming from separate branches, created simultaneously or at different time scales, have a space-like interval between them and cannot talk to (let alone collide with) one another.

\medskip

Our alternative explanation, the ``rattlesnake effect", RSE, is straightforward to reject/validate experimentally. 
All one needs to do is look into the {\em internal composition}\/ of jets selected for an abnormally large hadron multiplicity.\footnote{After this paper was completed, we learned of a forthcoming CMS publication \cite{CMS_RSE} in which the possibility of two-pronged structure of ellipticity generating events is put under scrutiny.}
To this end, one should study the hadron yield in jets with varying jet radii $R$ and employ the existing well-developed tools~\citm{jetography}{jetcorrs} to reveal the jet substructure.
\medskip

\begin{figure}[ht]
   \centering%
   \includegraphics[width=0.7\textwidth]{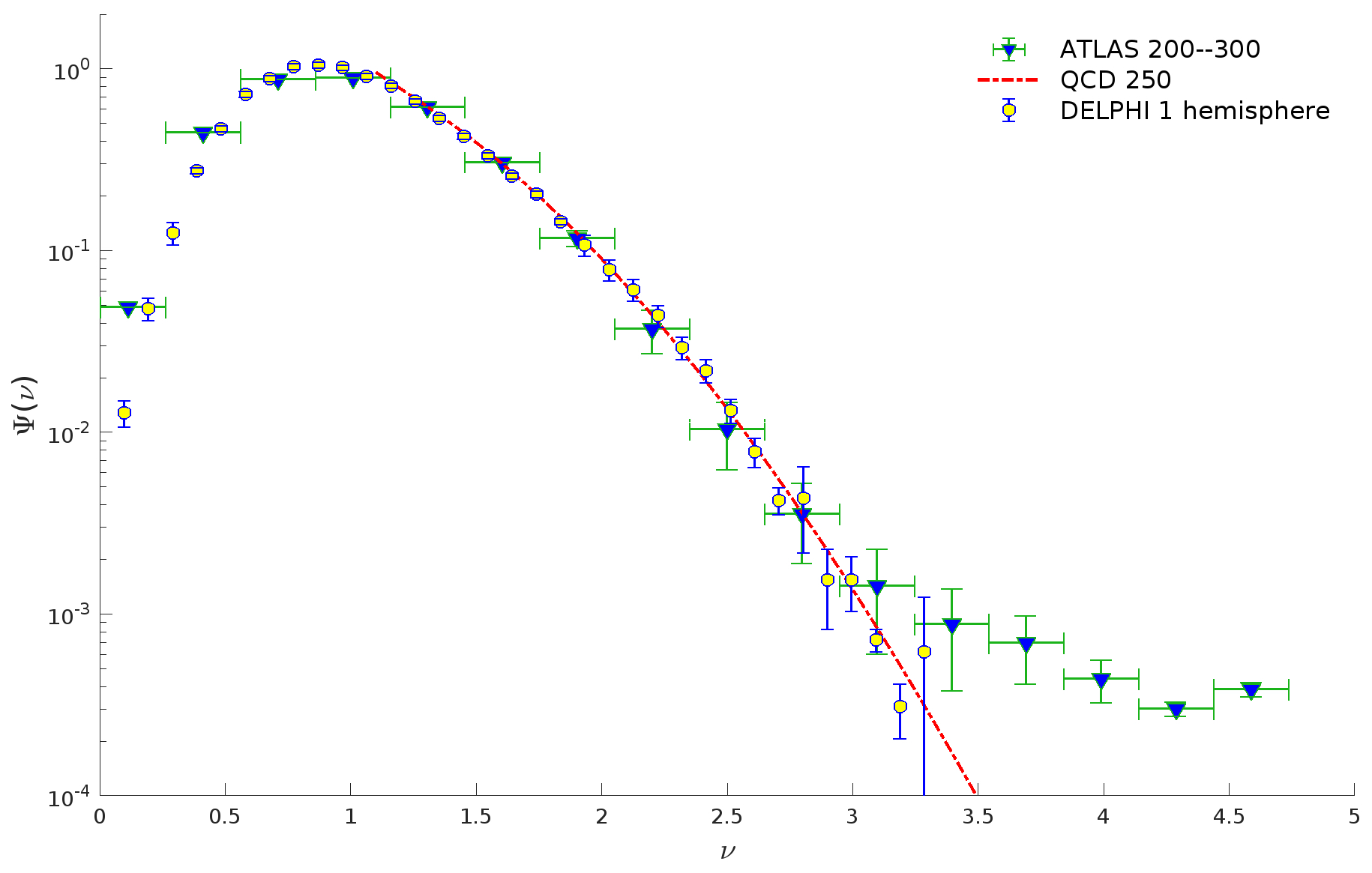}
  \vspace{-0.5cm}
\caption{\label{fig:3ATLAS-DELPHI} ATLAS and DELPHI jets with comparable hardness}
\end{figure}

To conclude, similar phenomena, namely, the slowing of the falloff of the P-KNO distribution and the funny behaviour of $v_2^*$ should manifest themselves in \ee annihilation as well, when/if the data on rare multiplicity fluctuations with $n/\lrang{n}>3$ become available.  As we show in Fig.~\ref{fig:3ATLAS-DELPHI}, presently available \ee data run out of statistics just where the enhanced tail should start to become apparent.  One may hope for confirmation/refutation of the RSE hypothesis from resurrected LEP or future \ee collider data.

It is important to stress that the RSE is  
most pronounced in the multiplicity fluctuation pattern inside a single-quark jet.
RSE is not supposed to manifest itself in the P-KNO distribution of hadron multiplicities in full $\ee$ annihilation events
(both hemispheres). 

As we have seen above, while discussing the energy share in Sec.~\ref{sec:share}, the high-multiplicity bias instigates the gluon subjet to take off maximum energy when $\rho<1$, that is when the gluon cascade is a more efficient particle producer than its parent. In the case of a \qq\ pair as a primary source of multiple hadroproduction, at modern (and foreseeable) energies $\rho_{\ee}=2\rho_q \simeq 4/3\>>\>1$. As a result, gluon radiation does its part in particle multiplication, but a single-gluon cascade is no longer motivated to take on full responsibility for it. 

\clearpage

\par \vskip 1ex
\noindent{\large\bf References}
\begin{enumerate}

\item\label{LC}
Lewis Carroll,  ``Alice's Adventures in Wonderland", 1865.

\item\label{D93}
Yu.L.  Dokshitzer,  
 ``Improved QCD treatment of the KNO phenomenon'',  \\
 \pl{305 }{295}{1993}.   
 % 295-301

 \item\label{CMS}
A.~Hayrapetyan \textit{et al.} [CMS],
 ``Observation of enhanced long-range elliptic anisotropies inside high-multiplicity jets in pp collisions at $\sqrt{s}=13$ TeV", 
\prl{133}{142301}{2024}.
% Phys. Rev. Lett. 133 (2024) 142301

 \item\label{ATLAS-2019}
G. Aad \textit{et al.} [ATLAS], 
``Properties of jet fragmentation using charged particles measured with the ATLAS detector in $pp$ collisions at $\sqrt{s}=13$ TeV,''
\pr{100}{052011}{2019}.

\item\label{AMPolyakov} 
A.M.  Polyakov,  
``{A Similarity hypothesis in the strong interactions. 1. Multiple hadron production in e+ e- annihilation}", 
\spj{32}{296}{1971};  \\
``{Similarity hypothesis in strong interactions. 2. Cascade formation of hadrons and their energy distribution in e+ e- annihilation}",
\ib{33}{850}{1971}.

\item\label{KNO}
Z.  Koba,  H.B.  Nielsen and P.  Olesen, 
``{Scaling of multiplicity distributions in high-energy hadron collisions}", 
\np{40}{317}{1972}.

\item\label{BCM}       
 A. Bassetto,  M. Ciafaloni and  G. Marchesini,    
``{Jet Structure and Infrared Sensitive Quantities in Perturbative QCD}", 
   \prep{100}{201}{1983}.
%    pages = "201--272",
   
 \item\label{DFK}         
   Yu.L.  Dokshitzer,  V.S. Fadin and V.A.  Khoze, 
 ``{Double Logs of Perturbative QCD for Parton Jets and Soft Hadron Spectra}", 
    \zp{15}{325}{1982}.  
    
% \href{book}
 {https://www.lpthe.jussieu.fr/\~{}yuri/BPQCD/BPQCD.pdf}

\item\label{Book} 
       Yu.L. Dokshitzer,  V.A. Khoze,  A.H. Mueller and S.I. Troyan, 
       {\em Basics of Perturbative QCD}, ed. J.\ Tran Thanh Van, 
       (Editions Fronti{\`e}res, 1991).

       https://www.lpthe.jussieu.fr/~yuri/BPQCD/BPQCD.pdf

\item\label{CMSsoft}
V.~Khachatryan \textit{et al.} [CMS],
``Charged Particle Multiplicities in $pp$ Interactions at $\sqrt{s}=0.9$, 2.36, and 7 TeV,''
\jhep{01}{079}{2011}
%doi:10.1007/JHEP01(2011)079
%[arXiv:1011.5531 [hep-ex]].

\item\label{BW84}    
    B.R. Webber,     ``Average Multiplicities in Jets",
    \pl{143}{501}{1984}.

\item\label{ESW}
R.K.~Ellis, W.J.~Stirling and B.R.~Webber,
``QCD and collider physics'',
Camb.~Monogr.~Part.~Phys.~Nucl.~Phys.~Cosmol.~\underline{8} (1996) 1
%Cambridge University Press, 2011.
%ISBN 978-0-511-82328-2, 978-0-521-54589-1
%doi:10.1017/CBO9780511628788

\item\label{CMW}
 S. Catani,  G. Marchesini and B.R. Webber,  
 ``QCD coherent branching and semiinclusive processes at large x", 
 \np{349}{635}{1991}.
% CMW coupling

\item\label{DKTSpec}
Yu.L. Dokshitzer,  V.A. Khoze and S.I. Troian,  
``Specific features of heavy quark production. LPHD approach to heavy particle spectra", 
\pr{53}{89}{1996}.  
% 89-119 • e-Print: hep-ph/9506425 [hep-ph]
  \item\label{MW84}
E.D.  Malaza and B.R.  Webber,  
 ``QCD Corrections to Jet Multiplicity Moments", 
\pl{149}{501}{1984}.
% Phys.Lett.B 149 (1984) 501-503 

\item\label{MW86}
E.D.  Malaza and B.R. Webber,  
 ``Multiplicity Distributions in Quark and Gluon Jets", 
\np{267}{702}{1986}.
% Nucl.Phys.B 267 (1986) 702-713 

\item\label{CT}
F.  Cuypers and K.  Tesima, ``Resummed multiplicity moments",
\zp{54}{87}{1992}.

\item\label{AHM}
A.H. Mueller, 
``Multiplicity and Hadron Distributions in QCD Jets: Nonleading Terms",
\np{213}{85}{1983}.

\item\label{EAO}
Yu.L. Dokshitzer  and S.I. Troyan,  
``Asymptotic Freedom and Local Parton--Hadron Duality", 
Proceedings of the XIX Winter School of the LNPI,  vol. 1,  page 144.  Leningrad, 1984.

\item\label{Malaza}
E.D. Malaza, ``Multiplicity distributions in quark and gluon jets to $\cO{\as}$",\\
\zp{31}{143}{1986}.

%\item\label{OPALqg}   OPAL g/q ratio
% \item\label{DELPHIqg}  DELPHI g/q ratio

\item\label{DELPHI2005}
J. Abdallah \textit{et al.} [DELPHI], 
``{Charged particle multiplicity in three-jet events and two-gluon systems}",
\epj{44}{311}{2005}.  
 %   eprint = "hep-ex/0510025",

\item\label{OPAL2002}
    G. Abbiendi \textit{et al.} [OPAL],
   ``Particle multiplicity of unbiased gluon jets from $e^{+} e^{-}$ three jet events", 
%    eprint = "hep-ex/0111013",
\epj{23}{597}{2002}.

\item\label{CLEO}
M.S. Alam  \textit{et al.} [CLEO],
 ``{Study of gluon versus quark fragmentation in $\Upsilon \to  g g \gamma$ and $\ee\to\qq\gamma$ events 
 at $\sqrt{s}$ = 10 \GeV}", 
 %   eprint = "hep-ex/9701006",
\pr{56}{17}{1997}.

\item\label{ALEPHgq}    
D. Buskulic  \textit{et al.} [ALEPH], 
    ``{Quark and gluon jet properties in symmetric three-jet events}",
    \pl{384}{353}{1996}.
 
 \item\label{OPAL1992} 
 P.D. Acton  \textit{et al.} [OPAL], 
 ``{A Study of charged particle multiplicities in hadronic decays of the Z0}",
  \zp{53}{539}{1992} .

\item\label{DELPHI} 
P. Abreu  \textit{et al.} [DELPHI],
``Production of charged particles, K0(s), K+-, p and Lambda in Z --\ensuremath{>} b anti-b events and in the decay of b hadrons",
\pl{B 347}{447}{1995}.

\item\label{L3} 
B. Adeva \textit{et al.} [L3],
``Studies of hadronic event structure and comparisons with QCD models at the Z0 resonance'',
\zp{55}{39}{1992}.
%Z.Phys.C 55 (1992) 39-62

\item\label{TASSO} 
W. Braunschweg  \textit{et al.} [TASSO], 
``Charged Multiplicity Distributions and Correlations in \ee\ Annihilation at PETRA Energies", 
\zp{45}{193}{1989}.

\item\label{LuLe}
Yu.L. Dokshitzer,  S.I. Troian  and V.A. Khoze, 
    ``{Multiple hadroproduction in hard processes with nontrivial topology. (In Russian)}", 
    \sjnp{47}{881}{1988} ; \\
P. Eden,  G. Gustafson and V.A. Khoze, 
``{On particle multiplicity distribution in three jet events}",
\epj{11}{345}{1999}.
% Eur. Phys. J. C11 (1999) 345, hep-ph/9904455

 \item\label{anti-kt}
 M. Cacciari, G.P. Salam and G. Soyez, 
 ``{The anti-$k_t$ jet clustering algorithm}",
\jhep{04}{063}{2008}.
%Journal of High Energy Physics, Volume 2008, JHEP04(2008).

% A RECENT STUDY OF ATLAS KNO 
\item\label{KNO ATLAS}
G.R. Germano,  F.S. Navarra, G. Wilk and Z. Wlodarczyk,  
 ``{Emergence of Koba-Nielsen-Olsen scaling in multiplicity distributions in jets produced at the LHC}", 
\pr{110}{034026}{2024}.
%  eprint = "2406.04856",

 \item\label{Gardner}
 Parker Gardner on behalf of the CMS collaboration, 
``Observation of QCD collectivity inside high-multiplicity jets in pp collisions with the CMS experiment", 
Talk at Quark Matter,  Sept. % 3rd-9th, 
2023.

https://cds.cern.ch/record/2862457/files/HIN-21-013-pas.pdf

\item\label{CMS 2017}
V. Khachatryan \textit{et al.} [CMS],
``Evidence for collectivity in pp collisions at the LHC", \\
\pl{765}{193}{2017}.
% Phys.Lett.B 765 (2017) 193-220
% 0.3 < p_t < 3 GeV was chosen to minimize correlations from jets while studying MB events
%  e-Print: 1606.06198 [nucl-ex]

\item\label{Texas}
A. Baty, P. Gardner and W. Li, 
``{Novel observables for exploring QCD collective evolution and quantum entanglement within individual jets}", 
\prc{107}{064908}{2023}.
%    eprint = "2104.11735",

\item\label{ADKT85}
Ya.I. Azimov \textit{et al.}, % , Yu.\ L.\ Dokshitzer, V.\ A.\ Khoze and S.\ I.\ Troyan, \\
``Similarity of Parton and Hadron Spectra in QCD Jets", \\
\zp{27}{65}{1985}.
% Z.Phys.C 27 (1985) 65

\item\label{CMS_RSE}
The CMS Collaboration, 
``Unveiling the dynamics of long-range correlations in
high-multiplicity jets through substructure engineering in
pp collisions at CMS",
CMS PAS HIN-24-024.

\item\label{jetography}
G.P. Salam, "{Towards Jetography}",
   % eprint = "0906.1833",
  \epj{67}{637}{2010}.

\item\label{jetsubstr}
M. Dasgupta, A. Fregoso, S. Marzani and G.P. Salam,
"{Towards an understanding of jet substructure}",
  %  eprint = "1307.0007",
\jhep{09}{029}{2013}.

\item\label{jetcorrs}
    D. Adams \textit{et al.},
    "{Towards an Understanding of the Correlations in Jet Substructure}",
   % eprint = "1504.00679",
    \epj{75}{409}{2015}.

\item\label{YDBrems}
Yu.L. Dokshitzer,``{Bremsstrahlung}",
Contribution to: 11th NATO Advanced Study Inst.\ on Techniques and Concepts of High-Energy Physics,
    H.B. Prosper, B. Harrison and M. Danilov (ed),
    NATO Sci. Ser. \underline{C 566} (2001) 51.

\item\label{Nemes}
Gerg\"o Nemes, 
``New asymptotic expansion for the Gamma function", 
{\em Archiv der Mathematik}~\underline{95} (2010) 161.

\end{enumerate}

\section*{Appendices}

\appendix

\renewcommand{\theequation}{\thesection.\arabic{equation}}
% #########################################################
\mysection{Running coupling\label{AppCoupl}}
\setcounter{equation}{0}

We employ the 2-loop coupling satisfying
\begin{subequations}\label{2loopbeta}
\beq
\frac{d}{d\ln Q^2} \frac{\as(Q)}{4\pi} = - \beta_0 \left( \frac{\as(Q)}{4\pi} \right)^2 - \beta_1\left( \frac{\as(Q)}{4\pi} \right)^3
\eeq
with
\beq
 \beta_0 = \frac{11}3N_c-\frac23 n_f = 11 - \frac23 n_f  ,  \quad \beta_1 = \frac{34}3N_c^2 -2C_F n_f - \frac{10}3N_c  n_f =  {102} -  \frac{38}{3}n_f .
\eeq
\end{subequations}
This yields the differential equation 
\beq
 \frac{d}{d\ln Q} \left( \frac{2\pi}{\as(Q^2)}\right) =  \beta_0 + \frac12\beta_1  \left(\frac{\as(Q)}{2\pi}\right)
\eeq
with the approximate  solution
\beq
  \frac{\as(Q)}{2\pi} = \left( \beta_0 Y + \frac{\beta_1}{2} \ln Y + \mbox{const} \right)^{-1},
\eeq
where $Y=\ln(Q/\LQCD)$. This form, where one keeps the term $\ln Y$ in the denominator without  expanding in $\ln Y/Y$, is convenient in one respect:
it allows one to easily relate the desired value of the coupling with the $ \LQCD$  parameter.
For example,  should we want $\as(M_Z)$ to be equal to a given value $\alpha_Z$,  we simply solve the equation
\begin{subequations}\label{alZ}\beq
  \alpha _Z  = {2\pi} \left({ \beta_0 \ln\frac{Q}{\LQCD} + \frac{\beta_1}{2} \left[ \ln\ln \frac{Q }{\LQCD} -\ln \ln \frac{M_Z}{\LQCD} \right]}\right)^{-1} , 
\eeq
\beq
 \LQCD = M_Z\cdot \exp \frac{2\pi}{\beta_0 \alpha_Z} .
\eeq
\end{subequations}
By the way,  both the second loop term and not expanding the denominator have a little impact on the results. 

We remind the reader that since multiplicities are driven by soft-gluon radiation,  we employ the {\em physical} coupling, also known as the ``bremsstrahlung" or CMW \cite{CMW}, scheme.

% \vspace{1 true cm}

%%%%%%%%%%%%%%%%%%%##%
\mysection{Accuracy of the asymptotic P-KNO model \label{AppDLA}}
\setcounter{equation}{0}

Multiplicity moments are generated by Taylor expansion of the Laplace transform about the origin:
\beq
  \Phi(-\beta) = \sum_{k=0}^\infty  \frac{\beta^k}{k!}\, g_k. 
\eeq
The DLA moments of the gluon jet multiplicity distribution come from the corresponding recurrence relation \citd{DFK}{Book}
\beq\label{eq:Phi5}
 \Phi(-\beta) = 1 + \beta + \frac43 \frac{\beta^2}{2!} + \frac94 \frac{\beta^3}{3!} 
 + \frac{208}{45} \frac{\beta^4}{4!} + \frac{2425}{216} \frac{\beta^5}{5!} + \ldots
\eeq
Moments of high rank $k\gg1$ are determined by the singularity of $ \Phi$ positioned at $z=-\beta/C =1$:
\beq\label{eq:PhiBook}
\Phi^{\mbox{\scriptsize as}}(-\beta) = 2\left[ \frac{z}{(1-z)^2} -\frac13\ln(1-z) \right] +\cO{1}.
\eeq
Expanding this formula near $z=0$ produces the asymptotic moments \eqref{eq:gkDLA}.
Eq.\ \eqref{eq:PhiBook} was derived for $z$ close to 1 to describe high-rank moments.
Nevertheless, one can attempt to apply it to all moments, including low-rank ones.
To do that, it suffices to adjust \eqref{eq:PhiBook} to ensure the correct normalization, $g_0=1$, by fixing the $\cO{1}$ term:
\beeq\label{eq:Phias}
\Phi^{\mbox{\scriptsize as}}(-\beta) &=& 1+ 2\left[ \frac{z}{(1-z)^2} -\frac13\ln(1-z) \right] 
= 1+ 2\sum_{k=1}^\infty \left[k + \frac{1}{3k} \right] \left(\frac{\beta}{C} \right)^k, \\
\label{eq:ekasB}
   g_{k}^{\mbox{\scriptsize DLA}} &=& \frac{2k!}{C^k} \left(k + \frac{1}{3k} \right).
\eeeq
For a quark jet or jet ensemble with effective ``source strength" $\rho$,
the Laplace transform becomes
\beq
  \Phi^{(\rho)}(-\beta) = \sum_{k=0}^\infty  \frac{\beta^k}{k!}\, g_k^{(\rho)}
  = \left[\Phi\left(\frac{-\beta}{\rho}\right)\right]^\rho.
\eeq
Substituting the expansions \eqref{eq:Phi5} and \eqref{eq:Phias}
one can compare the first few 
moments coming from the exact solution and the large-$k$ asymptotic formula for any jet ensemble.

\subsection{Gluon jet}
\medskip

\begin{center}
\begin{tabular}{c||c|c||c|c}
&  recurrence & numeric   & asymptotic &numeric   \\ \hline
$g_1$    &  1 &   1.000 &  $\frac8{3C}$  &  1.045 \\[1mm]
$g_2$   & $ \frac43$ & 1.333 & $\frac{26}{3C^2}$ &  1.330\\[1mm]
 $g_3$  &  $\frac94$      &2.250 & $\frac{112}{3C^3}$  & 2.244   \\ [1mm]
 $g_4$  & $\frac{208}{45}$   &  4.622  & $\frac{196}{C^4}$ &4.616 \\[1mm]
$g_5$  & $\frac{2425}{216} $  & 11.227  & $\frac{1216}{C^5}$ & 11.218
\end{tabular}
\end{center}
%\end{table}
\smallskip

  At large $k$ the remainder is proportional to $g_k/k^3$.

\subsection{Quark jet}
For quark jet phenomenology, instead of the canonical 4/9 we use $\rho$ = 2/3.
This gives the following table
\medskip

\begin{center}
\begin{tabular}{c||c|c||c|c}
&  recurrence & numeric   & asymptotic &numeric   \\ \hline
$g_1$    &  1 &   1.000 &  $\frac{8}{3C} $  &   1.045  \\[1mm]
$g_2$   & $ \frac32$ & 1.500 & $\frac{85}{9C^2}$ & 1.449 \\[1mm]
 $g_3$  &  $\frac{49}{16}$   &3.063 & $\frac{1376}{27C^3}$  &  3.064  \\ [1mm]
 $g_4$  & $\frac{319}{40}$   &  7.975  & $\frac{27352}{81C^4}$ & 7.953\\[1mm]
$g_5$  & $\frac{3243}{128} $  & 25.336 & $\frac{668498}{243C^5}$ & 25.380
\end{tabular}
\end{center}

\subsection[$e^+e^-$]{\boldmath $e^+e^-$}  
For $e^+e^-$ phenomenology, we use $\rho$ = 4/3.
This gives the following table
\medskip

\begin{center}
\begin{tabular}{c||c|c||c|c}
&  recurrence & numeric   & asymptotic &numeric   \\ \hline
$g_1$    &  1 &   1.000 &  $\frac{8}{3C} $  &   1.045  \\[1mm]
$g_2$   & $ \frac54$ & 1.250 & $\frac{149}{18C^2}$ & 1.270 \\[1mm]
 $g_3$  &  $\frac{121}{64}$   &1.891 & $\frac{854}{27C^3}$  &  1.901  \\ [1mm]
 $g_4$  & $\frac{1079}{320}$   & 3.372  & $\frac{93059}{648C^4}$ & 3.382\\[1mm]
$g_5$  & $\frac{14227}{2048} $  & 6.947 & $\frac{1466089}{1944C^5}$ & 6.958
\end{tabular}
\end{center}

In conclusion, we observe that the asymptotic approximation is already quite good even at $k= 1$, for all relevant values of $\rho$.

% \vspace{1 true cm}

\mysection{\boldmath Factors $\chi(k)$, $\cF(k)$ \label{AppGam}}
\setcounter{equation}{0}

% #########################################################

\subsection{$\Gamma$--miracles}

Gerg\"o Nemes proposed the following approximate formula \cite{Nemes}
\beq
{\displaystyle \Gamma (z)\approx {\sqrt {\frac {2\pi }{z}}}\left({\frac {1}{e}}\left(z+{\frac {1}{12z-{\frac {1}{10z}}}}\right)\right)^{z}} .
\eeq
A simpler (though less accurate at large $z$) approximation
\begin{subequations}
\beq\label{Gammasimp}
{\displaystyle \Gamma (z) =  {\sqrt {\frac {2\pi }{z}}} \left( {\frac {1}{e}}\left(z+{\frac {1}{12z}}\right)\right)^{z}} \cdot \left[ 1 + \cO{\frac{10^{-3}}{z^3}} \right]
\eeq
does a fantastic job for small (but important!) values of the argument:
\beeq
 \Gamma (1) &\simeq & \frac{\sqrt  {2\pi }}{e} \cdot \left(1+\frac{1}{12}\right)  \approx 0.998982  =  1-\cO{10^{-3}} ,  \\ 
 \Gamma (2) &\simeq & \frac{\sqrt  {\pi }}{e^2} \cdot \left(2+{\frac {1}{24}} \right)^{2}  \approx 0.999898 = 1-\cO{10^{-4}}  .
\eeeq
\end{subequations}
To improve the approximation of the interpolating factor $\chi(k)$ let us use
\beeq\label{Gammashift}
 \Gamma(1+z) & \approx & 
 \sqrt{2\pi(1+z)} e^{-(1+z)} (1+z)^{z} \left(1 + {\frac {1}{12(1+z)^2}}\right)^{1+z} .
\eeeq

\subsection{Ratio of  $\Gamma$ functions}   
First we apply eq.(\ref{Gammashift}) to $\Gamma(1+\gamma k)$: 
\begin{subequations}\label{3Gammas}
\beeq
 \Gamma(1+\gamma k) &=&  \frac{\sqrt {2\pi (1+\gamma k)}}{e} e^{-\gamma k} (1+\gamma k)^{\gamma k}
  \left( 1 + \frac{1}{12(1+\gamma k)^2}\right)^{1+\gamma k} \nonumber \\
 &=&   \frac{\sqrt {2\pi (1+\gamma k)}}{e} \gamma^{\gamma k}  \left( \frac{k}{e}\right)^{\gamma k}\, \left[ 1 + \frac{1}{\gamma k}\right]^{\gamma k}
 \cdot  \left( 1 + \frac {1}{12 (1 + \gamma k)^2} \right)^{1+\gamma k} 
\eeeq
From here one easily gets
\beeq
\Gamma(1+k) &=&  \frac{\sqrt {2\pi (1+  k)}}{e}   \left( \frac{k}{e}\right)^{  k}\, \left[ 1 + \frac{1}{ k}\right]^{k}
 \cdot  \left( 1 + \frac {1}{12 (1 +  k)^2} \right)^{1+k} , \\
\Gamma(1+(1\!-\!\gamma)k) &=&  \frac{\sqrt {2\pi (1+(1\!-\! \gamma) k)}}{e} (1\!-\! \gamma)^{(1\!-\! \gamma) k}  \left( \frac{k}{e}\right)^{(1\!-\! \gamma) k}\, \left[ 1 + \frac{1}{(1\!-\! \gamma) k}\right]^{(1\!-\! \gamma) k} \nonumber \\
 && \cdot  \left( 1 + \frac {1}{12 (1 + (1\!-\! \gamma) k)^2} \right)^{1+ (1\!-\! \gamma) k} .
\eeeq
\end{subequations}

Constructing the ratio of the $\Gamma$ functions (\ref{3Gammas}) one obtains
\beeq\label{3Gratio}
&& \frac{\Gamma(1+k)}{\Gamma(1+\gamma k)\Gamma(1+(1\!-\!\gamma) k)}  =
\frac{e}{\sqrt{2\pi}}
 \sqrt{\frac{1+k}{(1+\gamma k)(1+ (1\!-\!\gamma)k)}}   \cdot
 \bigg[ \gamma^\gamma (1\!-\!\gamma)^{1\!-\!\gamma}\bigg]^{-k}  \nonumber \\
{ }\quad  &&\cdot \left[ 1 + \frac{1}{\gamma k}\right]^{ -\gamma k}  
\times \frac{  \left( 1 + \frac {1}{12 (1 +  k)^2} \right)^{1+k}} { \left( 1 + \frac {1}{12 (1 + \gamma k)^2} \right)^{1+\gamma k}    \left( 1 + \frac {1}{12 (1 + (1\!-\! \gamma) k)^2} \right)^{1+ (1\!-\! \gamma) k}}.
\eeeq
Here we have dropped two factors whose ratio is close to 1 for both small and large $k$:
\[
\frac{\left[ 1 + \frac{1}{ k}\right]^{ k} } {\left[ 1 + \frac{1}{(1\!-\! \gamma) k}\right]^{ (1\!-\! \gamma) k}} = \left.  \frac11 \right|_{k=0} =  \left.  \frac ee \right|_{k\to \infty} \> = 1\bigg|_{\gamma \to 0}.
\]
We might simplify eq.(\ref{3Gratio}) even further by disregarding the two  factors in the ``1/12" correction terms:   % in (\ref{3Gratio}),
\beq\label{eq:R2}
R_2 = \frac{  \left( 1 + \frac {1}{12 (1 +  k)^2} \right)^{1+k}} {  \left( 1 + \frac {1}{12 (1 + (1\!-\! \gamma) k)^2} \right)^{1+ (1\!-\! \gamma) k}}. 
\eeq 
In this way, we would lose control of the relative correction $\cO{k^{-1}}$ in the tail while preserving the DLA limit ($\gamma\to 0$)  and the correct normalization (the $k=0$ moment).
As a function of moment $k$, the ratio \eqref{eq:R2} 
departs from $R_2(0)\!=\!1$ by decreasing with a mild slope $R'(0)= -\gamma/12$, hits the minimal value
$R_{\min}= 1-\frac{\gamma}{12(1+\sqrt{1\!-\!\gamma})^2}$ at $k=\frac{1}{\sqrt{1\!-\!\gamma}}$ and then returns to unity
as $1-\frac{\gamma}{(1\!-\!\gamma)k}$.
The deviation of this ratio from unity might be as large as 8\%\ in the unrealistic $1\!-\!\gamma \ll 1$ region.  
In reality,  this deviation does not exceed 1\%. 
So  the approximation $R_2=1$ seems reasonable. 
 
The simplified $\chi$ factor becomes
\beq\label{chi-full}
  \chi(k) =  \sqrt{\frac{1+k}{(1+\gamma k)(1+ (1\!-\!\gamma)k)}}  
  \>\frac{e \left[ 1 + \frac{1}{\gamma k}\right]^{ -\gamma k}}{\sqrt{2\pi}} 
 %  \frac{\left[ 1 + \frac{1}{\gamma k}\right]^{- \gamma k} }{\sqrt{(1+\gamma k) (1\!-\!\gamma)}}  
\cdot   \left(  {1 + \frac {1}{12 (1 + \gamma k)^2}} \right)^{-(1 + \gamma k)}
\eeq
At the point responsible for proper normalization of the P-KNO function,  the moment $k=0$,  
another miracle occurs in eq.(\ref{chi-full}) which also applies to $\gamma=0$ (the limiting P-KNO distribution or `DLA'):
\beq\label{eq:miracle3}
  \chi(k=0) =  \chi(\gamma=0) = \frac{e}{\sqrt{2\pi}} * \frac{12}{13} = 1.001019  \qquad {\mbox{sic!}}   % 27823
\eeq

\subsection{An alternative approach to the ratio of $\Gamma$ functions and $\cF(k)$}

With the help of Nemes approximation \eqref{Gammasimp} it is straightforward to derive another relation that covers factorials, exponents {\bf and} powers of $k$:
\begin{subequations}\label{eq:cFversion}
\beeq\label{Gamma2R}
\big[\gamma^{ \gamma }(1\!-\!\gamma)^{(1\!-\!\gamma) }\big]^{k } \cdot \frac{ \Gamma (1+ k)}{ \Gamma (1+ \gamma k)} &=&  {\Gamma (\half + (1\!-\!\gamma) k)} \cdot \cF.
\eeeq
Here $\cF$ is a new slowly changing prefactor in place of $\chi$ \eqref{eq:chidef}
\beeq\label{eq:cFdef2}
\cF (k,\gamma) &=&   \frac1{\sqrt{2\pi}}   \sqrt{\frac{1+k}{1 +\gamma k} }
\cdot \frac{\cR(1, k)}{\cR(1, \gamma k) \cR(\half,(1\!-\!\gamma)k) },
\eeeq
where
\beq
 \cR(\Delta, k) \equiv e^{-\Delta} \left(1+\frac\Delta{k}\right)^k \cdot \left(1+ \frac 1{12(\Delta+k)^2}\right)^{\Delta+k}.
\eeq
\end{subequations}
\begin{figure}[hb]
\centering{
\includegraphics[width=0.7\textwidth]{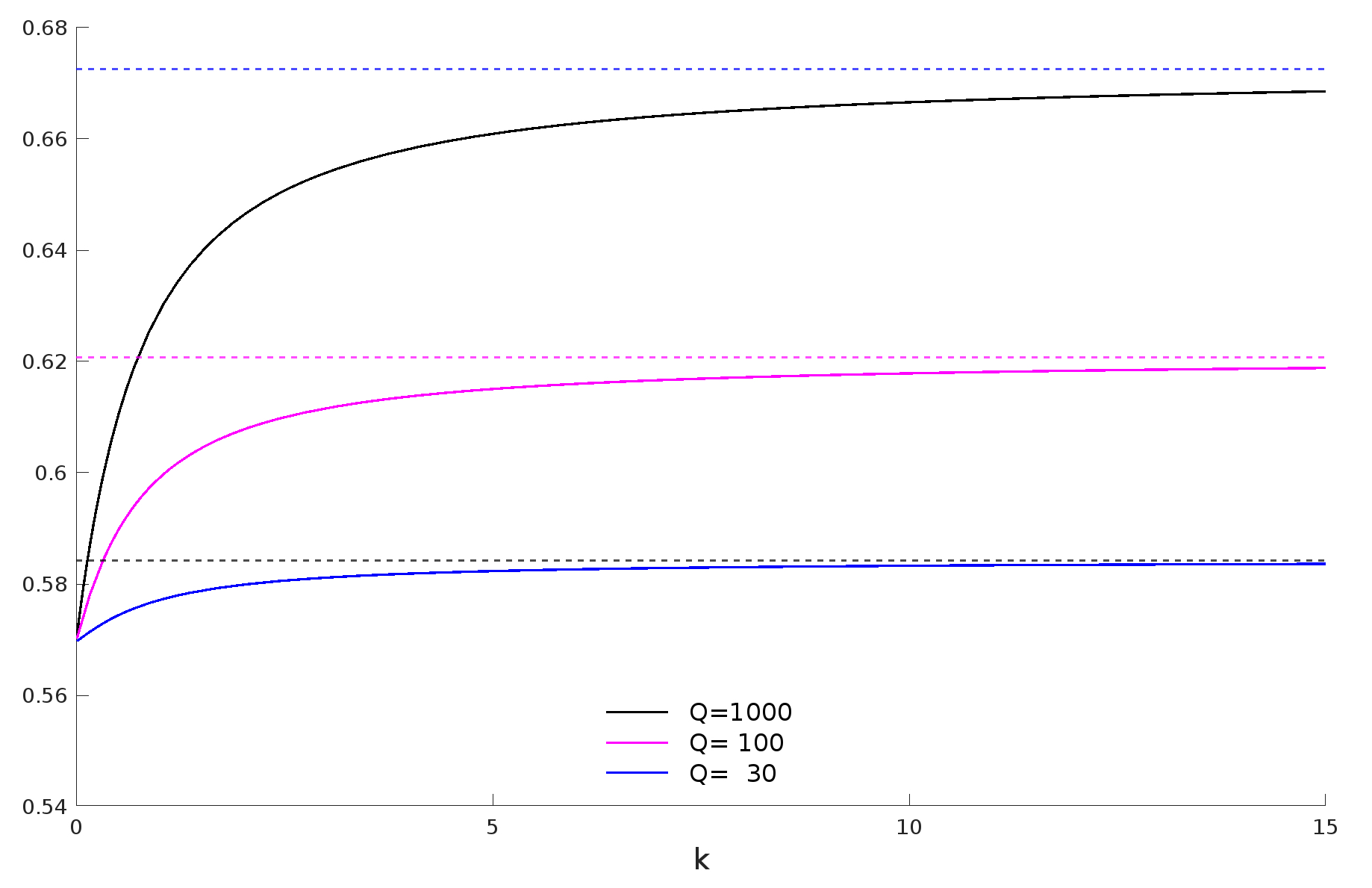}
\caption{\label{fig:cF}$\cF$ as a function of rank $k$ of multiplicity moments} }
\end{figure}
Figure~\ref{fig:cF} confirms that the factor $\cF(k)$ does not vary much over the whole range of $k$.

\vspace{1 true cm}

% #########################################################
\mysection{Derivation of the P-KNO tail \label{AppPsi}}
\setcounter{equation}{0}

In this Appendix we show how to restore $\Psi(\nu)$ from its moments.

The moments \eqref{eq:gkdef} read 
\beeq 
   g_k &=& g_{k}^{\mbox{\scriptsize DLA}}\cdot \frac{[\Gamma(1+\gamma)]^k}{\Gamma(1+k\gamma)}, \qquad   g_{k}^{\mbox{\scriptsize DLA}} = \frac{2 k!}{C^k}\left(k+ \frac1{3k}\right) \qquad (k\gg 1), \\
   g_k &=& \frac{2}{\Ct^k}\left(k+ \frac1{3k}\right) \cdot \frac{k!}{\Gamma(1+\gamma k)}, \qquad \mbox{with} \quad \Ct \equiv \frac{C}{\Gamma(1+\gamma)}. 
\eeeq
Use the relation
\[
 \frac{k!}{\Gamma(1+\gamma k)} =  \frac{\Gamma(1+k(1\!-\!\gamma))}{\left[ \gamma^\gamma (1\!-\!\gamma)^{1\!-\!\gamma}\right]^{k} }  \cdot  \chi(k)
\]
with the finite factor $ \chi(k)$ from \eqref{eq:chidef}, which decreases slowly above $k\sim 1$ (see Appendix \ref{AppGam}).
We drop the factor $\chi$ for now and when it comes to finding the distribution $\Psi(\nu)$, we will recall and evaluate it at a characteristic point $k=\ksd(\nu)$.

\subsection{Constructing the source function $\Psi(\nu)$}
\subsubsection{$\chi$ version}

\noindent
The moments take the form
\begin{subequations}\label{eq:gkall}
\beq
g_k =2\left(k+ \frac1{3k}\right)\frac{\Gamma(1+k(1\!-\!\gamma))}{D^k} \,\chi(k). % \qquad D\equiv 
\eeq
with $D(\gamma)$ defined by \eqref{eq:Ddef}.

Observing that
\[
 k\cdot\Gamma(1+k(1\!-\!\gamma))
=\frac{\big([ 1+k(1\!-\!\gamma)] -1\big)\, \Gamma(1+k(1\!-\!\gamma)) }{1\!-\!\gamma} ,
\]
and
\[
\frac1k \Gamma(1+k(1\!-\!\gamma))= (1\!-\!\gamma)\Gamma(k(1\!-\!\gamma)),
\]
we get a sum of Gamma-functions with shifted arguments (modulo the factor $\chi$)
\beq\label{eq:gk3}
  {g}_k =
  \frac{ 2\mu}{ D^{k}}\Big[ \Gamma(2 + k\mu^{-1} ) - \Gamma(1 + k\mu^{-1}) + \frac{1}{3\mu^2} \Gamma(k\mu^{-1}) \Big].
\eeq
\end{subequations}
Now we look for a function $\Psi_\tau$ ($\tau=2,\, 1,\,0$) whose Mellin moments reproduce 
\beq\label{eq:GD}
 \Gamma(\tau + k\mu^{-1} ) \cdot D^{-k} .
\eeq
Taking $\Psi_\tau$ in the form
\beq
 \Psi_\tau(x) = \psi_0 \cdot D (Dx)^{\epsilon_\tau} e^{-[Dx]^\mu} ,
\eeq
and introducing $z=(Dx)^\mu$ as integration variable,
\beeq\label{eq:Gint}
  \int_0^\infty dx\, x^k \Psi_\tau (x) &=&  \psi_0 D^{-k} \int_0^\infty dy\, y^{k+\epsilon_\tau} e^{-y^\mu} 
  = \frac{\psi_0}{\mu} D^{-k}  \int_0^\infty dz \, z^{[k+\epsilon_\tau -(\mu-1)]/{\mu}} e^{-z} \nonumber  \\ 
  &=&   \left[ \frac{\psi_0}{\mu}\right]\cdot  \frac{\Gamma(k\mu^{-1}+ (\epsilon_\tau + 1)\mu^{-1} )} {D^k} .
\eeeq

\noindent
Compared with (\ref{eq:GD}) the result is as follows
\begin{subequations}\label{eq:GammaSol}
\beeq
 \mu &=& \frac1{1\!-\!\gamma} ,  \\
\psi_0 &=& \mu ,  \\
 \epsilon_\tau &=& \tau\cdot \mu-1  .
\eeeq
\end{subequations}
Recalling the factor $\chi(k)$, the final answer for the multiplicity distribution in a gluon jet becomes
\beq\label{eq:Psi_Psi}
\Psi(\nu) =   \frac{ 2\mu^2}{\nu } \left( [D\nu]^{2\mu} 
- [D\nu]^{\mu} + \frac{1}{3\mu^2}  \right) e^{- [D\nu]^{\mu} }  \cdot \chi(\ksd) .
\eeq
At the point $\gamma=0$ \eqref{eq:Psi_Psi} turns into the DLA answer.
%\end{subequations}

%%%%%%%%%%%%%%
 \subsubsection{$\cF$ version} 
Alternatively, we may use the representation \eqref{eq:cFversion} for the ratio of the Gamma functions.

Invoke the DLA factor,
\[
2\left(k+\frac{1}{3 k}\right)\cdot \frac{ \Gamma (1+ k)}{ \Gamma (1+ \gamma k)},
\]
and use machinery similar to that of \eqref{eq:gkall}. 

Observing that
\[
 k\cdot\Gamma(\half+k(1\!-\!\gamma))
=\frac{\big([ \half+k(1\!-\!\gamma)] -\half\big)\, \Gamma(\half+k(1\!-\!\gamma)) }{1\!-\!\gamma} 
= \mu \bigg[\Gamma(\threehalf+k(1\!-\!\gamma)) -\half \Gamma(\half+k(1\!-\!\gamma)) \bigg] ,
\]
we get a sum of Gamma-functions with shifted arguments:
\beq\label{eq:gk32}
  {g}_k =
  \frac{ 2\mu}{ D^{k}}\left[ \Gamma(\threehalf + k\mu^{-1} ) - \left( \frac12 - \frac{1}{3\mu k}\right) \Gamma(\half + k\mu^{-1}) \right] e^{-[D\nu]^{\mu}} \cdot {\cF}.
\eeq
We did not play with the last ``1/3" correction term $\cO{k^{-2}}$ since the perspective of dealing with $\Gamma$ of a negative argument did not look attractive. Anyway, its contribution is hardly noticeable.

This leads to an alternative form of the answer:
\beq\label{eq:newPsi}
\Psi(\nu) =   \frac{ 2\mu^2}{\nu } \left( [D\nu]^{\threehalf\mu} 
- \frac12[D\nu]^{\half\mu} + \frac{1}{3\mu} [D\nu]^{-\half\mu}  \right) e^{- [D\nu]^{\mu}} \cdot \cF(\ksd) .
\eeq
This one may look less pretty than \eqref{eq:Psi_Psi} but can come in handy.
Here, the leading power of $\kappa=[D\nu]^\mu$ has decreased from 2 to $\threehalf$ because a $1/\sqrt{\kappa}$ factor was taken into account from $\chi$ to determine the optimal argument of the function $\Gamma$ in \eqref{eq:cFform}.  

The two expressions are formally identical asymptotically, $k\to \infty$.
However, the difference between \eqref{eq:Psi_Psi} and \eqref{eq:newPsi} exists and has a reason and a certain meaning. 
The point is, what values of $\nu$ are we interested in. 

The key parameter that determines which of the two is better suited is the value of $\gamma\cdot k$. It is the very parameter that formed the basis for the MDLA approach.

\smallskip
Equations \eqref{eq:Psi_Psi}, \eqref{eq:newPsi} solve the quest.

\subsection{Characteristic moment rank $\ksd$ \label{App:charmom}}

To finalise the evaluation of the P-KNO distribution, we need to determine the argument of the prefactors $\chi(k)$ and $\cF(k)$ in \eqref{eq:Psi_chi} and \eqref{eq:Psi_cF} correspondingly.

\noindent
\begin{minipage}{0.6\textwidth}
In order to adequately determine the characteristic moment rank $\ksd$ that is the core element of our back-and-forth exercise in the Laplace transformation, one needs the prefactor, $\chi$ or $\cF$, to be roughly constant. Neither of them is precisely so, but their slow change with $k$, either when $k\,\to\,\infty$ (the case of $\chi$) or in the origin ($\cF$) is guaranteed to leave the height and shape of the tail unaffected. 
\end{minipage}
\begin{minipage}{0.4\textwidth}
\centering
\begin{tabular}{c|c|c}

factor & at $k=0$  & at $k\to\infty$  \\[1 mm] \hline \\
\vspace{1 mm}

% & & & \\
$\chi(k)$ & $\displaystyle{ \frac{12}{13} \frac{e}{\sqrt{2\pi}} } \approx 1.0010$  & $\displaystyle{\frac1{\sqrt {2\pi \gamma(1\,-\,\gamma )\cdot k }}}$  \\[1 mm]
$\cF(k)$ & $\displaystyle{\sqrt{\frac{3e}{ 8\pi}}}\approx 0.5696$ & $\displaystyle{\frac1{\sqrt{2\pi\gamma}}} $ 
\end{tabular}
\end{minipage}

\medskip

The characteristic moment rank $\ksd(\nu)$ can be determined from the position of the Steepest Descent point 
of the integration over $\ln z$ in \eqref{eq:Gint}. 
Analysing the interplay between the exponential falloff and increase of the integrand driven by the leading power of $\kappa$, we deduce
$\ksd +\epsilon_2 +1 = \mu z$:
\begin{subequations}\label{eq:ksd_both}
\beeq\label{eq:ksd2}
 	 \ksd(\nu) \>&=& \>\mu \,\left( [D\nu]^\mu -2\right) \>=\>\mu \,\bigg( \kappa(\nu) -2\bigg), \\
     \label{eq:ksd32}
    	 \ksd(\nu) \>&=& \>\mu \,\left( [D\nu]^\mu -\frac32\right) \>=\>\mu \,\left( \kappa(\nu) -\frac32\right),
\eeeq
\end{subequations}
for the two representations. 

The Steepest Descent point is shifted in the $\cF$ version, and the characteristic moment $\ksd$ in \eqref{eq:ksd32} 
now has $\threehalf$ in place of the 2 in \eqref{eq:ksd2}.

Both prescriptions are valid for the tail of the distribution and work well as long as $\nu\ga 1$.
\noindent

Numerically, $\ksd$ is rather large. If $\as$ were very small, the tail would have been exponential and $\ksd\simeq C\nu\sim 5\div 8$ for $\nu=2\div3$. 
In reality $\gamma\simeq 0.4$,
$ D\simeq 1.5, \quad \mu\simeq 5/3$,
and $\ksd(\nu)$ reaches $20\div30$ and more for realistic values of $\nu$.

At the same time, trouble hits when one attempts to continue the tail $\Psi(\nu)$ down towards the maximum of the distribution, since both expressions \eqref{eq:ksd_both} vanish at some point around $\nu\!\approx\!{1}$.
Below this point $\chi$, $\cF$ become complex, setting the limit of applicability of the tail formule \eqref{eq:Psi_Psi}, \eqref{eq:newPsi}.

So we also tried a simplified version
\beq\label{eq:ksd_simp}
 \ksd \>=\> \mu\cdot\kappa \>=\> \mu(\gamma)\bigg[D(\gamma)\nu\bigg]^{\mu(\gamma)} 
\eeq
Formally speaking, it is not as accurate as \eqref{eq:ksd_both}, but does the job quite well.

\subsection{Comparison}
\stepcounter{figure}

\noindent
\begin{minipage}{0.48\textwidth}
In Fig.~\thefigure\ we compare  different choices of the $\ksd$ characteristic moment rank to show how they 
%alternative representations \eqref{eq:Psi_fin} 
compete in practice. 

The tails are indistinguishable. 
Meanwhile, the ``$\threehalf$" version \eqref{eq:Psi_cF} extends slightly further below $\nu=1$ than its counterpart \eqref{eq:Psi_chi}. The vertical lines mark the domain of applicability of these respective formulae.

The simplified version of the $\ksd$ choice \eqref{eq:ksd_simp} applied to the "$\threehalf$" representation \eqref{eq:Psi_cF}
is very close to its full version \eqref{eq:ksd32}. This is explained by the fact that, as we saw in Fig.~\ref{fig:cF}, the factor $\cF(k)$ varies very little over the relevant range of moment ranks.
\end{minipage}
\begin{minipage}{0.52\textwidth}
\centering
   \centering%
  \includegraphics[width=0.95\textwidth]{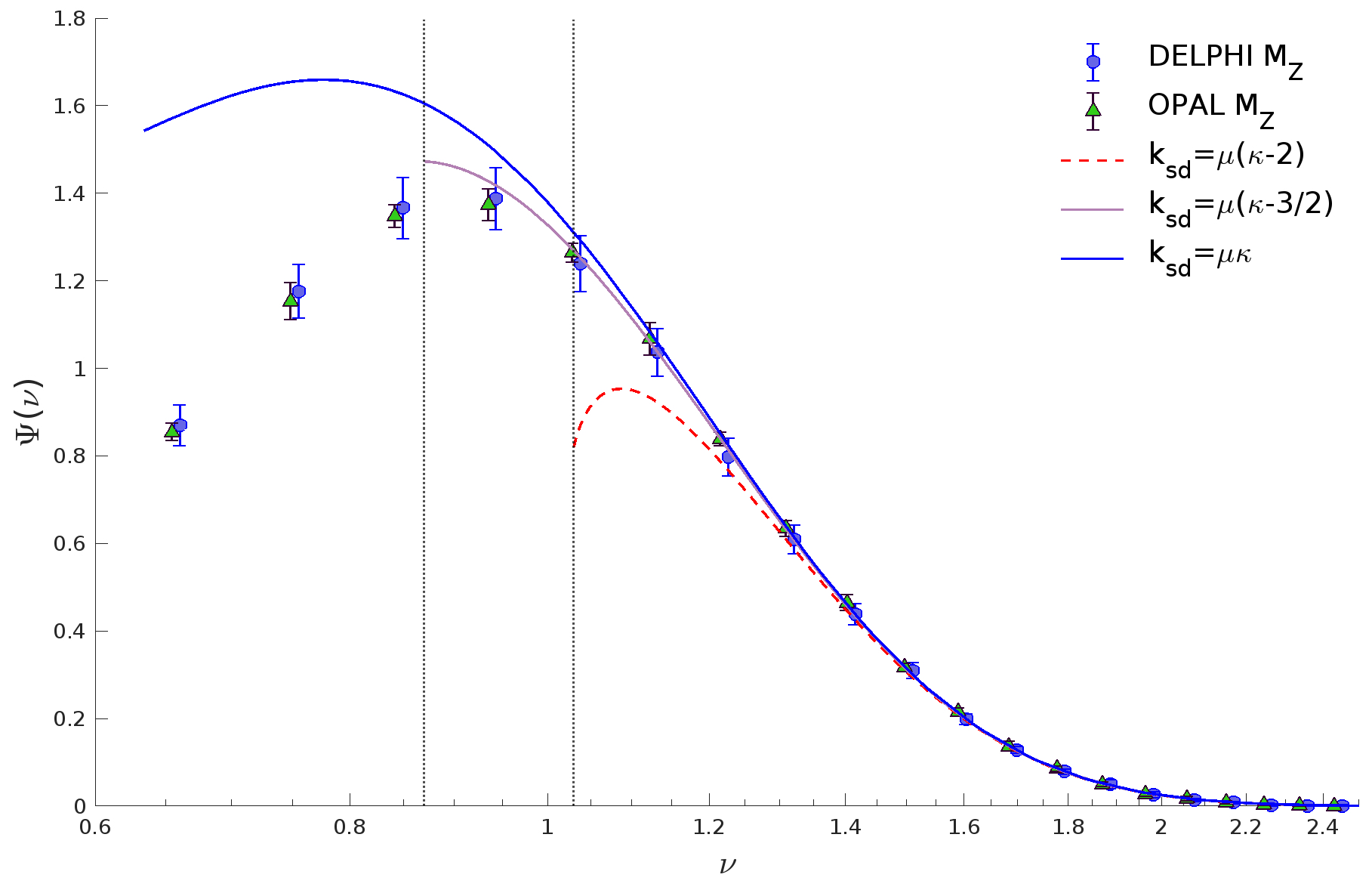} % \epsfig{file=2OPDE 32vs2.png,  width= 0.8\textwidth, angle=0} 
Figure \thefigure
\end{minipage}
\bigskip

Keeping this in mind, we chose to use the simplified prescription for the value of the characteristic moment rank $\ksd$ to use in practice. 
All plots of this paper where theoretical predictions are compared with experimental observations were made within this prescription: the version of the distribution \eqref{eq:Psi_cF} with $\ksd$ given by \eqref{eq:ksd_simp}.
\smallskip

We make this note not because the temptation to peek down below 1 is legitimate within our approach but rather because it raises a question to be addressed in the future.
\vspace{1 true cm}

% #########################################################

\mysection{Constructing $\Psi^{(\rho)} $ in the general case $\rho\neq 1$ \label{AppRho}}
\setcounter{equation}{0}

We represent $\Psi$ as
\[
\Psi(x) = \frac1x \,e^S, \qquad S = \ln(x\Psi(x)) .
\]

\subsection{Laplace transform $\nu\to\beta$}
Evaluate the Laplace transform:
\begin{subequations}
\beeq\label{eq:expon}
\Phi(\beta) &=& \int_0^\infty \frac{dx}{x}\, \exp{( S(x) -\beta x)}. 
\eeeq
The Steepest Descent in $\ln x$ gives
\beeq\label{eq:Sx}
S_x(x_{sd}(\beta)) &= & \beta \, x_{sd}(\beta), \qquad S_x(x)\equiv  \frac{d S(x)}{d\ln x} .
\eeeq
\eqref{eq:Sx} determines $x_{sd}$ as a function of $\beta$ .

The second derivative of the exponent \eqref{eq:expon} reads
\beq
 S_{xx} - \beta x = S_{xx} - S_x,
\eeq
where we used \eqref{eq:Sx}.
The answer for the Laplace image of $\Psi$ becomes
\beeq\label{eq:Image}
\Phi(\beta) &\simeq& \sqrt\frac{2\pi}{S_x -S_{xx}}  \exp \left\{ S(x_{sd}) - S_x (x_{sd}) \right\}.
\eeeq
\end{subequations}

\subsection{Inverse Laplace transform $\beta \to \nu$}

Now we construct the Laplace image of the P-KNO distribution for an ensemble with ``relative strength" $\rho$ as\footnote{In the book \cite{Book} the factor $1/\rho$ was missed from the argument of the function $\Phi$ of \eqref{eq:Phirho}. This resulted in an erroneous statement about the tail of the distributions.}
\beq\label{eq:Phirho}
\Phi^{(\rho)}(\beta)= \left[\Phi\left(\frac{\beta}{\rho}\right)\right]^\rho.
\eeq

Redefining the integration variable $\beta'/\rho \to \beta$
\beeq\label{eq:Texp}
\Psi^{(\rho)}(\nu) &=& \int \frac{d\beta'}{2\pi i} \, \exp \left(\beta'\nu + \rho\cdot [S-S_x]_{x_{sd}(\beta'/\rho)} \right) 
\cdot\left[ \sqrt\frac{2\pi}{S_x-S_{xx}} \right]^\rho  \nonumber \\
&=& \rho\int \frac{d\beta}{2\pi i} \, \exp \left\{\rho\left(\beta\nu +  [S-S_x]\Big|_{x_{sd}(\beta)} \right)\right\} 
\cdot\left[ \sqrt\frac{2\pi}{S_x-S_{xx}} \right]^\rho .
\eeeq

Now look for the stationary point in $\beta$:
\beq\label{eq:stp}
\nu + \frac{d}{d\beta}[S-S_x] =  \nu + \frac{d \ln {x_{sd}(\beta)}}{d\beta}\,[S_x -S_{xx}] .
\eeq

Using the S.D.\ equation \eqref{eq:Sx} we derive
\beeq
S_x(x_{sd}(\beta)) &= & \beta x_{sd}(\beta)\nonumber \\
\frac{d}{d\beta} S_x(x_{sd}(\beta)) = S_{xx} \frac{d \ln x_{sd}}{d\beta} &= & \frac{d}{d\beta} [ \beta x_{sd}(\beta)] =x_{sd} + \beta x_{sd}\frac{d\ln x_{sd}}{d\beta} \\
\label{eq:sp2}
\frac{d\ln {x_{sd}(\beta)}}{d\beta} &=& \frac{x}{S_{xx} - \beta x_{sd}}.
\eeeq

Substituting into \eqref{eq:stp} leads to
\beq\label{eq:stp2}
\nu + \frac{d}{d\beta}[S-S_x]  =  \nu - \frac{ S_{xx} -S_x }{S_{xx} - \beta x}\cdot x =0.
\eeq
Substituting \eqref{eq:Sx} one more time, we get an expected "comeback":% $\beta_{sp} \cdot x_{sd}(\beta_{sp})$.
\beq\label{eq:nux}
\eqref{eq:stp2} \>\Longrightarrow\>  \frac{\nu}{x} = \frac{ S_{xx} -S_x }{S_{xx} - \beta x_{sd}} \>=\>\frac{ S_{xx} -S_x }{S_{xx} - S_x}=1 .
\eeq

We also need the second derivative of the exponent of \eqref{eq:Texp},
\[
 T \equiv \rho\left(\beta\nu +  [S-S_x]\Big|_{x_{sd}(\beta)} \right) \,:
\]
\beq\label{eq:Tbb}
 T_{\beta\beta} = \rho\frac{d}{d\beta}\left\{\nu - \frac{ S_{xx} -S_x }{S_{xx} - \beta x}\cdot x \right\}
 = \rho\frac{d}{d\beta}\left\{\nu - x \right\}= - x \rho\frac{d\ln {x_{sd}(\beta)}}{d\beta} 
 = -\rho \frac{x^2}{ S_{xx} -S_x }
\eeq
Making use again of
\[
\beta\nu = \beta x \cdot \frac\nu{x} \stackrel{\eqref{eq:Sx}}{=} S_x \cdot\frac\nu{x} 
\stackrel{\eqref{eq:nux}}{=}S_x \cdot 1,
\]
one finally arrives at
\beq
  T = \rho \big( \beta\nu +  [S-S_x] \big) = \rho \cdot S .
\eeq

\subsection{Assembling the pieces}

Assembling the Gaussian factors we arrive at
\beq
\Psi^{(\rho)}(\nu)= \frac\rho{2\pi} \sqrt{\frac{2\pi}{T_{\beta\beta}}} \left[ \sqrt\frac{2\pi}{S_x-S_{xx}} \right]^\rho \cdot e^T
\stackrel{\eqref{eq:Tbb}}{=} \sqrt\rho \>\frac{\big[\nu\Psi(\nu) \big]^\rho}{\nu} \cdot  \left[ \frac{2\pi}{S_x-S_{xx}} \right]^{(\rho-1)/2}
\eeq

Up to now our manipulations were general. 
\smallskip

Now we invoke the structure of $\Psi(\nu)$ to derive
 \beeq
 S&=& -\kappa + 2\ln\kappa + \cO{\kappa^{-1}}, \quad S_x = -\mu\kappa \left(1-\frac2\kappa \right), \quad S_{xx} = -\mu^2\kappa , \nonumber \\
 S_x-S_{xx} &=& \mu(\mu-1)\kappa + 2\mu \>=\> \gamma \mu^2\kappa + 2\mu = \mu^2(\gamma \kappa +2(1\!-\!\gamma)).
 \eeeq
Invoking $\Psi$ from \eqref{eq:Psi_Psi} the answer take the final form
\beq\label{eq:VeryFinalRho2}
\Psi^{(\rho)}(\nu)\>= \>\frac{ \sqrt\rho \mu}{\nu} 
\left( 2\mu\left[\kappa^2-\kappa + \frac1{3\mu^2}\right]\>\chi(\mu\kappa) \right)^\rho\cdot 
e^{-\rho\kappa}\cdot\left( \frac{2\pi}{\gamma \kappa +2(1\!-\!\gamma)} \right)^{(\rho-1)/2} , \qquad \kappa = [D\nu]^\mu .
\eeq
For realistic values of $\gamma\sim 0.4$, we have $\kappa\sim (1.5\nu)^{1.7}$ and the sum in square brackets in the pre-exponent of $\Psi$ is positively definite for $\nu > 0.6$, so no problem for $\nu\ga 1$.

 The alternative version \eqref{eq:newPsi} permits us to descend a little below $\nu=1$ before failing.

\subsection{Analysis of the $\beta$ plane}

Returning to the Laplace image, let us discuss a few interesting points.

Since $\Psi(x)$ falls with $x\to +\infty$ faster than exponentially, the resulting image is regular at any finite $\beta$ in the complex plane. 
This means that $\Phi(\beta)$ has an essential singularity at $\beta=\infty$.

It is straightforward to see by examining the S.D.\ equation
\begin{subequations}
\beeq
 \label{eq:SDeq}
   -\beta x - \mu (Dx)^\mu + 2\mu &=& 0  
\eeeq 
which  determines indirectly the dependence of the S.D.\ point $x_{sd}$ on $\beta$ as  
\beeq
 \label{eq:SDdep}
  \mu\, \frac{ [Dx]^\mu - 2 }{x}\> &=& \> -\beta ,  \qquad x=x_{sd}(\beta).  
\eeeq
\end{subequations}
In general, this relation cannot be solved analytically.
Meanwhile, by taking $x_{sd}$ large, one can solve it approximately:
\[
\mu [Dx]^{\mu-1}\approx - D\beta, \qquad \kappa \equiv [Dx]^{\mu} = \big(- \mu^{-1}D\beta\big)^{\mu/(\mu-1)}
\propto \big(- \beta\big)^{1/\gamma} .
\]
This exercise demonstrates that the function is multi-valued.
When exponentiated, it leads to oscillations at infinity which become wild when the coupling decreases to zero.

In this way a regular function tries to mimic a pole that has to be there in the DLA limit $\gamma=0$.

\subsubsection{Explicit expression for $\Phi(\beta)$}
To see whether our solution reproduces the transition to DLA, and how well it does it, we first have to write down $\Phi$ explicitly.

We have derived above
\beeq\label{eq:Image2}
\Phi(\beta) &\simeq& \sqrt\frac{2\pi}{S_x -S_{xx}}  \exp \left\{ S(x_{sd}) - S_x (x_{sd}) \right\}
\eeeq
with
\[
  S(x) = \ln (x\,\Psi(x) ).
\] 
Having invoked the function $\Psi$ from \eqref{eq:Psi_Psi}, we obtained above
\beeq
S_x &=& -\mu (\kappa -2) \\
S_{xx} &=& -\mu^2 \kappa \\
S_x -S_{xx} &=&  \mu^2( \gamma\kappa + 2(1\!-\!\gamma))
\eeeq
where in the pre-exponent we kept the leading second power of $\kappa$ only.

Now,
\[
\sqrt\frac{2\pi}{S_x -S_{xx}} = \frac{\sqrt{2\pi}}{\mu \sqrt{\gamma\kappa + 2(1\!-\!\gamma) }}
\]
and \eqref{eq:Image2} becomes
\beeq\label{eq:PhiFinal}
\Phi(\beta) &=& \frac{\sqrt{2\pi}\mu}{\sqrt{\gamma\kappa + 2(1\!-\!\gamma) }}\cdot 2\left[\kappa^2-\kappa+\frac1{3\mu^2}\right]
% \Psi(x_{sd}(\beta) 
\cdot e^{(\mu-1)\kappa-2\mu}\chi(\ksd), \qquad \kappa\equiv \kappa (x_{sd}(\beta)).
\eeeq

\beeq\label{eq:Sx2}
S_x(x_{sd}(\beta)) &= & \beta \, x_{sd}(\beta), \qquad S_x(x)\equiv  \frac{d S(x)}{d\ln x} .
\eeeq

\subsubsection{DLA}
At $\gamma=0$ \eqref{eq:SDdep} turns into 
\beq
\frac{  Cx-2}{x}  = -\beta \>  \Longrightarrow \quad  x_{sd} = \frac2{C+\beta}.
\eeq
Substituting in \eqref{eq:PhiFinal} we recover the second-order DLA pole almost perfectly normalized:
\beeq
  \Phi^{[DLA]}(\beta) &\simeq& \sqrt{2\pi}\cdot 2\Big[ \left(\frac{2C}{C+\beta}\right)^2 +\>\ldots \Big] 
  {e^{-2}} \chi(0) \>=\> \frac{2C^2}{(C+\beta)^2} \times \frac{4\sqrt{\pi}}{e^2}, 
   \\
\label{eq:4rpi}    \frac{4\sqrt{\pi}}{e^2} & \approx& \>  0.96 \qquad  \mbox{in place of 1.} 
\eeeq

\subsection{Mellin moments of $\Psi^{(\rho)}$ \label{App:rhomom}}

Invoking the general expression for the P-KNO distribution,
\beeq
\label{eq:PsiRHOApp}    
  \Psi^{(\rho)}(\nu)  &=&  \,\frac{{\sqrt{ \rho}}} {\nu}\,   
\bigg[\> \nu \Psi(\nu) \> \bigg]^\rho
      \cdot  \left( \frac{ \sqrt{2\pi} \, (1-\gamma)} {\sqrt{\gamma[D\nu]^\mu + 2\left[\threehalf\right](1\!-\!\gamma) }}
        \right)^{\rho-1}  ,
 \eeeq
we can use either of the representations,
\begin{subequations}\label{eq:Psi_finApp}
\beeq\label{eq:Psi_chiApp}
 \Psi(\nu) &=&  \frac{2\mu^2}{\nu } \, \left( [D\nu]^{2\mu} -[D\nu]^{\mu} +\frac1{3\mu^2}\right)  e^{-[D\nu]^\mu}   \cdot  \chi(\ksd) ; \\
 \label{eq:Psi_cFApp}
\Psi(\nu) &=&   \frac{ 2\mu^2}{\nu } \left( [D\nu]^{\threehalf\mu} 
- \frac12[D\nu]^{\half\mu} + \frac{1}{3\mu} [D\nu]^{-\half\mu}  \right) e^{-[D\nu]^\mu} \cdot \cF(\ksd) .
 \eeeq 
\end{subequations}
Employing the short notation $\kappa=[D\nu]^\mu$,
\begin{subequations}
\beeq\label{eq:chiverApp}
\bigg[\> \nu \Psi^{(\rho)}(\nu) \> \bigg]_\chi ^\rho &=& \kappa^{2\rho}\cdot e^{-\rho\kappa}  \cdot
\left[ 2\mu^2\chi (\mu\kappa)  \left( 1 - \kappa^{-1} +\frac{ \kappa^{-2}}{3\mu^2} \right) \right]^\rho \cdot G_\chi, \\
\label{eq:cFverApp}
\bigg[\> \nu \Psi^{(\rho)}(\nu) \> \bigg]_\cF ^\rho &=& \kappa^{\threehalf\rho}\cdot e^{-\rho\kappa}  \cdot
\left[ 2\mu^2\cF(\mu\kappa)  \left( 1 -\frac{\kappa^{-1}}2 +\frac{ \kappa^{-2}}{3\mu} \right) \right]^\rho \cdot G_\cF.
\eeeq
\end{subequations}
Here $G$ are the Gaussian multipliers:
\begin{subequations}
\beeq
G_\chi &=& \sqrt\rho  \left( \frac{ \sqrt{2\pi} \, \mu^{-1}} {\sqrt{\gamma\kappa + 2\mu^{-1} }}
        \right)^{\rho-1} , \\
 \label{eq:Gauss_cF}
G_\cF &=& \sqrt\rho  \left( \frac{ \sqrt{2\pi} \, \mu^{-1}} {\sqrt{\gamma\kappa + \threehalf\mu^{-1} }}.
        \right)^{\rho-1}
\eeeq
\end{subequations}
To calculate the Mellin integral,
\beq
g_{k}^{(\rho)}  \propto \int_0^\infty \frac{d\nu}{\nu}\, \nu^{k} \> \big[\nu\Psi^{(\rho)}(\nu)\big] ,
\eeq
we introduce $t=\rho \kappa$ as integration variable, 
\[
\nu = \frac1D \left(\frac{t}{\rho}\right)^{1/\mu}, \quad \frac{dt}{t} = \mu\, \frac{d\nu}{\nu}. % \qquad \nu = 
\]

\subsubsection{$\chi$ version. DLA}
For the $\chi$ version \eqref{eq:chiverApp} the Mellin integral produces
\begin{subequations} \label{eq:gqmm}
\beeq
g_{k}^{(\rho)} \>& = \> &\mu^{-1}\, \frac{(2\mu^2\chi / \rho^2)^{\rho}} { [D\rho^{1/\mu}]^{k} }\> \Gamma\big(k\mu^{-1} + 2\rho \big)
 \cdot \left( 1-\frac{\rho} {t_0} + \frac{\rho^2}{3\mu^2 t_0^2}\right)^\rho \cdot G_\chi 
\eeeq
with $t_0$ the steepest descent point
\beq \label{eq:gqmSD}
t_0 \>\simeq\>  k \mu^{-1}  + 2 \rho .
\eeq
\end{subequations}
This representation is useful to analyse the DLA limit of small coupling, $\gamma \kappa \equiv \gamma[D\nu]^\mu \ll 1$.
In this regime, the factor $G_\chi$ reduces to a constant and does not play a role in determining the dependence of the multiplicity moments on the rank $k$. Setting $\gamma=0$ ($\mu=1, D=C$) we get
\[
G_\chi \approx \sqrt\rho\, {\pi}^{\half(\rho-1)}, \qquad t_0 \simeq k   + 2\rho,
\]
\def\DLA{\mbox{\scriptsize DLA}}
\beeq\label{eq:DLAE30}
g_{\DLA}^{(\rho)}(k) \>&\simeq& \> \frac{2^\rho}{C^k \, \rho^{k+2\rho}} \>\Gamma(k+2\rho)\cdot 
\left( 1-\frac{\rho} {t_0} + \frac{\rho^2}{3 t_0^2}\right)^\rho \cdot \sqrt\rho\, {\pi}^{\half(\rho-1)} \nonumber \\
&=& \frac{2\cdot  \rho^{-k}} {C^k}\cdot \Gamma(k+2\rho) \> \left( 1-\frac{\rho} {t_0} + \frac{\rho^2}{3 t_0^2}\right)
\cdot \left[ 2\sqrt\pi\left( 1-\frac{\rho} {t_0} + \frac{\rho^2}{3 t_0^2}\right) \right]^{\rho-1} {\rho^{-2\rho +\half}}.
\eeeq
This expression {\em almost}\/ passes the check for $\rho=1$, at which point it has to reproduce the DLA gluon jet moments \eqref{eq:ek}.
Indeed, observing that 
\[
\Gamma(k+ 2\rho) = (k+2\rho-1)\cdot \Gamma(k+ 2\rho-1),
\]
we get
\[
(k+2\rho-1)\cdot \left( 1-\frac{\rho} {k+2\rho} + \frac{\rho^2}{3 (k+2\rho)^2}\right) \simeq k\cdot  \left( 1 + (\rho-1) + \frac{\rho}{k^2}\left[1+\frac\rho3\right] \right).
\]
Setting $\rho=1$ we reproduce the gluon multiplicity moments,
\[
g_{\DLA}(k) = \frac{2\, k\, k!}{C^k} \bigg(1+\cO{k^{-2}} \bigg),
\]
{\em modulo}\/ the $\cO{k^{-2}}$ correction term.

Evaluating the answer on the basis of the steepest descent and stationary phase approximations left its trace: the leading and first subleading $1/k$ terms are reproduced, but the $1/k^2$ correction is not.

\medskip
For the sake of completeness, in the following tables the values of the first few multiplicity moments as predicted by 
eq.\eqref{eq:DLAE30} are compared with the exact values that follow from the DLA recurrence relations.

\paragraph{\boldmath quark jet ($\rho=\twothird$)}
\begin{center}
\begin{tabular}{c||c|c||c}
&  recurrence & numeric   & eq. (E.30)    \\ \hline
$g_0$    &  1 &   1.000 &  1.147 \\[1mm]
$g_1$    &  1 &   1.000 &  1.054 \\[1mm]
$g_2$   & $ \frac32$ & 1.500 & 1.538  \\[1mm]
 $g_3$  &  $\frac{49}{16}$ & 3.063  &3.113  \\ [1mm]
 $g_4$  & $\frac{319}{40}$ & 7.975  &  8.090  \\[1mm]
$g_5$  & $\frac{3243}{128} $ & 25.336 & 25.708
\end{tabular}
\end{center}

\paragraph{\boldmath $e^+e^-$ ($\rho = \fourthird$)}
%\medskip
\begin{center}
\begin{tabular}{c||c|c||c}
&  recurrence & numeric   & eq. (E.30)    \\ \hline
$g_0$    &  1 &   1.000 &  1.199 \\[1mm]
$g_1$    &  1 &   1.000 &     1.154 \\[1mm]
$g_2$   & $ \frac54$ & 1.250 & 1.394 \\[1mm]
 $g_3$  &  $\frac{121}{64}$   &1.891  &  2.057 \\ [1mm]
 $g_4$  & $\frac{1079}{320}$   & 3.372   & 3.603\\[1mm]
$g_5$  & $\frac{14227}{2048} $  & 6.947  & 7.324
\end{tabular}
\end{center}

The moment $g_0$ (normalization of the distribution) is understandably off. Otherwise, the agreement looks reasonable given that the tail-oriented approach developed in this paper was not aimed at describing low-rank multiplicity moments accurately.

\subsubsection{$\cF$ version. RSE squeezing}

In the opposite regime $\gamma \kappa \gg 1$ the $\cF$ version of the distribution  better suits the task.
Moreover, in this situation the Gaussian multiplier \eqref{eq:Gauss_cF} affects the answer:
\beq
G_\cF \>\simeq\> \left(\frac{t}{\rho}\right)^{-\half(\rho-1)} \cdot \sqrt\rho \left[ \sqrt{2\pi\gamma^{-1}} \,\mu^{-1} \right]^{\rho-1}.
\eeq
The result follows
\beeq \label{eq:gqmm2}
g_{k}^{(\rho)} \>& = \> &\mu^{-1}\,\frac{(2\mu^2\cF/ \rho)^{\rho}} { [D\rho^{1/\mu}]^{k} }\> \Gamma\big(k\mu^{-1} + \rho +\half \big)
 \cdot \left( 1-\frac{\rho} {2t_0} + \frac{\rho^2}{3\mu t_0^2}\right)^\rho \cdot\left[ \sqrt{2\pi\gamma^{-1}} \,\mu^{-1} \right]^{\rho-1}.
 %\nonumber \\
% &=& \frac{\rho^{-k} }{D^{k}} \cdot \rho^{\gamma k} \cdot { \Gamma\big({(1-\gamma)k +   \rho +\half}\big)} 
% \cdot \left( 1-\frac{\rho} {2t_0} + \frac{\rho^2}{3\mu t_0^2}\right)^\rho 
\eeeq
with $t_0$ the steepest descent point, now positioned at
\[
t_0 \>\simeq\>  k  \mu^{-1} + \rho +\half.
\]
We have used this form of the multiplicity moments of a quark jet in the discussion of the typical gluon energy in RSE:
\beeq
g_{q-m}^{(\rho)} \>& \propto\> & \frac{\rho^{m-q} }{D^{q-m}} \cdot \rho^{\gamma (q-m)} \cdot { \Gamma\big({(1-\gamma)(q-m) + %\threehalf 
 \rho +\half}\big)} 
 \cdot \left( 1-\frac{\rho} {2t_0} + \frac{\rho^2}{3\mu t_0^2}\right)^\rho 
\eeeq
This version is not suited for evaluating low rank moments as it is applicable in the asymptotic regime $\gamma k\gg 1$. 

\vspace{1 cm}
% #########################################################

\mysection{Multiplicity and correlations in RSE}

Since we have a toy model, let's play. 
Let us estimate in a more quantitative manner what snake-tongue configurations, RSE, bring into the game. 
\medskip

\def\RSE{\mbox{\scriptsize RSE}}

Ascribing to the gluon and quark subjets the hardness scales
\beq
Q_g = z_g Q, \quad Q_q = z_q Q = (1-z_g) Q\,,
\eeq
with $Q$ representing the full hardness of the CMS jet, $Q=2*550\,\GeV*\sin (R/2)=430\,\GeV$ (resulting in $\gamma=0.38$),
gives the following rough  estimate of the total RSE multiplicity:
\beq
  \lrang{N_{\RSE}(Q)} = N_q\big((1-z_g)^2 Q\big) + N_g\big(z_g^2 Q\big) \simeq
 \left[ (1-z_g)^{\gamma} +\rho_q^{-1}\,z_g^{\gamma} \right]    \cdot N_q\big( Q\big) .
\eeq
The snake multiplicity relative to a single-quark jet increases steadily with $z_g$ until it reaches its maximum value of almost 2 at very large $z_g\simeq \twothird$.
% At $z_g = \third$ as in Fig.~16 the multiplicity gain is also not small: a factor of 1.39. 
For $z_g = \quart$ which point, according to the estimate in Fig. 13, the RSE reaches when $\nu\simeq3$, the overall snake-generated multiplicity of hadrons is already $80\%$ greater than $n_q$.
The excess multiplicity yield is welcome: It is what the rattlesnakes were called for.

\bigskip

A simple analysis of inclusive measurement of two particles allows splitting the second factorial multiplicity moment into two parts,
\beq
 \lrang{N_{\RSE}(N_{\RSE}-1)} = \left. \Big[ \lrang{N_{\RSE}(N_{\RSE}-1)} - 2N_qN_g   \Big]\right|_{same}
 \>+\> \left. \Big[ 2N_qN_g   \Big]\right|_{away} .
\eeq
The away-side correlation stays large ($\ga 30\%$) over the entire range of $z_g$ and reaches its  maximal value
\beq
\frac{\mbox{away}}{\mbox{total}} \>=\> \frac{2N_q\,N_g}{\lrang{N_{\RSE}}^2} \>=\> \frac12  
\eeq
at $z_g\simeq \quart$ ($\nu=3$).

 \end{document}